\documentclass[useAMS,usenatbib]{mn2e} 
\usepackage{natbib}
\usepackage{graphicx}
\usepackage{setspace}
\usepackage{mathtools}
\usepackage{pdflscape, longtable}
\usepackage{breqn}
\usepackage[toc,page]{appendix}
\renewenvironment{thebibliography}[1]{%
\begin{oldthebibliography}{#1}%
\setlength{\itemsep}{-.3ex}%
}%
{%
\end{oldthebibliography}%
}

\newcommand{\UCHII}{UCH\,\textsc{ii}}
\newcommand{\HII}{H\,\textsc{ii}}

\def \arcsec {$^{\prime\prime}$}
\def \arcmin {$^\prime$}

\usepackage{multirow}

\usepackage[labelfont=it,textfont={bf,it}]{caption}
 
\title[The JCMT Gould Belt Survey: Evidence for radiative heating and contamination in the W40 complex.]{The JCMT Gould Belt Survey: Evidence for radiative heating and contamination in the W40 complex.}

\author[D. J. Rumble]{D. Rumble$^{1}$, J. Hatchell$^{1}$, K. Pattle$^{2}$, H. Kirk$^{3}$, T. Wilson$^{1}$, J. Buckle$^{4, 5}$, D.S. Berry$^{6}$, \newauthor H. Broekhoven-Fiene$^{7}$, M.J. Currie$^{6}$, M. Fich$^{8}$, T. Jenness$^{6, 9}$, D. Johnstone$^{6, 3, 7}$, \newauthor J.C. Mottram$^{10, 11}$, D. Nutter$^{12}$, J.E. Pineda$^{13, 14, 15}$, C. Quinn$^{12}$, \newauthor C. Salji$^{4, 5}$, S. Tisi$^{8}$, S. Walker-Smith$^{4, 5}$, J. Di Francesco$^{3, 7}$, M.R. Hogerheijde$^{10}$, \newauthor D. Ward-Thompson$^{2}$, P. Bastien$^{16}$, D. Bresnahan$^{2}$, H. Butner$^{17}$, \newauthor M. Chen$^{7}$, A. Chrysostomou$^{18}$, S. Coude$^{16}$, C.J. Davis$^{19}$, E. Drabek-Maunder$^{20}$, \newauthor A. Duarte-Cabral$^{1}$, J. Fiege$^{21}$, P. Friberg$^{6}$, R. Friesen$^{22}$, G.A. Fuller$^{14}$, \newauthor S. Graves$^{6}$, J. Greaves$^{12}$, J. Gregson$^{23, 24}$, W. Holland$^{25, 26}$, G. Joncas$^{27}$, \newauthor J.M. Kirk$^{2}$, L.B.G. Knee$^{3}$, S. Mairs$^{7}$, K. Marsh$^{12}$, B.C. Matthews$^{3, 7}$, G. Moriarty-Schieven$^{3}$, \newauthor C. Mowat$^{1}$, J. Rawlings$^{28}$, J. Richer$^{4, 5}$, D. Robertson$^{29}$, \newauthor E. Rosolowsky$^{30}$, S. Sadavoy$^{31}$, H. Thomas$^{6}$, N. Tothill$^{32}$, S. Viti$^{28}$, G.J. White$^{23, 24}$, \newauthor J. Wouterloot$^{6}$, J. Yates$^{28}$, M. Zhu$^{33}$\\
$^{1}$Physics and Astronomy, University of Exeter, Stocker Road, Exeter EX4 4QL, UK\\
Remaining affiliations are listed at the end of the paper
\\
}
\begin{document}

\date{Accepted 2015. Received 2015.}

\pagerange{\pageref{firstpage}--\pageref{lastpage}} \pubyear{2015}

\maketitle 

\label{firstpage}

\begin{abstract}

We present SCUBA-2 450\,$\micron$ and 850\,$\micron$ observations of the W40 complex in the Serpens-Aquila 
region as part of the James Clerk Maxwell Telescope (JCMT) Gould Belt Survey (GBS) of nearby 
star-forming regions. 
We investigate radiative heating by constructing temperature maps from the ratio of SCUBA-2 fluxes 
using a fixed dust opacity spectral index, $\beta$ = 1.8, and a beam convolution kernel to achieve a 
common 14.8\arcsec\ resolution. 
We identify 82 clumps ranging between 10 and 36\,K with a mean temperature of 20$\pm$3\,K. Clump temperature is strongly correlated with proximity to the external OB association and there is no evidence that the embedded protostars significantly heat the dust. We identify 31 clumps that have cores with densities greater than 10$^{5}$cm$^{-3}$. Thirteen of these cores contain embedded Class 0/I protostars. Many cores are associated with bright-rimmed clouds seen in \emph{Herschel} 70\,$\micron$ images. From JCMT HARP observations of the $^{12}$CO 3\hbox{--}2 line, we find contamination of the 850\,$\micron$ band of up to 20 per cent. We investigate the free-free contribution to SCUBA-2 bands from large-scale and ultracompact \HII\ regions using archival VLA data and find the contribution is limited to individual stars, accounting for 9 per cent of flux per beam at 450\,$\micron$ or 12 per cent at 850\,$\micron$ in these cases. 
We conclude that radiative heating has potentially influenced the formation of stars in the Dust Arc sub-region, favouring Jeans stable clouds in the warm east and fragmentation in the cool west.

\end{abstract}



\section{Introduction}


Understanding the impact of heating, via feedback, is of vital importance for the wider 
inquiry into what mechanisms govern the behaviour of molecular clouds \citep{Jeans:1902dz}. 
Feedback occurs, via internal mechanisms, from radiative heating by the stellar photosphere and accretion luminosity \citep{calvet98} of young stellar objects (YSOs). Molecular outflows and shocks \citep{Davis:1999ly} may also radiatively heat a cloud to a lesser extent. External sources of heating include photons produced by stars, which can drive strong stellar winds \citep{Canto:1984dq, Ziener:1999kl, Malbet:2007zr} and H\,\textsc{ii} regions \citep{Koenig:2008jo, Deharveng:2012fk}, as well as the interstellar radiation field \citep*[ISRF;][]{Mathis:1983dq, Shirley:2000fk, Shirley:2002kx}. Simulations, including those by \cite{Bate:2009uq}, \cite{Offner:2009pt}, and \cite{Hennebelle:2011ly}, have suggested that internal radiative feedback can suppress 
cloud fragmentation, leading to higher mass star-formation, whereas observations by 
\cite{Rumble:2015vn} provided evidence that external heating can influence the 
evolution of star-forming clouds. 

\begin{figure}
\begin{centering}
\includegraphics[scale=0.42]{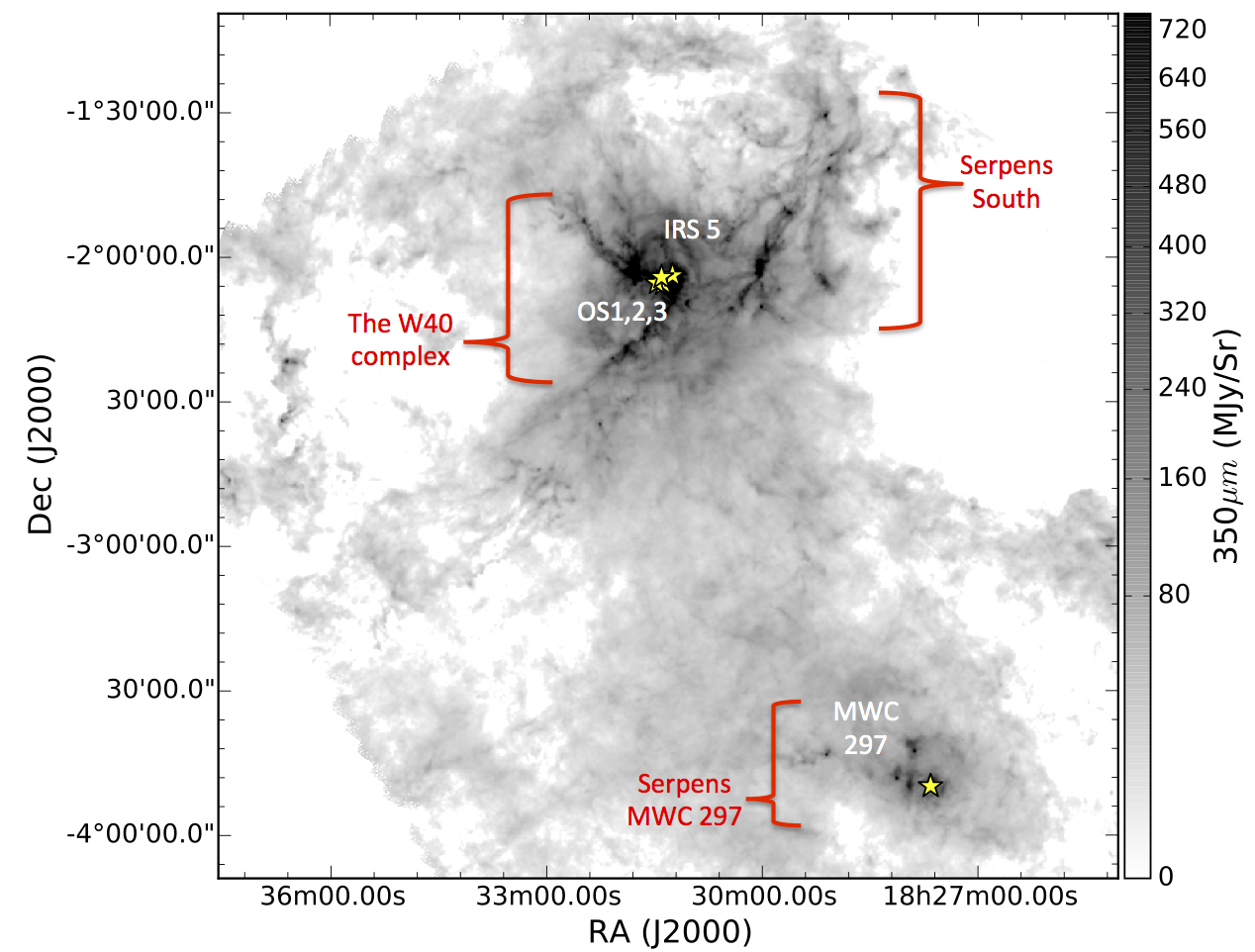}
\caption{\emph{Herschel} 350\,$\micron$ map of the Aquila Rift including the W40 complex, Serpens South and Serpens MWC 297. OB stars are marked with yellow crosses and labelled Bontemps et al. (2010).} 
\label{fig:findingchart}
\end{centering}
\end{figure} 


Fluxes of cool YSOs observed at longer wavelengths may appear on the Rayleigh-Jeans tail of the continuum, where temperature cannot be calculated. Use of the shorter SCUBA-2 450\,$\micron$ band allows for dust temperatures up to 35\,K to be reliably calculated for an opacity modified grey-body model fit to a flux ratio. Flux ratios have been calculated by \cite{Mitchell:2001ve}, \cite{Reid:2005ly}, \cite{Hatchell:2013ij}, and \cite{Rumble:2015vn} for Submillimetre Common-User Bolometer Array (SCUBA, \citeauthor{Cunningham:1994bh} \citeyear{Cunningham:1994bh}) and SCUBA-2 \citep{Holland:2013fk} bands. The flux ratio method does not compromise on the high resolution of the JCMT (14.6\arcsec) but does introduce an inherent degeneracy between temperature and the dust opacity index, $\beta$, requiring an assumption in either case \citep{Shetty:2015zr}.

This study uses data from the JCMT Gould Belt Survey (GBS) of nearby star-forming 
regions \citep{WardThompson:2007ve} to measure dust temperatures. The full survey 
maps all major low- and intermediate-mass star-forming regions within 0.5\,kpc observable 
from the JCMT with the continuum bolometer array SCUBA-2 (\citeauthor{Holland:2013fk} 
\citeyear{Holland:2013fk}). The JCMT GBS provides some of the most sensitive maps of 
star-forming regions where $A_V > 3$ with a target sensitivity of 3\,mJy\,beam$^{-1}$ at 
850\,$\micron$ and 12\,mJy beam$^{-1}$ at 450\,$\micron$. The 9.8\,\arcsec\ (450\,$\micron$) 
and 14.6\,\arcsec\ \citep[850\,$\micron$;][]{Dempsey:2013uq} resolutions of the JCMT 
allow for detailed study of structures such as filaments and protostellar envelopes down 
to the Jeans length.

We focus on the W40 complex (presented in Figure \ref{fig:findingchart}). The neighbouring Serpens South filament is thought to be part of the Aquila Rift ($255\pm55\hbox{ pc}$ \citealt{Straizys:2003nx}). \cite{Bontemps:2010fk} and \cite{Maury:2011ys} therefore conclude a physical association with Serpens South on account of proximity. However, \cite{Kuhn:2010kl} calculates a distance of 600$\hbox{ pc}$ via fits to the X-ray luminosity function. \cite{Shuping:2012ly} construct SEDs from IR data of bright objects in the W40 complex and estimate a distance between 455$\hbox{ pc}$ and 536$\hbox{ pc}$. We use a mean distance based on these calculations of $500\pm50\hbox{ pc}$, following \cite{Radhakrishnan:1972fk}, and \cite{Mallick:2013kx}. The W40 complex is therefore assumed to be spatially separated from the Serpens South region \citep{Straizys:2003nx, gutermuth08}.

The W40 complex is a site of high-mass star-formation associated with a cold molecular cloud 
\citep{Dobashi:2005uq} and includes a blistered H\,\textsc{ii} region \citep{Westerhout:1958uq} 
powered by an OB association \citep{Zeilik:1978qf, Smith:1985bv}. The OB association is 
comprised of IRS/OS1a (O9.5), IRS/OS2b (B4) and IRS/OS3a (B3) and an associated stellar cluster of 
pre-main-sequence (PMS) stars that are detected in the X-ray by \cite{Kuhn:2010kl}. OS1aS 
is the primary ionising source of the H\,\textsc{ii} region that was detected in the radio via 
free-free emission \citep{Vallee:1991zr}. The OB association drives the formation of the larger 
nebulosity Sh2-64 \citep{Sharpless:1959hc}. The W40 complex is detected in the submillimetre 
by the \emph{Herschel} Space Telescope \citep{Andre:2010kx, Menshchikov:2010kl, Konyves:2010oq, 
Bontemps:2010fk, Konyves:2015uq} with three significant filaments (W40-N, W40-S and 
the Dust Arc). \cite{Rodney:2008ij} present a further review of the W40 complex. 


In addition to the SCUBA-2 data for Aquila, we make use of JCMT HARP $^{12}$CO 3\hbox{--}2 
line emission observed over an area 30 times larger than that observed by \cite{van-der-Wiel:2014vn}, 
and remove the contribution from the CO line emission to the 850\,$\micron$ SCUBA-2 maps. We also make use of archival 
VLA 21\,cm data (45\arcsec\ resolution; \citeauthor{Condon:1998kx} \citeyear{Condon:1998kx}) and assess the impact of the larger scale free-free emission contribution to the SCUBA-2 bands. Furthermore, we use archival AUI/NRAO 3.6\,cm data alongside the \cite{Rodriguez:2010bs} 3.6\,cm catalogue of compact radio sources to assess the impact of smaller scale free-free emission contribution to the SCUBA-2 bands. Finally, we complement our findings with 70\,$\micron$ observations from the \emph{Herschel} archive.  



This paper is structured as follows. In Section 2, we describe the observations of the Serpens-Aquila region with SCUBA-2 and HARP. In Sections 3 and 4, we describe the methods by which the contributions of CO line and free-free continuum emission was removed from SCUBA-2 observations. In Section 5, we apply our method for producing temperature maps. In Section 6, we identify clumps from SCUBA-2 data and calculate their properties. In Section 7, we discuss the evidence for radiative feedback influencing the evolution of clumps and the formation of stars in the W40 complex.


\section{Observations and Data Reduction}

\subsection{SCUBA-2}

\begin{figure*}
\begin{centering}
\includegraphics[scale=1.1]{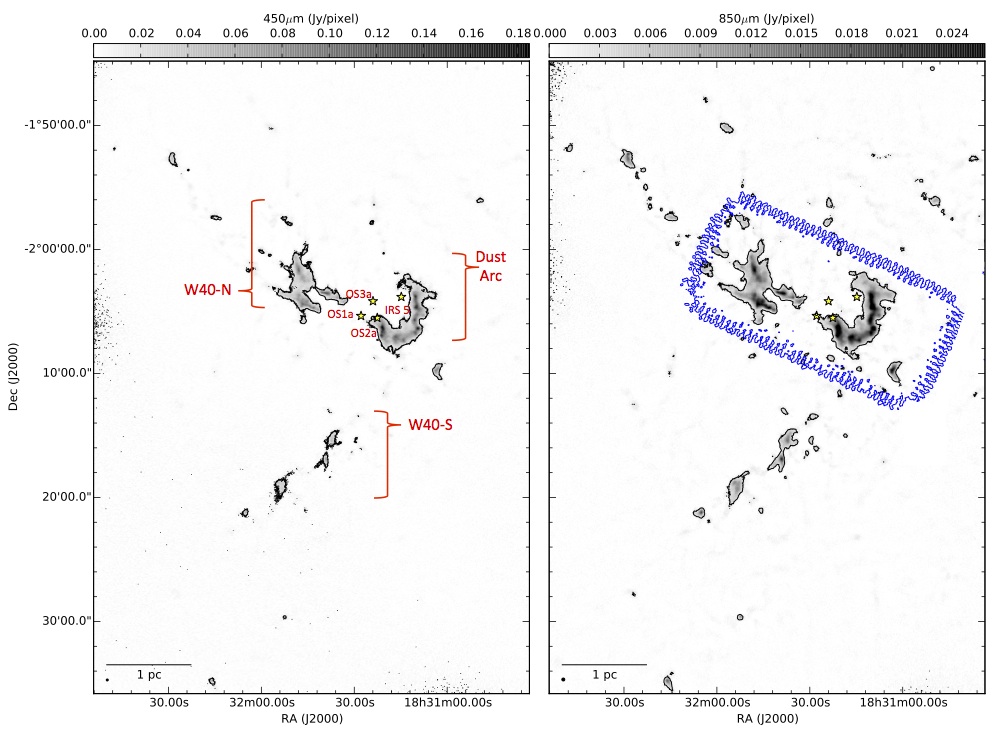}
\caption{SCUBA-2 450\,$\micron$ (\emph{left}) and 850\,$\micron$ (\emph{right}) data. The SCUBA-2 850\,$\micron$ data have had contaminating CO emission covering an 6\arcmin43\arcsec\ $\times$17\arcmin54\arcsec\  area removed (region outlined in blue, see Section 3.1 for more details). The resulting maps have been further filtered to remove structure above 4\arcmin\ in size (see Section 2.1). The contours show 5\,$\sigma$ levels in both cases: levels are at 0.0173\,Jy/ 2\arcsec\ pixels and 0.0025\,Jy/ 3\arcsec\ pixels  at 450\,$\micron$ and 850\,$\micron$ respectively.} 
\label{fig:maps}
\end{centering}
\end{figure*} 

Aquila was observed with SCUBA-2 \citep{Holland:2013fk} between the 21st of April and 5th of July 2012 as part of the JCMT GBS MJLSG33 SCUBA-2 Serpens Campaign. Four separate fully sampled 30\arcmin\ diameter circular continuum observations (PONG1800 mapping mode, \citeauthor{Kackley:2010zr} \citeyear{Kackley:2010zr}) were taken simultaneously at 850\,$\micron$ and 450\,$\micron$, and subsequently combined into mosaics. The beam sizes in the two bands are 9.8\,\arcsec\ (450\,$\micron$) 
and 14.6\,\arcsec\ (850\,$\micron$). The 450 and 850\,$\micron$ maps for the entire Aquila W40 / Serpens South area covered by SCUBA-2 are shown in Appendix A (Figure \ref{fig:SCUBA-2}) along with the data reduction masks and variance maps. The spatially-filtered 450 and 850\,$\micron$ mosaics for the W40 region are shown in Figure \ref{fig:maps}; the 850\micron\ emission also has CO contamination removed (see Sect. 3). The dates, central positions and weather conditions of the observations are listed in Table \ref{tab:obs}.

\begin{table*}
\centering
\caption{SCUBA-2 observations of Aquila}
\label{tab:obs}
\begin{tabular}{llllll}
PONG   & RA Dec                 & \# of & Weather    & Observation dates                       & Mean standard deviation        \\
   & (J2000)                 & Obs.  & band(s)   &                                         & (Jy per pixel)       \\
   \hline
NE & 18:31:34.6 -01:54:05.30 & 4     & 1       & 21st, 23rd April, 3rd May 2012          & 4.7$\times$10$^{-4}$ \\
NW & 18:29:30.6 -01:47:30.30 & 4     & 1, 2    & 3rd, 4th, 5th May 2012                  & 4.5$\times$10$^{-4}$ \\
SE & 18:32:13.8 -02:24:12.30 & 6     & 1, 2, 3 & 8th May, 10th, 11th June, 5th July 2012 & 4.0$\times$10$^{-4}$ \\
SW & 18:30:09.8 -02:17:37.30 & 4     & 1       & 7th, 8th, 18th May 2012                 & 4.6$\times$10$^{-4}$
\end{tabular}
\end{table*}

The data were reduced as part of the GBS Legacy Release 1 (LR1, \citeauthor{Mairs:2015vn} 
\citeyear{Mairs:2015vn}) using an iterative map-making technique (\texttt{makemap} 
in {\sc smurf}, \citeauthor{Chapin:2013vn} \citeyear{Chapin:2013vn}), and gridded to a 3\arcsec\ 
pixel grid at 850\,$\micron$ and a 2\arcsec\ pixel grid at 450\,$\micron$. The iterations 
were halted when the map pixels, on average, changed by $<$0.1\,per cent of the 
estimated map rms noise. The initial reductions of each individual scan were coadded 
to form a mosaic from which a signal-to-noise ratio (SNR) mask was produced for each region. 
Masks were selected to include regions of emission in the automask reductions 
with SNRs higher than 3 with no additional smoothing. 

The final mosaic was produced from a second reduction using this mask to define 
areas of emission. Detection of emission structure and calibration accuracy (see below) 
are robust within the masked regions and are uncertain outside of the masked region 
\citep{Mairs:2015vn}. 

A spatial filter of 600\arcsec\ was used in the reduction, which means that within 
appropriately sized masks flux recovery is robust for sources with Gaussian Full 
Width Half Maximum (FWHM) sizes less than 2.5\arcmin. Sources between 
2.5\arcmin\ and 7.5\arcmin\ in extent will be detected, but both the flux and the 
size are underestimated because Fourier components with scales greater than 
5\arcmin\ are removed by the filtering process. Detection of sources larger than 
7.5\arcmin\ is dependent on the mask used for reduction.

The data were initially calibrated in units of pW and are converted to 
Jy per pixel using Flux Conversion Factors (FCFs) derived by \cite{Dempsey:2013uq} 
as FCF$_{\mathrm{arcsec}}$ = 2.34 $\pm$ 0.08\,pW$^{-1}$\,arcsec$^{-2}$ 
and 4.71 $\pm$ 0.5 Jy\,pW$^{-1}$\,arcsec$^{-2}$ at 850\,$\micron$ and 
450\,$\micron$ respectively. The calibration uncertainties on the standard FCFs 
are 3\% at 850\,$\micron$ and 11\% at 450\,$\micron$.  For ratios and temperatures 
derived from these data, it is not the uncertainties at each wavelength but the 
uncertainties on the calibration ratio that matter. Due to correlations between the 
450\,$\micron$ and 850\,$\micron$ FCF measurements, the errors do not propagate 
simply. For a single scan, the calibration ratio is FCF$_{450}$/FCF$_{850}$ = 2.04 
$\pm$ 0.49 (J. Dempsey, priv. comm.).  The SCUBA-2 mosaics for Aquila 
were made with at least four scans per region. Assuming these to be randomly 
drawn from the distribution of calibration ratios, the uncertainty on the ratio 
reduces to 2.04 $\pm$ 0.25 or a calibration uncertainty of 12.5\%.

The PONG scan pattern leads to lower noise in the map centre and overlap regions, 
while data reduction and emission artefacts can lead to small variations in the noise 
over the whole map. Due to varying conditions over the observing periods, the noise 
levels are not consistent across the mosaic. Typical noise levels are 3.5 mJy/pix or 
0.50 mJy/pix per 2\arcsec or 3\arcsec\ pixel at 450\,$\micron$ and 850\,$\micron$, 
respectively. 

After masking, we found that the SCUBA-2 data reduction process was removing less 
large scale structure at 450\,$\micron$, relative to 850\,$\micron$. As a result, a 
significant number of pixels had flux ratio values that would lead to unphysically high 
temperatures (defined as ratios higher than 9.5 which 
correspond to temperatures greater than 50\,K). Flux ratio and temperature 
are related through the `temperature equation' that is given below in Section 5.
The details of flux ratio preparation are given in Appendix A. To ameliorate 
this problem, a further spatial filter was applied to the data, the details of which 
are provided in Appendix B. In summary, a scale size of 4\arcmin\ at both 450 
and 850\,$\micron$ was found to be the optimal solution and the SCUBA-2 
maps were filtered accordingly. 

\subsection{HARP}

\begin{figure*}
\begin{centering}
\includegraphics[scale=0.95]{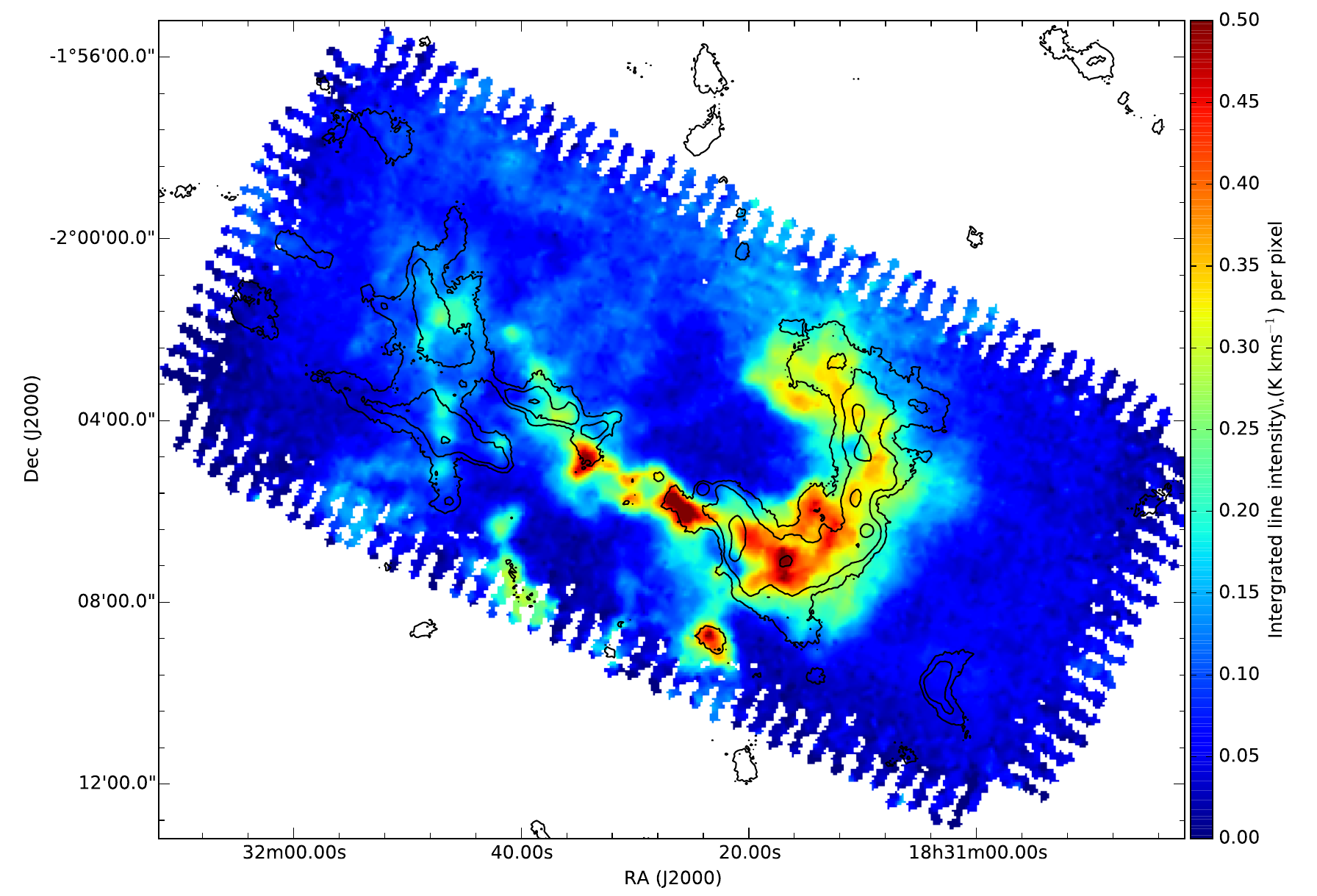} \includegraphics[scale=0.67]{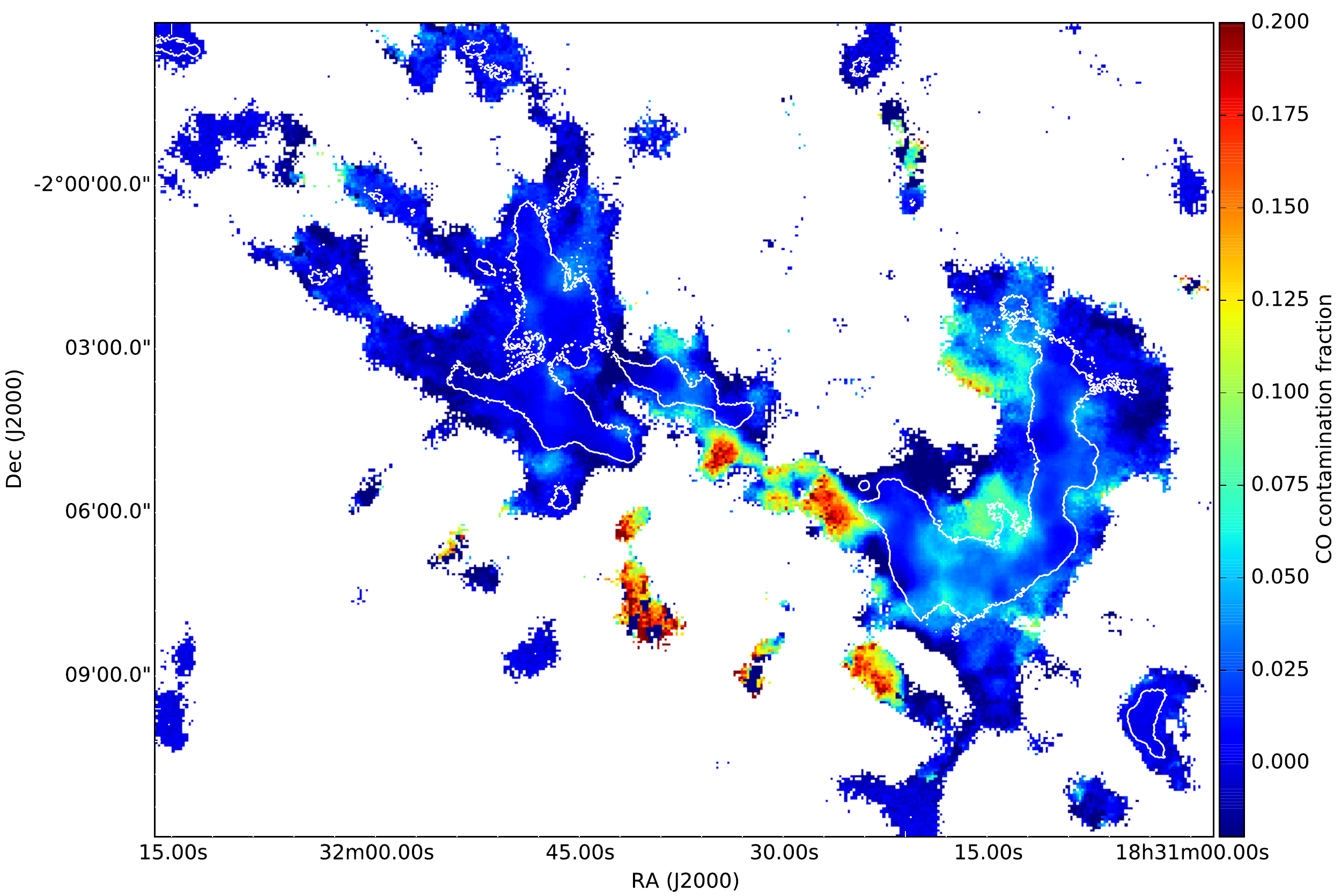}
\caption{\textbf{Upper} $^{12}$CO 3\hbox{--}2 integrated intensity map over the entire range (from -90 to +100\,km s$^{-1}$) of the central region of the W40 complex. Contours show SCUBA-2 850\,$\micron$ emission at the 5\,$\sigma$, 15\,$\sigma$ and 50\,$\sigma$ levels. \textbf{Lower} the fraction of SCUBA-2 850\,$\micron$ that can be attributed to $^{12}\textrm{CO}$ 3\hbox{--}2 345\,GHz line emission. The SCUBA-2 data are masked at 3\,$\sigma$ and the 5\,$\sigma$ level is shown by the white contour.} 
\label{fig:CO}
\end{centering}
\end{figure*}

Archival HARP $^{12}\textrm{CO}$ 3\hbox{--}2 data \citep{van-der-Wiel:2014vn} confirms 
the presence of red- and blue-shifted gas in the Dust Arc (Figure \ref{fig:maps}) but coverage 
is limited to a 2\arcmin\ $\times$ 2\arcmin\ region centred on the local peak of the submilimetre 
emission, and therefore we commissioned an extended survey of the W40 complex in 
$^{12}\textrm{CO}$ 3\hbox{--}2 that included the whole of the Dust Arc and W40-N, as 
presented in Figures \ref{fig:maps} and \ref{fig:CO} upper. Subsequent to our observations, 
\cite{Shimoikura:2015kx} published maps of the W40 complex from Atacama Submillimeter Telescope 
Experiment (ASTE) observations in $^{12}\textrm{CO}$ 3\hbox{--}2 and HCO$^{+}$ 
4\hbox{--}3 with a similar coverage, but at a lower effective resolution of 
22\arcsec\ compared to the JCMT (14.6\arcsec).

Aquila was observed with HARP (Heterodyne Array Receiver Programme, 
\citeauthor{Buckle:2009fk} \citeyear{Buckle:2009fk}) on the 4th of July 2015 as part of the 
M15AI31``active star-formation in the W40 complex" proposal. The main beam efficiency, 
$\eta_{\mathrm{MB}}$, taken from the JCMT efficiency archive is 0.61 at 345\,GHz. Two 
sets of four basket-weaving scan maps were observed over an approximately 7\arcmin 
$\times$18\arcmin\ area (position angle = 65${\degr}$) at 345.796\,GHz to observe the 
$^{12}$CO 3\hbox{--}2 line. A sensitivity of 0.3\,K was achieved on 1\,km s$^{-1}$ velocity 
channels in weather Grade 4 ($\tau_{225}$ = 0.16). Maps were referenced against an 
off-source position at RA(J2000) = 18:33:29.0, Dec.(J2000) = -02:03:45.4, which had been 
selected as being free of any significant CO emission in the \cite{Dame:2001qf} CO 
Galactic Plane Survey. 

The observed cube has two distinct velocity components at 5 and 10\,km\,s$^{-1}$ that 
are consistent with the observations of \cite{Zeilik:1978qf} and \cite{Shimoikura:2015kx}. 
A third component at 7\,km\,s$^{-1}$ is detected in  observations of the HCO$^{+}$ 4\hbox{--}3 
line. \cite{Shimoikura:2015kx} suggest that $^{12}$CO 3\hbox{--}2 is heavily affected by 
self-absorption by this third cloud component, making a full analysis of velocity structure 
of the W40 complex challenging.   

The data were first reduced using the {\sc smurf} \texttt{makecube} technique 
\citep{Jenness:2015fk}. An integrated intensity map, corrected for main beam efficiency, 
was produced by collapsing along the entire velocity range and subsequently run through 
the SCUBA-2 data reduction pipeline with the effect of filtering out scales larger than 
5\arcmin\ as well as regridding to 3\arcsec\ pixels. Figure \ref{fig:CO} upper presents the reduced 
$^{12}$CO 3\hbox{--}2 integrated intensity map for the W40 complex. 

The bulk of the $^{12}$CO 3\hbox{--}2 gas coexists with the brightest dust observed by SCUBA-2, but a bright filament of $^{12}$CO 3\hbox{--}2 emission is observed between W40-N and the Dust Arc with no corresponding SCUBA-2 emission (see Figure \ref{fig:CO}, upper panel). This is interpreted by \cite{Shimoikura:2015kx} as low density gas that has been swept up and heated by the expanding \HII\ region.


\subsection{YSO catalogues}

Alongside a known population of one late O star, three B stars and two Herbig AeBe stars \citep{Smith:1985bv, 
Shuping:2012ly} there is a young stellar cluster \citep{Kuhn:2010kl, Kuhn:2015fk}. The \emph{Spitzer 
Space Telescope} legacy programme `Gould's Belt: star-formation in the solar neighbourhood' (SGBS, PID: 30574, 
\citeauthor{Dunham:2015kx} \citeyear{Dunham:2015kx}) provides specific locations and properties of the 
YSOs. This catalogue is incomplete due to saturation of \emph{Spitzer} at the heart of the OB association 
and may be contaminated by the IR bright clouds in the nebulosity. Additional catalogues are required to verify 
and complete the YSO population. 


We create a new, conservative YSO catalogue of SGBS objects \citep{Dunham:2015kx} matched with \cite{Mallick:2013kx}'s \emph{Spitzer} catalogue, \cite{Maury:2011ys}'s MAMBO catalogue of submillimetre objects, and \cite{Kuhn:2010kl}'s 
X-ray catalogue. The SGBS objects are matched with the \cite{Mallick:2013kx} sources, except where the 
SGBS is saturated around the \HII\ region. In those cases, we turn to the \cite{Kuhn:2010kl} catalogue of 
K-band excess objects as a proxy list of Class II and III objects. By matching the \cite{Mallick:2013kx} and 
\cite{Kuhn:2010kl} sources, IR bright clouds which may have been misidentified as sources can be excised. 
These two sub-catalogues are subsequently merged, with any duplicates removed. The \cite{Maury:2011ys} 
catalogue of submillimetre objects, and the \cite{Rodriguez:2010bs} and \cite{Ortiz-Leon:2015vn} radio YSOs are 
added separately and are not examined for IR contamination. We include a classification where it is reported 
by an author; otherwise Class is determined by IR dust spectral index, $\alpha$, between 2-24\,$\micron$ 
(based on the boundaries of $\alpha_{\mathrm{IR}}$ = 0.3, -0.3 and -1.6 for Class 0/I, FS, II and III, respectively, 
as summarised by \citealt{evans09}). In lieu of a comprehensive YSO catalogue covering the whole of the W40 
complex, our composite catalogue, presented in Table \ref{tab:YSO}, allows a conservative analysis to be 
made of the global YSO distribution. 

\begin{table}
\centering
\caption{Our composite YSO catalogue, produced from the combined SGBS (Dunham et al. 2015), the Mallick et al. (2013) \emph{Spitzer} catalogue, the Maury et al. (2011) MAMBO catalogue, the Kuhn etal. (2010)
X-ray catalogue, and the Rodr$\mathrm{\acute{i}}$guez etal. (2010) catalogue of radio YSOs. The full catalogue is available online.}
\label{tab:YSO}
\begin{tabular}{lccccc}
Name$^{a}$                   & YSO  &	 $\alpha$$^{b}$	& T$_{\mathrm{bol}}$$^{b}$	 \\
                   & class  &	(2-24\micron)&      (K)       \\
\hline
  2MASS18303312-0207055   &II        &   -0.78   & 1400  \\
  2MASS18303314-0220581   &II        &  -1.93   & 1700  \\
  2MASS18303324-0211258    &II        &   -0.48  &  950   \\
  2MASS18303509-0208564    &II        &  -0.86   & 1200  \\
  2MASS18303590-0206492    &II        &   -1.45   & 1700  \\
\hline
\end{tabular}
\raggedright

${a}$ - 2MASS or CXO name where available. RA Dec coordinates (J2000) where not. \\ 
${b}$ -  T$_{\mathrm{bol}}$ and $\alpha$ values as published by \cite{Dunham:2015kx}. \\
\end{table}




\section{CO contamination of SCUBA-2 850\,$\micron$}


We use our HARP 345.796\,GHz observations of the $^{12}$CO 3\hbox{--}2 line emission 
to assess the impact of CO emission on the 850\,$\micron$ band, which it is known 
to contaminate \citep{Gordon:1995uq}.  

In other GBS regions, \cite{Davis:2000vn} and \cite{Tothill:2002ys} observed CO contamination of 
SCUBA data of up to 10\% whilst \cite{Hatchell:2009fk} have found contamination up 
to 20\%. \cite{Johnstone:2003ys}, \cite{Drabek:2012uq}, and \cite{Sadavoy:2013qf} record 
a minority of cases where CO emission dominates the dust emission (up to 90\%) in 
SCUBA-2 observations, with these regions hosting substantial molecular outflows in 
addition to ambient molecular gas within the clouds.

Given that CO contamination affects the 850\,$\micron$ band but not the 450\,$\micron$ band, 
an assessment of $^{12}$CO 3\hbox{--}2 line emission is vital for an accurate assessment of 
dust temperature with unaccounted CO emission producing artificially lower ratios and cooler 
temperatures (Equation \ref{eqn:temp}, see Section 5 below). We use \cite{Drabek:2012uq}'s method 
by which CO line integrated intensities can be converted into 850\,$\micron$ flux densities and 
directly subtracted from SCUBA-2 data.

\subsection{Contamination results}

Integrated intensity maps of $^{12}\textrm{CO}$ 3\hbox{--}2 emission are subtracted from 
the original SCUBA-2 850\,$\micron$ maps using a joint data reduction process before a 
4\arcmin\ filter is applied following the method outlined in Appendix C. The fraction of 
SCUBA-2 emission that can be accounted for by $^{12}\textrm{CO}$ 3\hbox{--}2 line 
emission is presented in Figure \ref{fig:CO} (lower panel). Contamination in W40-N is 
minimal with levels up to 5\%. The Dust Arc has significant contamination at a 
level of 10\%, reaching up to 20\% in some locations. 

Figure \ref{fig:COfilter} shows the distribution of flux ratios (see Equation \ref{eqn:temp} and the method 
given in Appendix A) with and without CO contamination contributing to the 850\,$\micron$ intensities, 
showing how even a small degree of CO contamination can have a significant effect on measuring 
temperatures in the cloud, e.g, the modal flux ratio increases from 6.8 to 7.8 when CO is subtracted. 
Furthermore, the FWHM of the distribution increases from 1.9 to 2.8. Subtracting CO from our maps 
increases the mean and standard deviation of temperature in regions where $^{12}\textrm{CO}$ 3\hbox{--}2 is 
detected, in comparison with temperatures derived from uncorrected maps. The distributions of flux 
ratios across the map, with and without the CO contamination, are compared and found to have a 
KS-statistic of 0.253, corresponding to 1.3\% probability that the two samples are drawn from the 
same parent sample. CO contamination in the W40 complex is having a significant impact on the 
distribution of flux ratios. 

\begin{figure}
\begin{centering}
\includegraphics[scale=0.35]{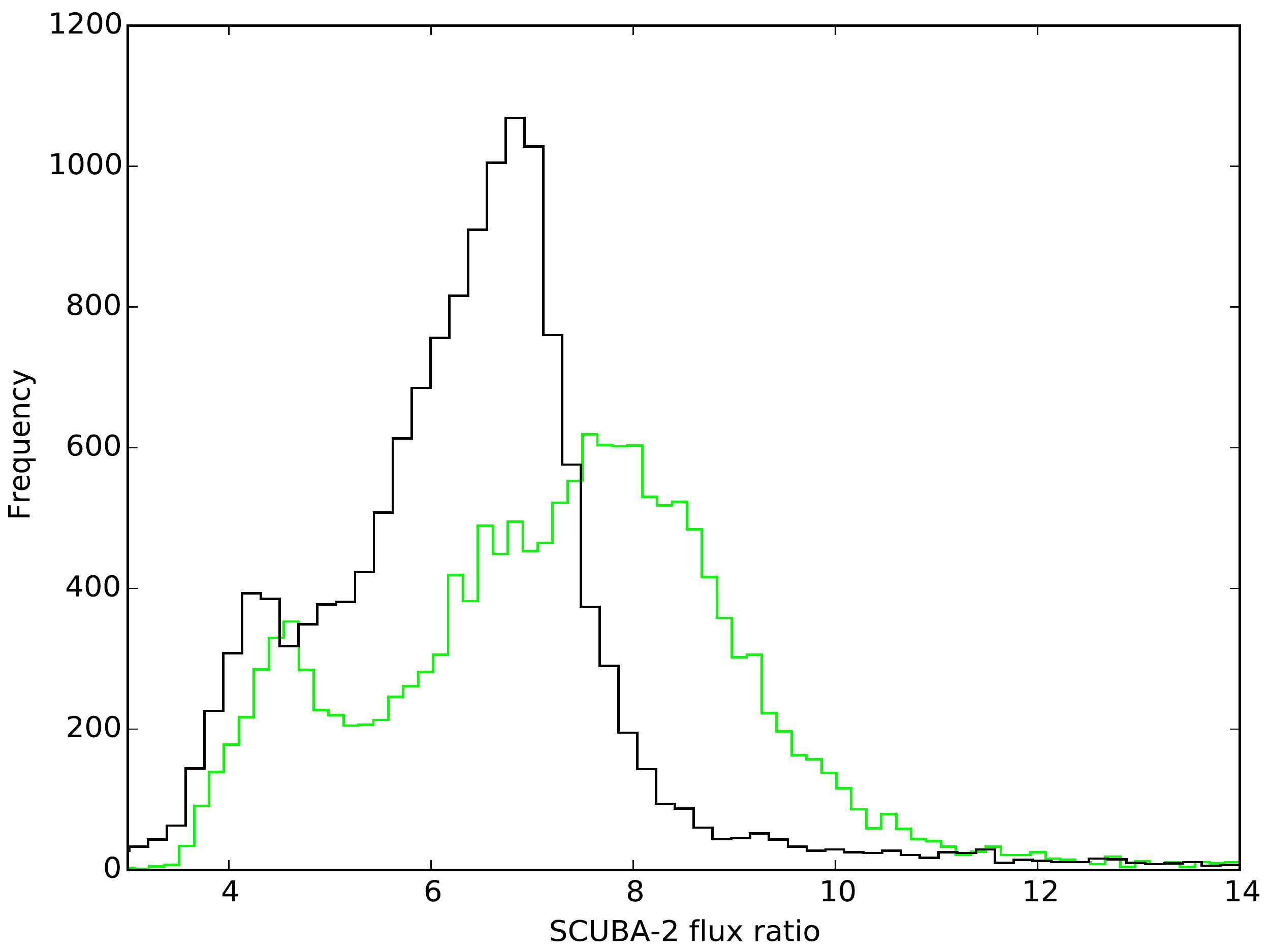}
\caption{The distribution of 450\,$\micron$/850\,$\micron$ flux ratio for the original (black) and CO-subtracted (light grey line, green in the online version) Aquila reductions with additional 4\arcmin\ spatial filtering. KS-statistics reveal a 1.3\% chance that the two data sets are drawn from the same distribution.} 
\label{fig:COfilter}
\end{centering}
\end{figure} 

\section{The free-free contribution to SCUBA-2 flux densities}

\begin{figure*}
\begin{centering}
\begin{tabular}{l}
\includegraphics[scale=0.35]{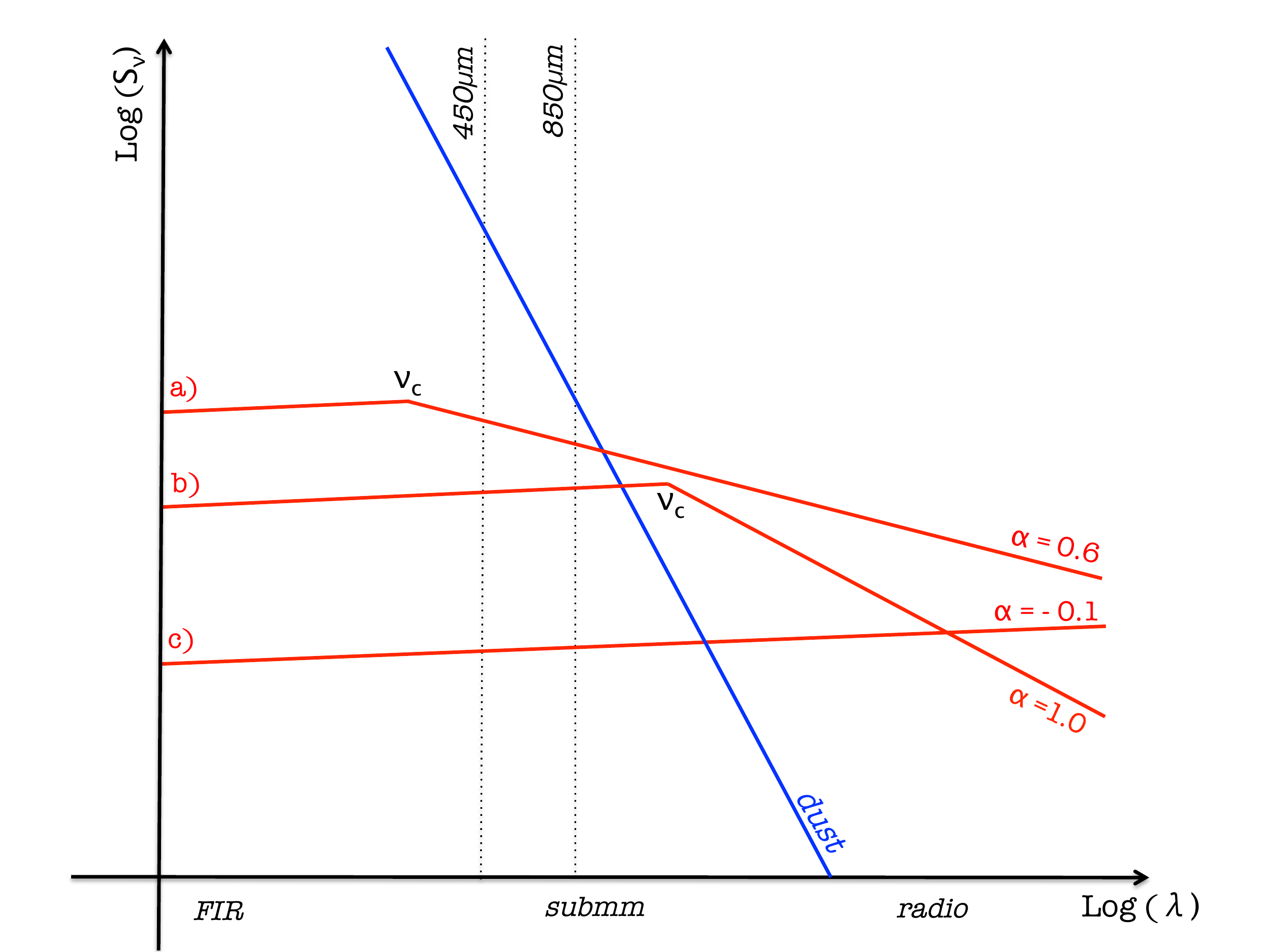}
\includegraphics[scale=0.35]{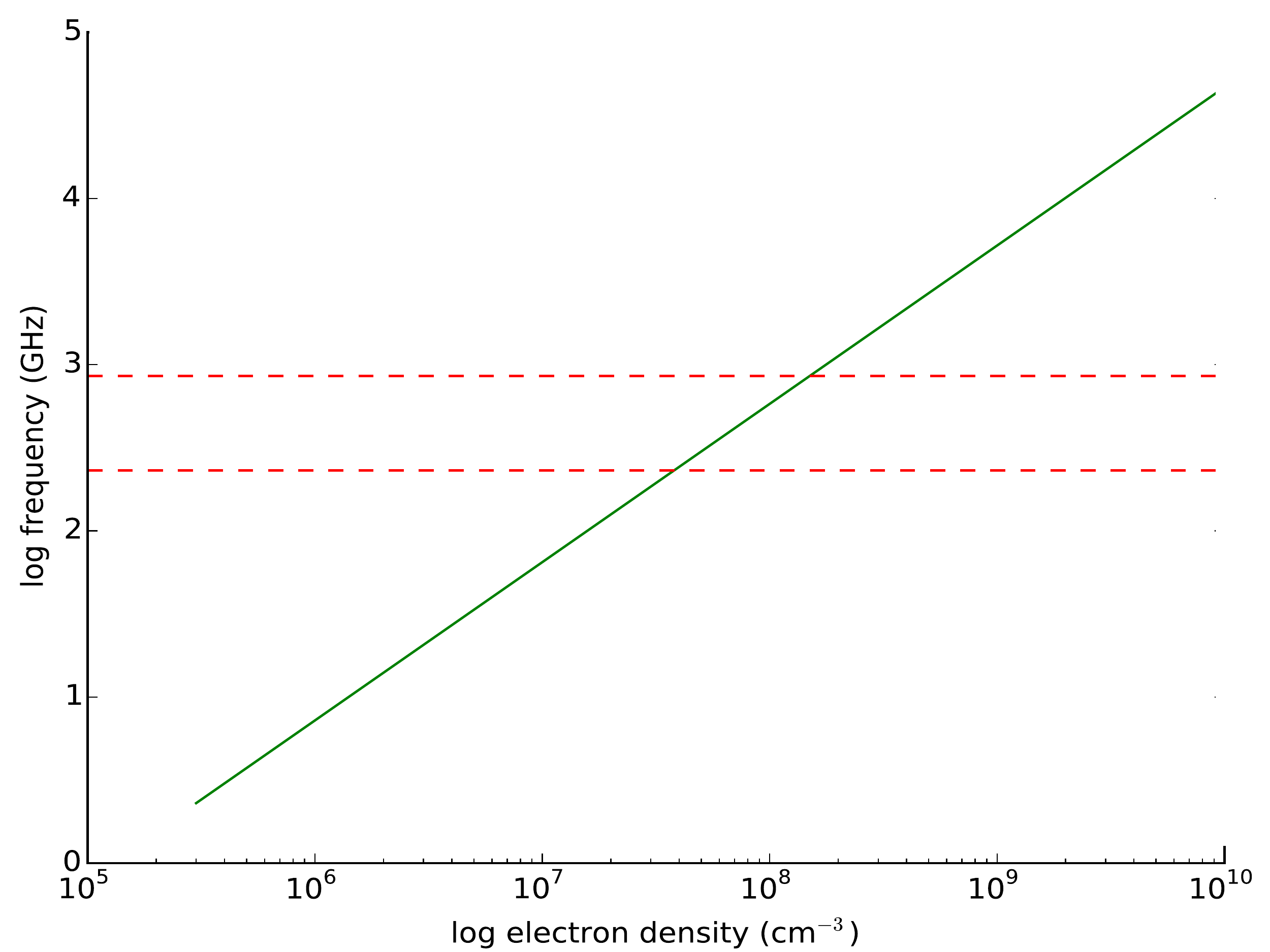}
\end{tabular}
\caption{\textbf{Left} a schematic of the SED shape for three hypothetical scenarios of free-free emission. 
\textbf{Case a)} an UCH\,\textsc{ii} with $\alpha_{\mathrm{ff}}$ = 0.6 has a turnover that occurs short-ward of the submillimetre regime, and as a result has a majority contributions to the 850\,$\micron$ band and a significant contribution to the 450\,$\micron$ band. 
\textbf{Case b)} a YSO emits free-free emission, $\alpha_{\mathrm{ff}}$ = 1.0, from a collimated jet. However the spectrum turns over to the optically thin regime long ward of submillimetre wavelengths, and consequently free-free emission contributes roughly equally to both SCUBA-2 bands. 
\textbf{Case c)} an H\,\textsc{ii} region has free-free emission from diffuse gas of $\alpha_{\mathrm{ff}}$ = -0.1 that outshines that from compact objects at long wavelengths. However, the flat spectrum means that at submillimetre wavelengths the emission is all but negligible.
\textbf{Right} free-free turnover as a function of launching electron density (as described by \citeauthor{Olnon:1975bh} \citeyear{Olnon:1975bh}  in Equation \ref{eqn:turnover}). Dashed lines indicate the submillimetre regime (1.3\,mm to 350\,$\micron$).}
\label{fig:freefreealpha}
\end{centering}
\end{figure*} 

We examine now the arguments for thermal Bremsstrahlung, or free-free, 
contributions to the SCUBA-2 data. We address questions regarding the source, 
strength, spectral index, and frequency of the turnover (from partially optically thick to optically 
thin) of free-free emission. We examine the various sources of free-free emission in 
the W40 complex and assess the magnitude of the free-free contributions to both SCUBA-2 
bands.

Free-free emission is typically observed from large-scale (1\,pc), optically thin, diffuse H\,\textsc{ii} regions with an approximately flat (-0.1) spectral index, $\alpha_{\mathrm{ff}}$ \citep{Oster:1961fk, Mezger:1967fk}. Free-free emission is also detected on smaller scales \citep{Panagia:1975uq} comparable to a protostellar core (0.05\,pc, \citealt{Rygl:2013ve}). 
On these smaller scales, free-free emission is produced by an ionised stellar wind \citep{Wright:1975kx, Harvey:1979qf} with an observed spectral index of $\alpha_{\mathrm{ff}}$ = 0.6 found by \cite{Harvey:1979qf}, \cite{Kurtz:1994cr}, and \cite{Sandell:2011dz} where emission is considered partially optically thick. Where the wind is collimated and accelerating, as in bipolar jets, the spectral index increases to $\alpha_{\mathrm{ff}} \simeq$ 1.0 \citep{Reynolds:1986cr}, as for example in AB Aur \citep{Robitaille:2014rr} and MWC 297 \citep{Rumble:2015vn}. In addition, non-thermal processes, such as gyrosynchrotron emission, have been known to lead to a negative spectral index (\citealt{Hughes:1991fk}, \citealt{Hughes:1995uq} and \citealt{Garay:1996kx}).

\subsection{Free-free emission in ultra compact H\,\textsc{II} regions}

At small scales, free-free spectral indices are steep at low frequencies until the emission 
undergoes a turnover, $\nu_{c}$, as it transitions from being partially optically thick to 
optically thin. Figure \ref{fig:freefreealpha} shows a schematic of the frequency behaviours 
of free-free emission, including this turnover. If the turnover occurs short-ward of submillimetre 
wavelengths, it is possible that free-free emission may contribute to the observed SCUBA-2 
fluxes. 

For a wind, the frequency of the free-free turnover is defined as a function of electron density, 
$n_{e}(r)$ = $n_0$ where $r$ $\leq$ $R$ and $n_e(r)$ = $n_{\mathrm{0}}(r/R)^{-2}$
where $r$ $\geq$ $R$, by \cite{Olnon:1975bh} as
\begin{equation}
\log_{10} \nu _{c} = - 0.516 + \frac{1}{2.1}\log_{10}\left ( \frac{8}{3}Rn_{0}^{2}T_{e}^{-1.35} \right ),  
\label{eqn:turnover}
\end{equation}
where $R$ is the launching radius of the wind (typically 10\,AU) and $T_{e}$ is the electron temperature 
(typically 10$^{4}$\,K). Figure \ref{fig:freefreealpha} (right) shows how $\nu_{c}$ is related to $N_{e,0}$. 
A minimum value of $N_{e,0}$ $\sim$ 2$\times$10$^{8}$\,cm$^{-3}$ is required for a turnover that occurs 
at wavelengths long ward of the submillimeter regime (1.3\,mm to 350\,$\micron$).

The electron density is not easily determined directly from observations. We assume that $n_{\mathrm{0}}$ 
is proportional to stellar mass and by association varies with spectral class. \cite{Sandell:2011dz} 
indicate that the free-free contribution to SCUBA-2 fluxes is significant for early B stars in their 
sample, but not for late B- and A-type stars. For early B-stars, a distinct point source is observed 
at submillimetre wavelengths. In the case of the B1.5Ve/4V \citep{Drew:1997qf} Herbig star MWC 297 
with an $\alpha_{\mathrm{ff}}$ = 1.03$\pm$0.02 \citep{Skinner:1993bh, Sandell:2011dz}, it was 
found that free-free emission contributed comparatively more to the 850\,$\micron$ band (82$\pm$4\,\%) 
than to the 450\,$\micron$ band (73$\pm$5\,\%) resulting in excessively low ratios \citep{Rumble:2015vn} 
that were misinterpreted as very low $\beta$ and potential grain growth by \cite{Manoj:2007ly}.

MWC 297 is the lowest mass star in the \cite{Sandell:2011dz} sample for which free-free radiation 
contributes at SCUBA-2 wavelengths and we therefore mark it as a limit of stellar type. We 
therefore assume that a free-free turnover can occur in the submillimetre regime only for spectral 
types earlier than B1.5Ve/4V. This gives an indication of whether the free-free emission from 
a massive star is likely to be optically thin or thick at submillimetre wavelengths.

\begin{figure}
\begin{centering}
\includegraphics[scale=0.5]{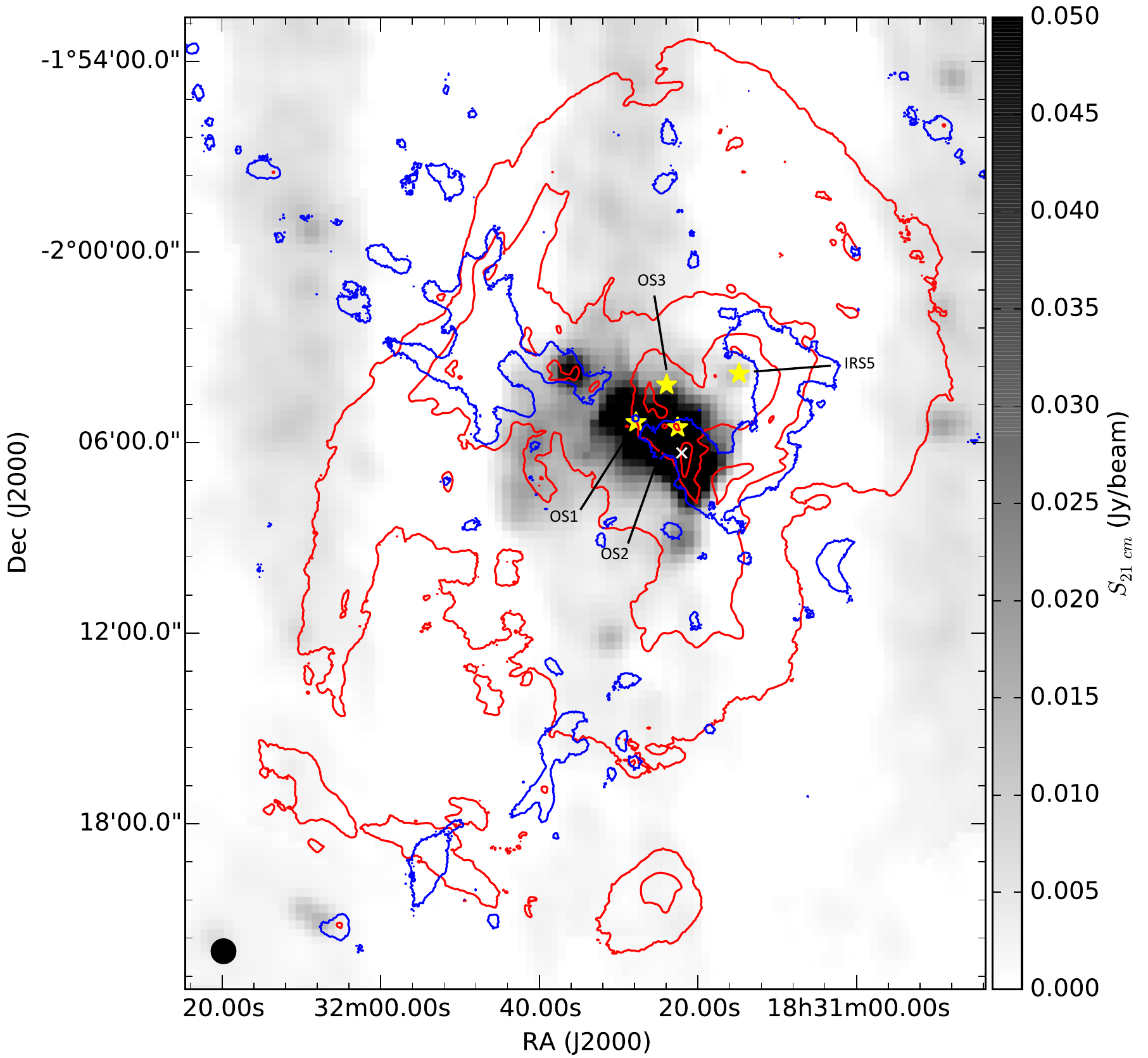}
\caption{Archival VLA 21\,cm NRAO VLA Sky Survey \citep{Condon:1998kx} continuum map of the W40 complex \HII\ region (45\arcsec\ resolution greyscale). Red: \emph{Herschel} 70\,$\micron$ contours of the nebulosity Sh2-64 at 300, 1200, 4800, 12000\,MJy/Sr. Blue: SCUBA-2 850\,$\micron$ contours of the dust cloud at the 5\,$\sigma$ level. Yellow stars indicate the locations of the OB stars, with the O9.5 star OS1 being the primary ionising object of the region. The white cross indicates the peak of the VLA 21\,cm continuum emission.} 
\label{fig:21cm}
\end{centering}
\end{figure} 


\begin{figure*}
\begin{centering}
\includegraphics[scale=0.8]{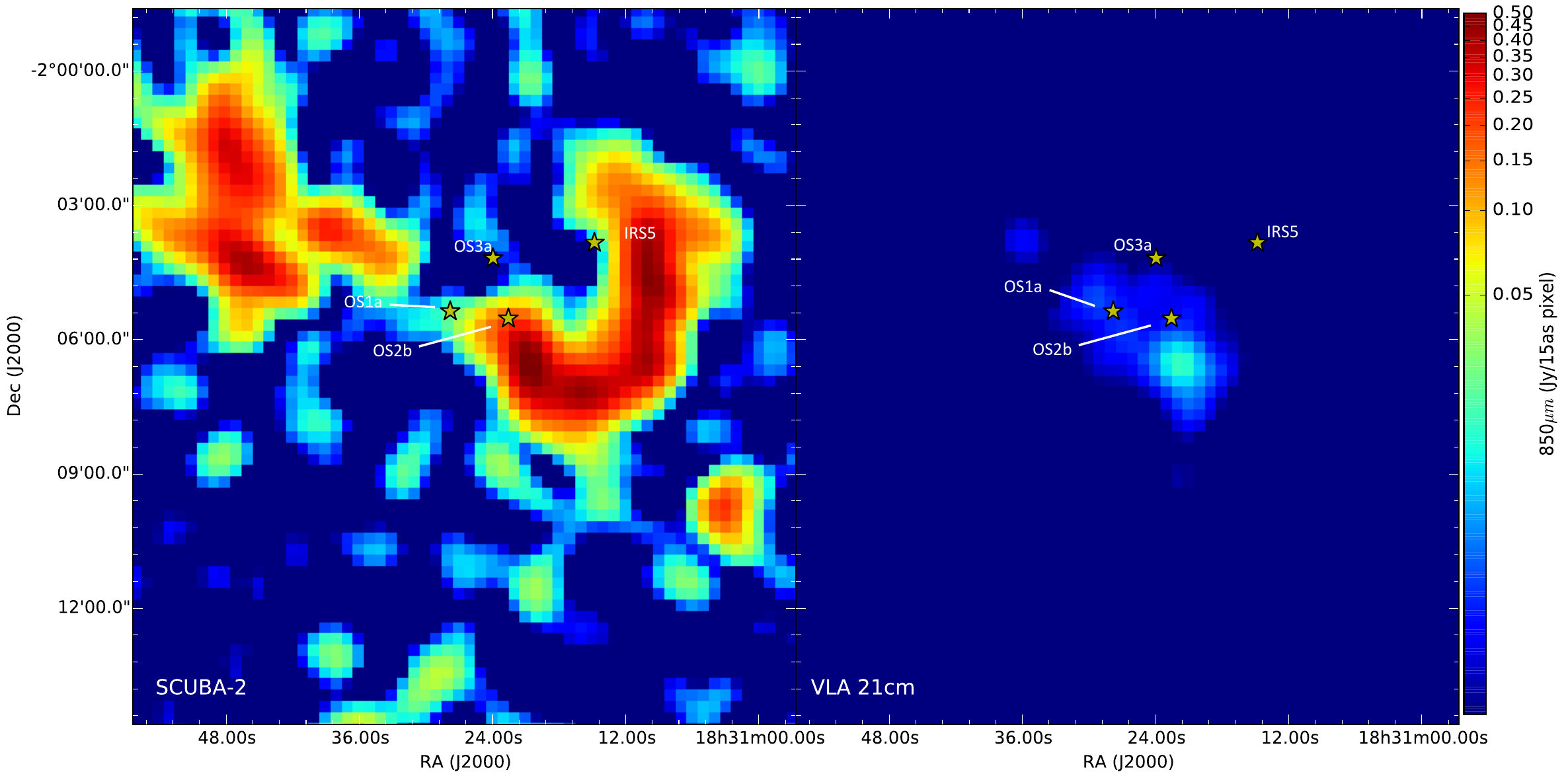}
\caption{The free-free contribution from large-scale \HII\ gas, modelled using archival VLA 21\,cm observations \citep{Condon:1998kx} assuming $\alpha_{\mathrm{ff}}$ = -0.1 (right), compared to SCUBA-2 dust emission at 850\,$\micron$ (left). Maps have common 15\arcsec\ pixels and 45\arcsec\ resolution. Markers indicate the locations of the OB stars.}
\label{fig:freefree21}
\end{centering}
\end{figure*} 

\begin{figure*}
\begin{centering}
\includegraphics[scale=0.7]{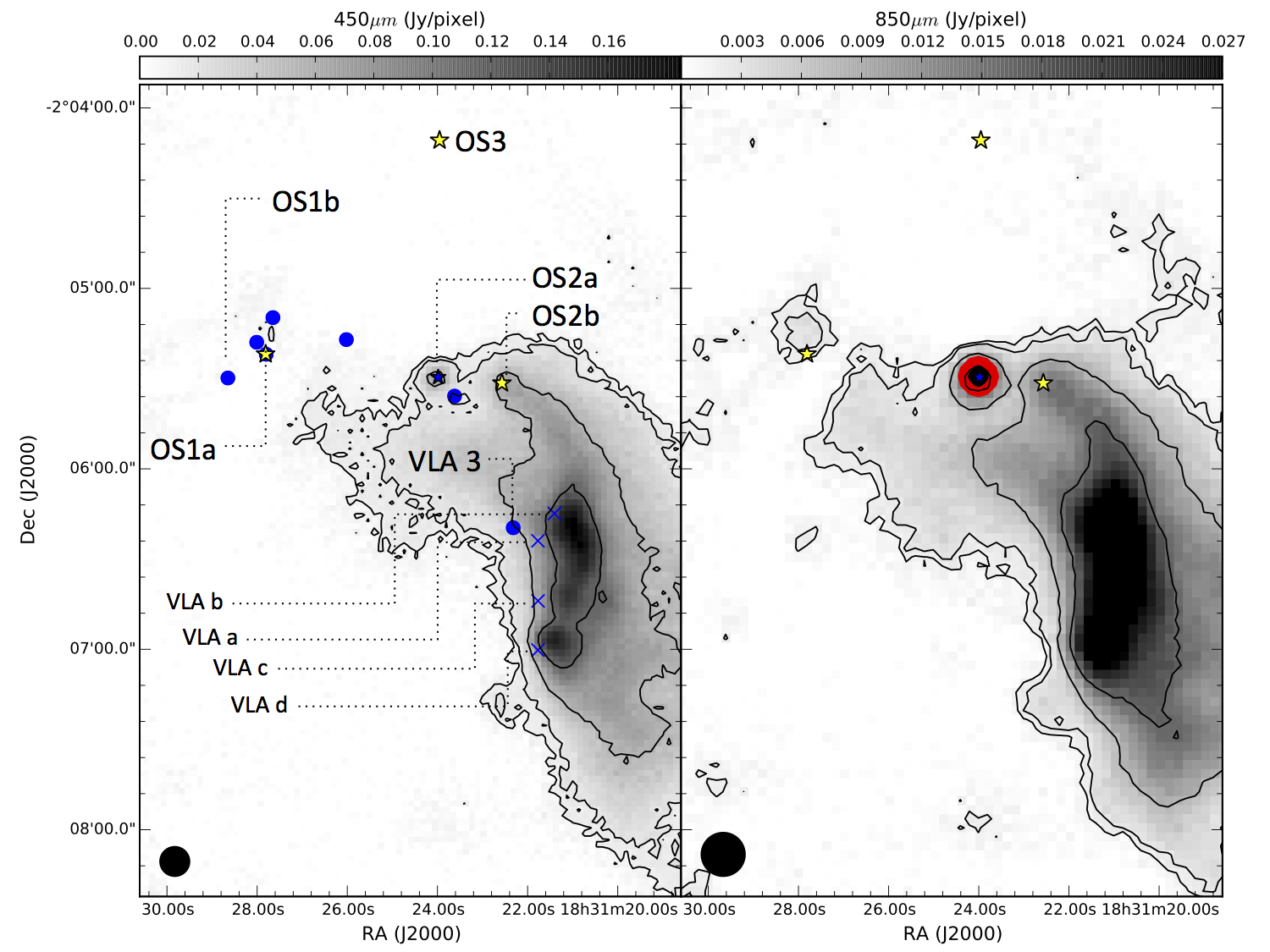}
\caption{The free-free contribution of the compact radio source OS2a (marked as a blue star) at 450\,$\micron$ (left) and 850\,$\micron$ (right), modelled as point sources with fluxes extrapolated from the Rodriguez et al. (2010) 3.6\,cm fluxes and assuming an $\alpha_{\mathrm{ff}}$ = 1.0. Yellow stars indicate the locations of the OB stars, blue filled circles the location of all the Rodriguez et al. (2010) and Ortiz-Le$\mathrm{\acute{o}}$n et al. (2015) compact radio source matches. Blue crosses mark the location of four peaks identified separately in the AUI/NRAO 3.6\,cm map (450\,$\micron$ only). Black contours trace SCUBA-2 data at 3$\sigma$, 5\,$\sigma$, 15\,$\sigma$ and 30$\sigma$. Red and black filled contours trace the optically thick free-free contribution at 3$\sigma$ and 5\,$\sigma$ (see Table \ref{tab:stars}).} 
\label{fig:freefree3_6}
\end{centering}
\end{figure*} 

\begin{figure}
\begin{centering}
\includegraphics[scale=0.5]{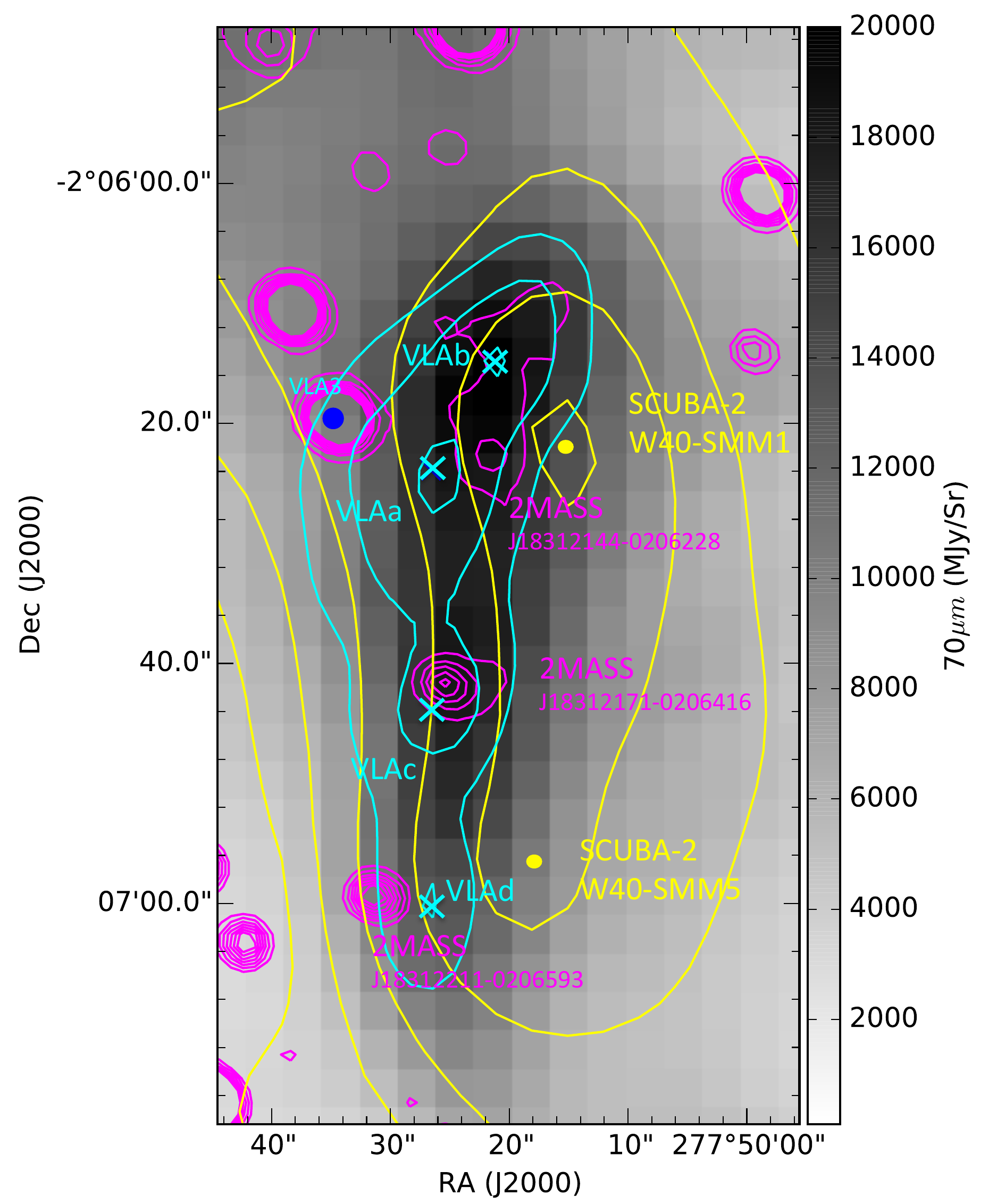}
\caption{Archival \emph{Herschel} 70\,$\micron$ data for W40-SMM1 and 5. Magenta contours (560, 570, 580, 590, 600, 610, 620 mag.) show several 2MASS point sources embedded within the eastern Dust Arc which is shown in the yellow SCUBA-2 850\,$\micron$ 5, 10, 20, 40, 60, 80$\sigma$ contours with circle markers at the peaks of the W40-SMM1 and 5 clumps (see Section 6). Cyan crosses show the four peaks in Archival AUI/NRAO 3.6\,cm map (contours at 0.01, 0.016 and 0.021\,Jy/beam). The Rodriguez et al. (2010) YSO `VLA3' is also shown.} 
\label{fig:SMM1}
\end{centering}
\end{figure}

\subsection{Free-free emission in the W40 complex}

Free-free emission in the W40 complex comes from a large-scale \HII\ region associated with the nebulosity  Sh2-64 \citep{Condon:1998kx}, observed in archival VLA 21\,cm continuum data presented in Figure \ref{fig:21cm}. \cite{Vallee:1991zr} initially measured the size the \HII\ region as 6\arcmin\ by 3\arcmin\ with a 1.7\arcmin\ diameter incomplete shell. \cite{Pirogov:2013ys} suggest a secondary \HII\ region powered by the B star IRS 5. However, we find no evidence for this in the 21\,cm emission (Figure \ref{fig:21cm}). 

 
Table~\ref{tab:stars} lists several small-scale \UCHII\ regions associated with bright NIR objects in the W40 complex. \cite{Rodriguez:2010bs} resolve 15 compact radio sources in 3.6\,cm emission (at 0.26\arcsec\ resolution) consistent with 2MASS sources and, by monitoring time-variability, are able to classify eight variable YSOs and seven non-variable \UCHII\ candidate regions. \cite{Ortiz-Leon:2015vn} expand this survey and observe the same region at 4\,cm and 6.7\,cm, detecting 41 radio objects, 15 of which are confirmed as YSOs and are presented in Figure \ref{fig:freefree3_6}. Both \cite{Rodriguez:2010bs} and \cite{Ortiz-Leon:2015vn} also identify non-compact radio sources without IR counterparts and these are interpreted as shock fronts from thermal jets that are likely formed by the local Herbig AeBe stars OS1b and OS2a/b. 


\subsection{Contribution results}

Building on the method outlined in \cite{Rumble:2015vn}, we model small- and large-scale free-free emission (separately). The contribution to SCUBA-2 maps is based either on a directly measured or an indirectly assumed value of $\alpha_{\mathrm{ff}}$.

For large-scale \HII\ emission, we extrapolate the archival VLA 21\,cm data presented in Figure \ref{fig:21cm} up to SCUBA-2 wavelengths of 450\,$\micron$ and 850\,$\micron$, assuming a spectral index of $\alpha_{\mathrm{ff}}$ = -0.1 as concluded by \cite{Rodney:2008ij}. The \textsc{findback} tool (see Appendix B) is used to remove structures larger than 5\arcmin\ (mimicking the SCUBA-2 data reduction process). Figure \ref{fig:freefree21} shows how the SCUBA-2 data are subsequently aligned and convolved to the larger resolution of the VLA 21\,cm data so the two data sets are comparable. The overall contribution of the large-scale \HII\ region to SCUBA-2 bands is very limited, at its peak contributing 5\% at 850\,$\micron$ and 0.5\% at 450\,$\micron$.

\cite{Ortiz-Leon:2015vn} calculate the free-free spectral index for the 14 YSOs, marked in Figure \ref{fig:freefree3_6}, 
that are detected in both their 4\,cm observations and the \cite{Rodriguez:2010bs} 3.6\,cm observations. The majority 
of YSOs have less than -0.1 $\alpha_{\mathrm{ff}}$ indicating non-thermal gyrosynchrotron emission that will not 
be bright at SCUBA-2 wavelengths and does not require further consideration. The OS1a cluster, OS2a, b and VLA-3 (J18312232-0206196) are all consistent with SCUBA-2 emission, as shown in Figure \ref{fig:freefree3_6}, and are 
considered for free-free contamination. These objects are summarised in Table \ref{tab:stars}. 

OS1 is a dense stellar cluster that includes the primary ionising object, OS1a(South), that is driving the \HII\ region. 
\cite{Ortiz-Leon:2015vn} calculate an $\alpha_{\mathrm{ff}}$ of -0.3$\pm$0.2 for this object which is consistent with 
an optically thin \HII\ region. The brightest radio source in the OS1 cluster is VLA-14 at 5.78\,mJy at 3.6\,cm (Table 
\ref{tab:stars}). It also has the most positive spectral index of $\alpha_{\mathrm{ff}}$ = 0.0$\pm$0.1. This corresponds 
to a peak 850\,$\micron$ flux of 0.17\,mJy, a value that falls well below the 1$\sigma$ noise level of that SCUBA-2 
band. We therefore conclude that there is no evidence that free-free emission from the OS1a cluster is contributing 
to the faint 850\,$\micron$ emission detected by SCUBA-2 (OS1a is not detected at 450\,$\micron$).     

\cite{Rodriguez:2010bs} and \cite{Ortiz-Leon:2015vn} calculate that VLA-3 has the most positive spectral index with 
$\alpha_{\mathrm{ff}}$ of 1.1$\pm$0.2. This index is consistent with a collimated jet source. \cite{Zhu:2006ee} and 
\cite{Shimoikura:2015kx} studied the $^{13}$CO~2\hbox{--}1, $^{12}$CO~1\hbox{--}0 and 3\hbox{--}2 line emission 
in this region and observed profiles symptomatic of outflows. However, we observe that the $^{12}$CO~3\hbox{--}2 
line in this region is highly extincted due to emission becoming optically thick at high densities, making reliable 
analysis of these features impractical. Figure \ref{fig:freefree3_6} shows that VLA-3 is heavily embedded within the 
Dust Arc; however, it is not associated with a strong point source in either SCUBA-2 bands in the same way that 
OS2a is. From this we conclude that free-free emission from YSO VLA-3 has turned over to optically thin at wavelengths 
longer than the submillimeter regime and does not provide a significant contribution to the SCUBA-2 bands. 

OS2a is a Herbig AeBe star that is detected as a strong point source by SCUBA-2 at both 450\,$\micron$ and 850\,$\micron$ (Figure \ref{fig:freefree3_6}). \cite{Rodriguez:2010bs} detect OS2a at 3.6\,cm, finding it to be variable and having evidence for jets through outflow knots. By contrast \cite{Ortiz-Leon:2015vn} detect emission at the location of OS2a, but do not report it as its SNR falls below their detection criteria (Ortiz-Leon. priv. comm.). Such behaviour is consistent with a variable object, and therefore it is not possible calculate a reliable $\alpha_{\mathrm{ff}}$. 

In order to make an estimate of the upper limit of the free-free contribution of OS2a, we model OS2a as a point source and extrapolate the \cite{Rodriguez:2010bs} 3.6\,cm flux up to 450 and 850\,$\micron$ based on an $\alpha_{\mathrm{ff}}$ of 1.0, consistent with indirect observations of local jet emission by \cite{Rodriguez:2010bs}. We make the optimistic assumption that OS2a is optically thick at SCUBA-2 wavelengths on the basis of the bright point source at that location that is observed at 450 and 850\,$\micron$ (Figure \ref{fig:freefree3_6}). The fluxes are subsequently convolved by the JCMT beam using its primary and secondary components for comparison with the SCUBA-2 data and presented in Table \ref{tab:freefree}. Using this method we calculate that the  free-free contribution for OS2a is at most 9\% at 450\,$\micron$ and 12\% at 850\,$\micron$. 

No radio or submillimeter point source has been observed at the location of OS2b, as shown in Figure \ref{fig:freefree3_6}. We therefore assume that any free-free radio emission must be faint and optically thin at SCUBA-2 wavelengths. This is consistent with its classification as a weak UV-photon-emitting B4 star \citep{Shuping:2012ly}. 








\begin{table*}
\caption{Summary of bright radio objects in the W40 complex and evidence for variability, jet emission and free-free opacity at SCUBA-2 wavelengths, from which a value of $\alpha_{\mathrm{ff}}$ can be estimated (if not previously calculated).  }
\begin{tabular}{|cccccccccc|}
\hline
Source	&	2MASS ID	& VLA ID$^{a}$ &	Type	$^{b}$	&	Time	$^{a}$				&	Jet$^{a}$?	&	SCUBA-2 		&	Optically		&	Spectral index				&	Distance$^{b}$  \\
		&		&		&			&	variable?	&			&	source?	&	 thick?	&	$\alpha_{\mathrm{ff}}$		&	(pc)		\\
\hline
\hline
OS 1a (North)	& J18312782-0205228	&	15 	&	Herbig AeBe	&	N	&	N	&	Y	&	Y	&	-0.3$\pm$0.2$^{c}$	&	536$^{+42}_{-95}$	\\
OS 1a (South)	& J18312782-0205228	&	-	&	O9.5			&	-	&	N	&	Y	&	-	&	-	&	536$^{+42}_{-95}$	\\
OS 1b		& J18312866-0205297	&	18	&	Class II		&	N	&	Y	&	N	&	N	&	-0.8$\pm$0.5$^{c}$	&	-				\\
OS 1c		& J18312601-0205169	&	8	&	Class II		&	Y	&	N	&	N	&	N	&	-0.6$\pm$0.2$^{c}$	&	-				\\
OS1d		& J18312766-0205097	&	13	&	Class II		&	Y	&	N	&	Y	&	N	&	0.1$\pm$0.2$^{c}$	&	-	
		\\	
OS 2a		& J18312397-0205295	&	7	&	Herbig AeBe	&	Y	&	Y	&	Y	&	Y	&	1.0	&	-				\\
OS 2b		& J18312257-0205315	&	-	&	B4			&	Y	&	Y	&	N	&	N	&	-0.1	&	455$^{+71}_{-59}$	\\
OS 3			& J18312395-0204107	&	-	&	B3*(binary)	&	-	&	-	&	N	&	-	&	-	&	454$^{+87}_{-48}$	\\
IRS 5		& J18311482-0203497	& 	1	&	B1			&	-	&	-	&	N	&	N	&	0.3$\pm$0.2$^{c}$	&	469$^{+217}_{-129}$	\\
-			&  J18312232-0206196	&	3 	&	Class II 	&	N	&	Y	&	N	&	N	&	1.1$\pm$0.2$^{c}$	&	-	\\
-			& -					&	14	&	-		&	N	&	N	&	N	&	N	&	0.0$\pm$0.1$^{c}$	&	-	\\
\hline
\end{tabular}\\
\raggedright
$^{a}$ Radio source ID and characterisation based on the findings of \cite{Rodriguez:2010bs}.\\
$^{b}$ Spectral classifications and distances calculated from Spectral Energy Distributions in \cite{Shuping:2012ly}.\\
$^{c}$ Free-free spectral index calculated by \cite{Ortiz-Leon:2015vn}.\\
\label{tab:stars}
\end{table*}%

\begin{table*}
\caption{Summary of significant free-free contributions to SCUBA-2 wavelengths from bright objects in W40. The uncertainty on flux density at 450\,$\micron$ is 0.017\,Jy/pix and 850\,$\micron$ is 0.0025\,Jy/pix.}
\begin{tabular}{@{}|c|ccccccccc|c|}
       & 3.6\,cm (Jy)	& \multicolumn{4}{c}{450\,$\micron$ (Jy)} & \multicolumn{4}{c}{850\,$\micron$ (Jy)} \\
Object &	VLA$^{a}$	& SCUBA-2       & Free-free      & Dust      & \%      & SCUBA-2      & Free-free      & Dust       & \% 	& $\alpha_{\mathrm{ff}}$     \\
\hline
\hline
OS2a   &	0.00240	& 1.83          & 0.16           & 1.67      & 9      & 0.558        & 0.069          & 0.489      & 12	&	1.0     \\
\hline
\end{tabular}\\
\raggedright
$^{a}$ VLA 3.6\,cm compact object fluxes \citep{Rodriguez:2010bs}. \\
\label{tab:freefree}
\end{table*}

\subsection{Additional free-free sources}

Observations by \cite{Rodriguez:2010bs} did not cover the four brightest peaks in the AUI/NRAO 3.6\,cm 
data that lie to the west of VLA-3, referred to here as VLAa, b, c, and d (Figure  \ref{fig:freefree3_6} and \ref{fig:SMM1}). 
These objects are orders of magnitude brighter than the \cite{Rodriguez:2010bs} sources and appear in 
close proximity to the peak 850\,$\micron$ emission and three 2MASS sources. Each 2MASS source is deeply 
reddened, consistent with an embedded YSO, suggesting that these objects could be young protostars. 
Examining the 70\,$\micron$ data (Figure \ref{fig:SMM1}), where the blackbody spectrum of a protostar is at 
its peak, FIR emission is brightest around the location of J18312144-0206228 and J18312171-0206416. 
Their respective alignment with VLAa /VLAc could be considered as an indicator of an \UCHII\ region around 
a massive protostar. This idea is consistent with the findings of \cite{Pirogov:2013ys} who observe CS 
2$\hbox{ --}$1 line emission and find evidence of infalling material linked to high-mass star-formation in 
the eastern Dust Arc.

Alternatively, we could be observing free-free emission from the shock/ionisation front from where the 
OS1a \HII\ region is interacting with the eastern Dust Arc, as proposed by \cite{Vallee:1991zr}. Using flux 
and distance in \cite{Kurtz:1994cr}'s Equation 4, we calculate that a Lyman photon density of 4.0$\times$10$^{46}$\,s$^{-1}$ 
is required to produce a total flux density of 0.167\,Jy for all four unidentified VLA sources at 3.6\,cm. 
We compare this value to the Lyman photon density produced by OS1a, a 09.5V star, which is the primary 
ionising source of the \HII\ region. We assume a minimum distance between OS1a and the filament of 
3\arcmin, consistent with \cite{Vallee:1991zr}, and calculate that the proposed ionisation front across the 
eastern Dust Arc would be exposed to, at most, 2.1\% of Lyman photons produced by OS1a at this distance. 
This percentage corresponds to a Lyman photon density of 1.67$\times$10$^{46}$\,s$^{-1}$, which is 
comparable to the flux observed given the approximate nature of this calculation.  

Given the speculation about the nature of these sources, we cannot reliably estimate a value of $\alpha_{\mathrm{ff}}$ 
for these objects. However, we do not observe significant SCUBA-2 peaks at the positions of these objects. 
We therefore conclude that any free-free emission observed here at 3.6\,cm is optically thin at SCUBA-2 
wavelengths and therefore regardless of their nature they produce no significant free-free contributions.





\section{Temperature mapping}

In this section we outline how the dust temperature is calculated from the ratio of SCUBA-2 fluxes at a common 
resolution, obtained using a convolution kernel, and how this method compares to our previous work with the dual-beam 
cross-convolution method described in \cite{Rumble:2015vn}. We present the temperature maps for the W40 complex 
and outline some of their notable features.

By taking the ratio of SCUBA-2 fluxes at 450 and 850\,$\micron$, the spectral index of the dust, 
$\alpha$, can be calculated as a power law of frequency ratio,  
\begin{equation}
\frac{S_{\mathrm{450}}}{S_{\mathrm{850}}} = \left (  \frac{\nu_{850} }{\nu _{450}}\right )^\alpha,  \label{eqn:alpha}
\end{equation}
where $\alpha$ can be approximated to 2 + $\beta$ (assuming the Rayleigh-Jeans limit). By assuming a full, opacity-modified Planck function, it is possible to show that $\alpha$ is derived from $\beta$ and the dust temperature, $T_{\mathrm{d}}$. As a result, Equation \ref{eqn:alpha} can be expanded into 
\begin{equation}
\frac{S_{\mathrm{450}}}{S_{\mathrm{850}}} = \left(\frac{850}{450} \right)^{3+\beta }\left(\frac{\exp(hc/\lambda _{\mathrm{850}}k_{\mathrm{b}}T_{\mathrm{d}})-1}{\exp(hc/\lambda _{\mathrm{450}}k_{\mathrm{b}}T_{\mathrm{d}})-1} \right)  \label{eqn:temp}
\end{equation}
to include both of these parameters \citep{Reid:2005ly}. By assuming a constant $\beta$ across a specific region, maps of temperature are produced for the W40 complex by convolving the 450\,$\micron$ map down to the resolution of the 850\,$\micron$ map using an analytical beam model convolution kernel (\citeauthor{Aniano:2011fk} \citeyear{Aniano:2011fk}, \citeauthor{Pattle:2015ys} \citeyear{Pattle:2015ys}, see Section 5.1 and Appendix A for details).

The relationship between wavelength and dust opacity is modelled as a power-law for a  specific dust opacity spectral index in the submillimetre regime,  
\begin{equation}
\kappa_{\lambda} = 0.012\times\left(\frac{\lambda}{850 \mu\mathrm{m}} \right )^{-\beta}\mathrm{cm^{2}/g},  
\label{eqn:beta}
\end{equation}
which is consistent with the popular OH5 model \citep{Ossenkopf:1994vn} of opacities in dense ISM, 
for a specific gas-to-dust ratio of 161 and $\beta$ = 1.8 over a wavelength range of 30\,$\micron$--1.3\,mm 
\citep{Hatchell:2005fk}. This value is consistent with \emph{Planck} observations \citep{Juvela:2015ys} and that used in other GBS papers (\citeauthor{salji:2015fk} \citeyear{salji:2015fk}, \citeauthor{Rumble:2015vn} \citeyear{Rumble:2015vn} and \citeauthor{Pattle:2015ys} \citeyear{Pattle:2015ys}), but less than the \cite{Planck-Collaboration:2011zr} who find a mean value of $\beta$ = 2.1 in cold clumps. 

%
The majority of the structure observed in the W40 complex is typical of the ISM and protostellar envelopes and our use of $\beta$ = 1.8 \citep{Hatchell:2013ij} reflects a standard approach, though $\beta$ has been found up to 2.7 in extended, filamentary regions (\citealt{Planck-Collaboration:2011zr}, Chen et al. submitted) and around 1.0 in discs \citep{Beckwith:1990fk, Sadavoy:2013qf, Buckle:2015vn}. 

A popular alternative to the SCUBA-2 flux ratio method discussed above is spectral energy distribution (SED) fitting, a method that has been widely used to derive both dust temperatures and $\beta$ from \emph{Herschel} \citep{Griffin:2010kx} SPIRE and PACS bands (\citealt{Shetty:2009uq}, \citealt{Bontemps:2010fk} and \citealt{Gordon:2010fk}). SED fitting can be limited by the emission model, the completeness of the spectrum, resolution and local fluctuations of $\beta$ \citep{Konyves:2010oq}. A detailed comparison between the SCUBA-2 flux ratio method and SED fitting is given in Appendix D. A comparison with the temperatures calculated by \cite{Maury:2011ys} and \cite{Konyves:2015uq} using SED fitting is given in Section 6.

\subsection{The convolution kernel}

The JCMT has a complex beam shape and different resolution at each of the 
SCUBA-2 bands and a common map resolution is required before the 
flux ratio can be calculated. Achieving common resolution of maps by using a 
kernel, as opposed to the cross-convolution with JCMT primary and secondary 
beams (see \citeauthor{Rumble:2015vn} \citeyear{Rumble:2015vn}), has the 
advantage of improved resolution flux ratio maps. The kernel method results in 
a temperature map with a resolution of 14.8\arcsec, approximately equal to the 
850\,$\micron$ map, whereas the resolution of the dual-beam method is 
19.9\arcsec. 

We apply the kernel convolution algorithm from \cite{Aniano:2011fk} using the 
\cite{Pattle:2015ys} adaptation to SCUBA-2 images. Details of the method are 
summarised in Appendix B and those authors' papers. Having achieved common 
resolution, ratio and temperature maps are produced following the method of 
\cite{Rumble:2015vn}. Details of the propagation of errors through the 
convolution kernel are also given Appendix B. 

\begin{figure*}
\begin{center}
\includegraphics[scale=1.]{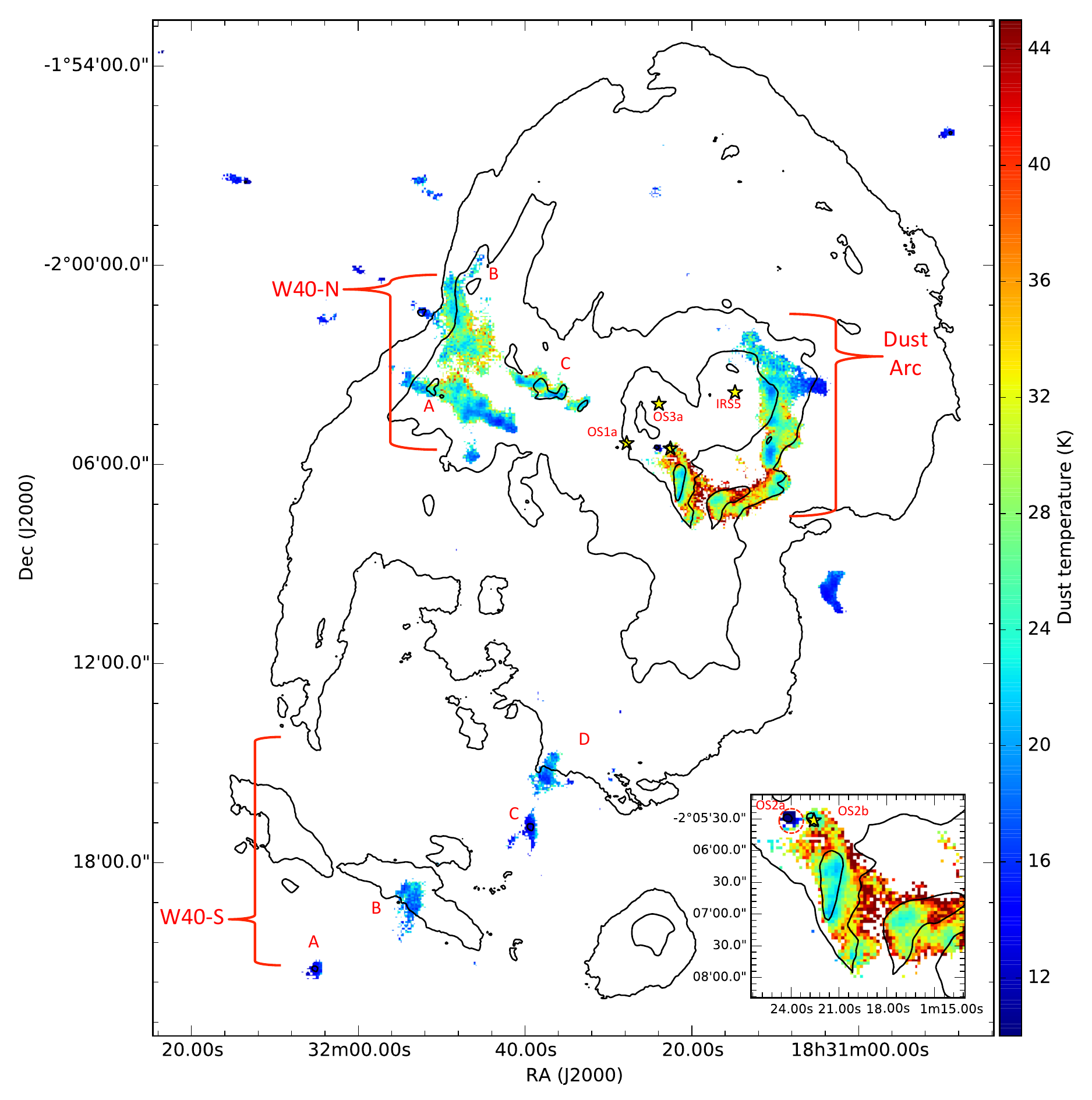}
\caption{Temperature map of the W40 complex with \emph{Herschel} 70\,$\micron$ contours at 300, 1200, 4800 and 12000\,MJy/Sr. Note that 850\,$\micron$ flux has been decontaminated for CO in all areas except W40-S. Temperatures are given at positions where the 850\,$\micron$ flux is at least five times the noise level and the fractional error on the temperature is less than 0.34. The insert shows a zoom in on the Eastern Dust Arc and the high mass stars OS2a/b.}
\label{fig:temp}
\end{center}
\end{figure*}

\subsection{Temperature and spectral index results}

\begin{table*}
\caption{A sample of submillimetre clumps and their respective SCUBA-2 and \emph{Herschel} fluxes. The full table is available online.}
\begin{tabular}{@{}lllllll}
Index$^{a}$	&	IAU object name$^{a}$	&	70\,$\micron$ intensity$^{b}$&		450\,$\micron$ flux$^{c}$	&	850\,$\micron$ flux$^{c}$	&	21\,cm intensity$^{d}$ &	Clump area\\
	&		&	(MJy/Sr)&		(Jy)	&	(Jy)	&	(Jy/pix)	&	(Pixels)\\
\hline
\hline
W40-SMM1	&	JCMTLSG J1831210-0206203	&	8304	&	93.50	&	10.44	&	0.124	&	759\\
W40-SMM2	&	JCMTLSG J1831102-0204413	&	3834	&	56.38	&	6.77	&	0.006	&	422\\
W40-SMM3	&	JCMTLSG J1831104-0203503	&	3622	&	71.64	&	9.26	&	0.005	&	807\\
W40-SMM4	&	JCMTLSG J1831096-0206263	&	1982	&	26.72	&	3.06	&	0.011	&	189\\
W40-SMM5	&	JCMTLSG J1831212-0206563	&	5007	&	41.98	&	4.73	&	0.091	&	350\\
W40-SMM6	&	JCMTLSG J1831106-0205413	&	2322	&	54.28	&	6.54	&	0.008	&	438\\
W40-SMM7	&	JCMTLSG J1831168-0207053	&	4854	&	62.94	&	6.87	&	0.013	&	514\\
W40-SMM8	&	JCMTLSG J1831468-0204263	&	2185	&	46.21	&	5.96	&	0.003	&	405\\
W40-SMM9	&	JCMTLSG J1831388-0203353	&	3533	&	32.30	&	3.77	&	0.015	&	313\\
W40-SMM10	&	JCMTLSG J1831038-0209503	&	344	&	24.63	&	3.57	&	0.004	&	389\\
\hline
\end{tabular}\\
\raggedright
$^{a}$Position of the highest value pixel in each clump (at 850\,$\micron$). \\
$^{b}$Mean \emph{Herschel} 70\,$\micron$ intensity.\\
$^{c}$Integrated SCUBA-2 fluxes over the clump properties. The uncertainty at 450\,$\micron$ is 0.017\,Jy/pix and at 850\,$\micron$ is 0.0025\,Jy/pix. There is an additional systematic error in calibration of 10.6\,\% and 3.4\,\% at 450\,$\micron$ and 850\,$\micron$, respectively.\\
$^{d}$Mean VLA 21\,cm intensity at 15\arcsec\ pixels.\\
\label{tab:results1}
\end{table*}

\begin{figure}
\begin{center}
\includegraphics[scale=0.6]{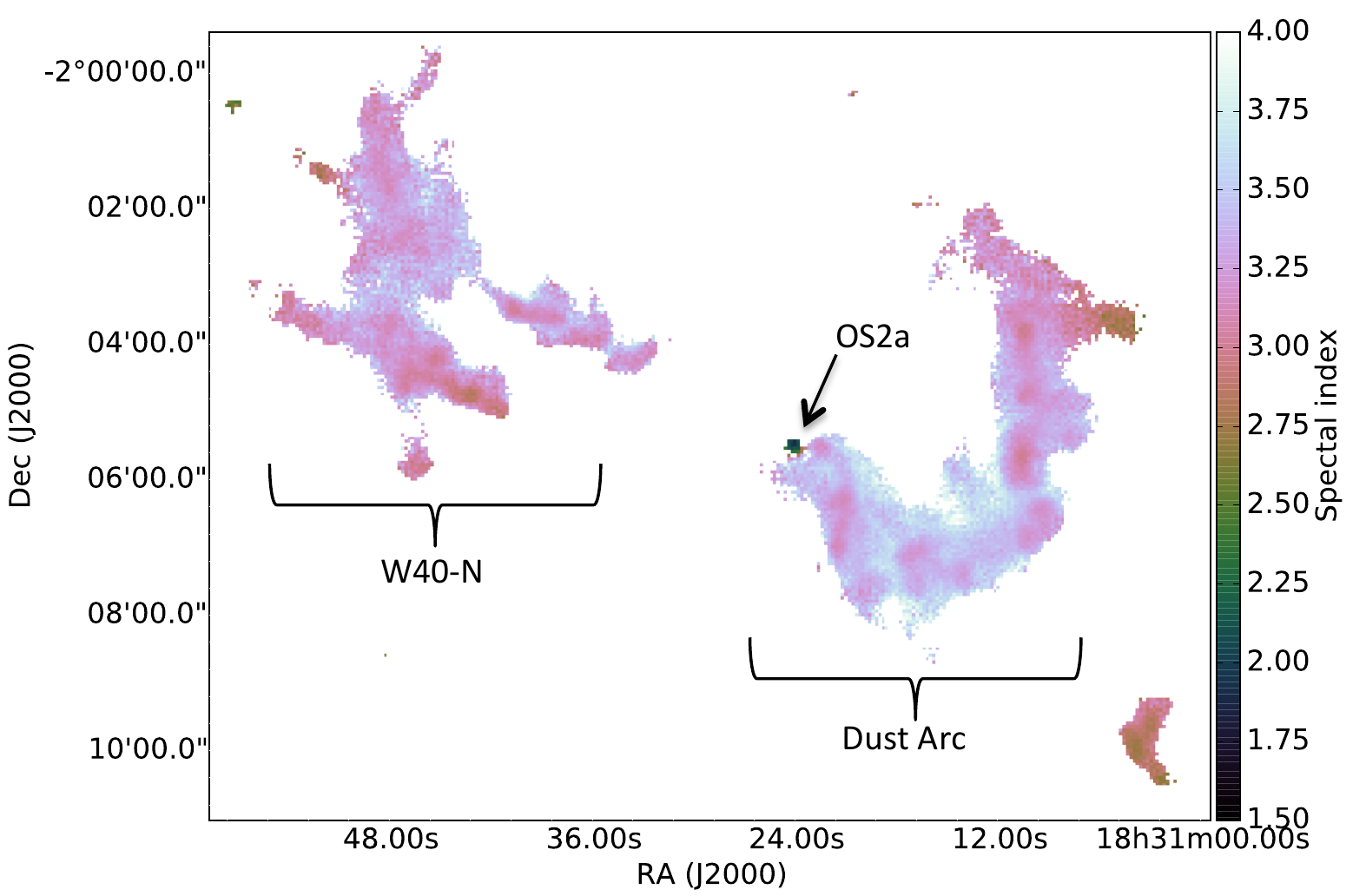}
\caption{SCUBA-2 spectral index $\alpha$ of W40-N and the Dust Arc. The spectral index for W40-S is not shown but is similar in value to W40-N.}
\label{fig:alpha}
\end{center}
\end{figure}

\begin{figure}
\begin{center}
\includegraphics[scale=0.35]{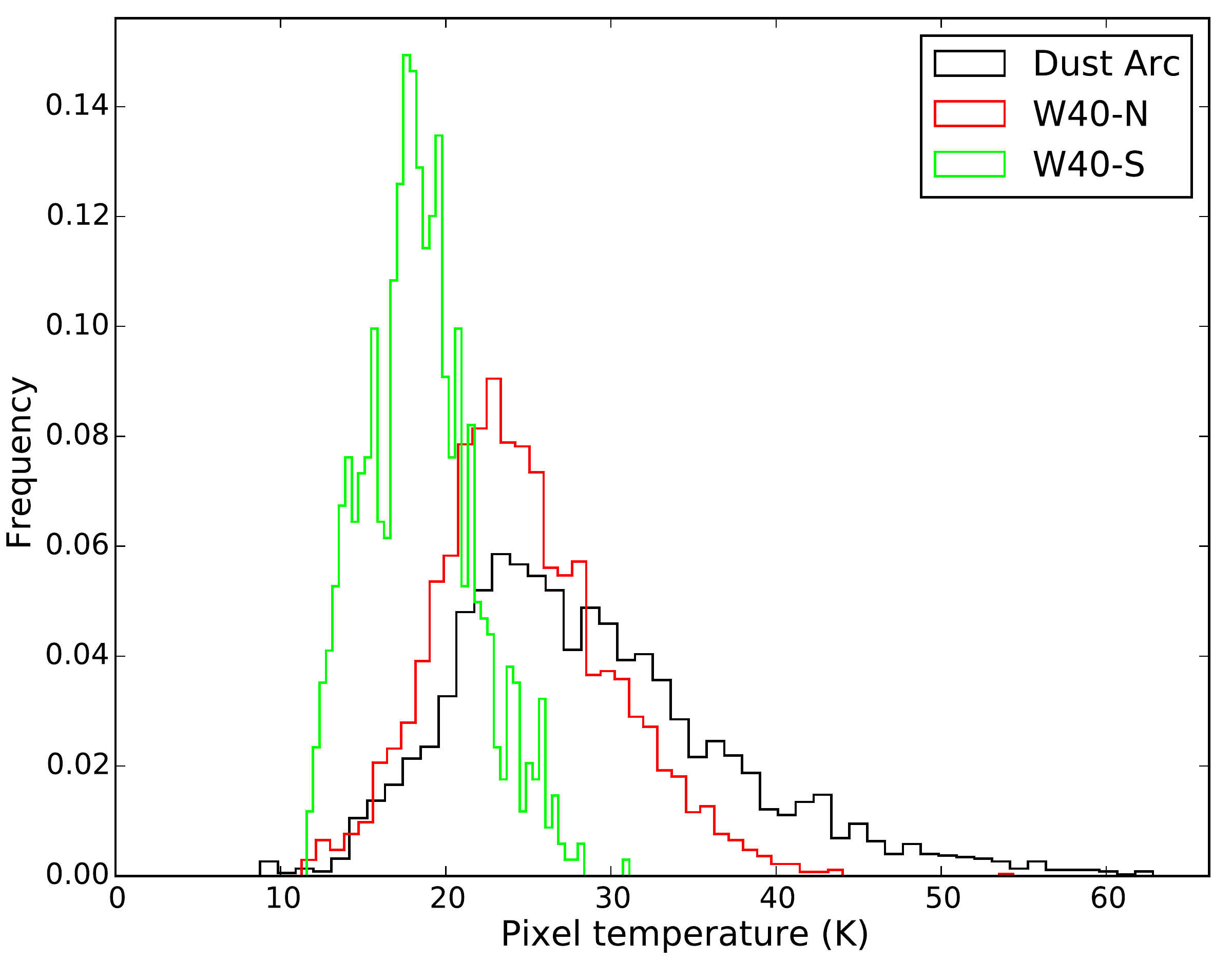}
\caption{The normalised distribution of pixel temperatures across the W40 complex clouds: Dust Arc (black, warmest), W40-N (red) and W40-S (green, coolest).}
\label{fig:temphist}
\end{center}
\end{figure}

The SCUBA-2 temperature and flux spectral index $\alpha$ of the W40 complex, 
calculated from post-CO, post-free-free reduced maps are presented in Figures 
\ref{fig:temp} and \ref{fig:alpha} respectively. The range of dust temperatures (9 to 
63\,K) in the W40 complex is presented in Figure \ref{fig:temphist} and is comparable 
to those calculated from SCUBA-2 data in NGC1333 by \cite{Hatchell:2013ij} and in 
Serpens MWC 297 by \cite{Rumble:2015vn}. 

We break the complex into major star-forming clouds. The Dust Arc, W40-N, 
and W40-S have mean temperatures of 29, 25 and 18\,K, respectively. 
Figure \ref{fig:temphist} presents the distribution of temperatures in 
these clouds. The highest temperature pixels (in excess of 50\,K) are found in 
the eastern Dust Arc whilst the lowest temperature pixels (9\,K) are associated 
with OS2a. 

A map of the SCUBA-2 flux spectral index (Figure \ref{fig:alpha}) is a more 
objective summary of the submillimetre SED. We find that $\alpha$ values are 
fairly constant across the filaments of the W40 complex with a mean $\alpha$ 
= 3.1$\pm$0.2, as expected for the ISM with $\beta$ = 1.8 and temperature 
approximately 20\,K. However, the value of $\alpha$ associated with OS2a (marked 
in Figure \ref{fig:temp}) is notably lower with a minimum value of $\alpha$ = 1.6$\pm$0.1. 

A low $\alpha$ has previously been explained by very low $\beta$, associated with 
grain growth \citep{Manoj:2007ly}, or low temperatures. \cite{Rumble:2015vn} 
demonstrated that lower spectral indices can also be caused by free-free emission 
contributing to SCUBA-2 detections, but in this case the free-free emission does not 
have a significant impact on $\alpha$. SCUBA-2 dust temperatures towards OS2a are 
some of the lowest in the whole region with values less than 9\,K (see Figure \ref{fig:temp} 
insert). Given a $\beta$=1.0, typical for circumstellar disks, $\alpha$ = 1.6 would require 
an unphysical temperature of less than 2\,K. Alternatively, an exceptionally low $\beta$ 
approaching zero would still require an excessively low temperature of less than 7\,K, 
comparable to the SCUBA-2 dust temperature of 9\,K (see Figure \ref{fig:temp} insert) 
observed at $\beta$=1.8. It is therefore unlikely that $\beta$ alone can explain these 
results. 

OS2a was detected by the VLA in \cite{Rodriguez:2010bs} in the autumn 2004 and 
noted as a variable radio source. Subsequent observations by \cite{Ortiz-Leon:2015vn} 
in summer 2011 failed to make a significant detection of OS2a, confirming the 
object as highly variable. The transient nature of OS2a could offer an alternative 
explanation for the exceptionally low dust spectral index observed by SCUBA-2 in 
the summer 2012. We note that \cite{Maury:2011ys} calculates a dust temperature of 
40$\pm$8\,K for OS2a from the 2\,$\micron$ - 1.2\,mm SED which incorporates 
observations from 2007 and 2009. Further work is required to fully address the nature 
of this source.  



\section{The SCUBA-2 clump catalogue}

\begin{figure}
\begin{center}
\includegraphics[scale=1.25]{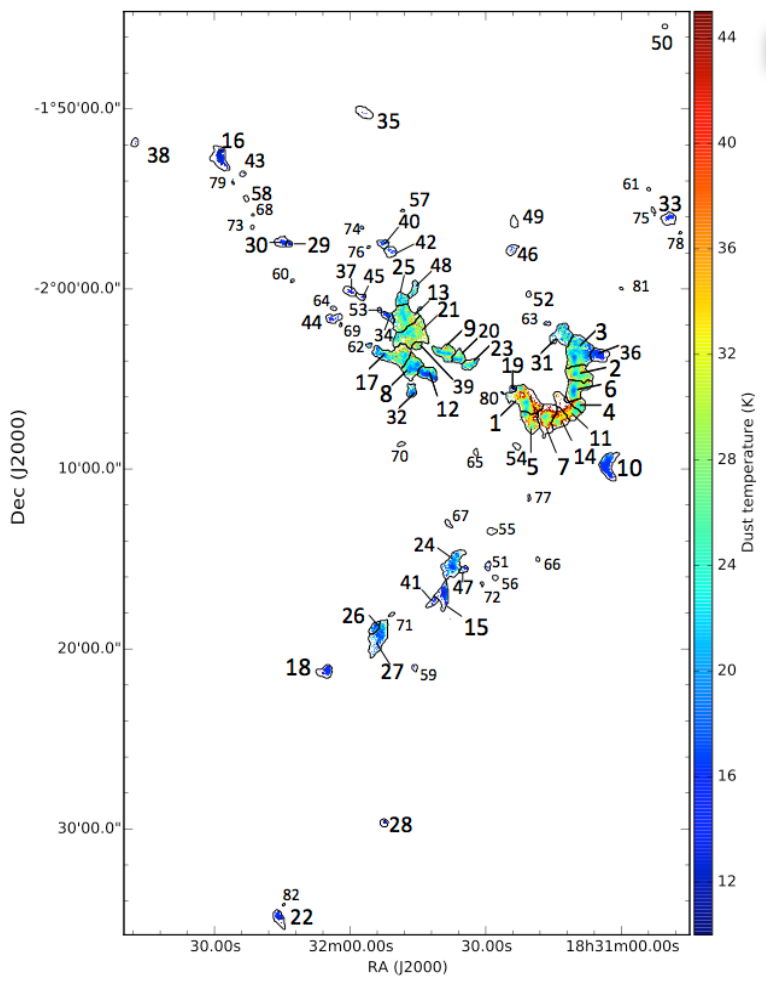}
\caption{Temperature map of the W40 complex with clumps identified in the SCUBA-2 CO subtracted, 4\arcmin\ filtered, and free-free subtracted 850\,$\micron$ data using the Starlink clump-finding algorithm \textsc{fellwalker} plotted as contours. Clumps are indexed in order of highest to lowest flux density, matching the order presented in Tables \ref{tab:results1} and \ref{tab:results2}.}
\label{fig:clumps}
\end{center}
\end{figure}

To analyse individual star-forming regions, we use the Starlink \textsc{fellwalker} 
algorithm \citep{Berry:2015uq} to identify clumps in the SCUBA-2 850\,$\micron$, CO 
subtracted, 4\arcmin\ filtered, free-free subtracted map. Details of \textsc{fellwalker} and 
the parameters used to refine clump selection are given in Appendix E.  We identify 82 
clumps in the W40 complex and their fluxes at 850\,$\micron$, as well as 70\,$\micron$, 
450\,$\micron$ and 21\,cm from \emph{Herschel}, SCUBA-2 and Archival VLA maps, 
respectively, are presented in Table \ref{tab:results1}. Clump positions in the W40 complex 
are presented in Figure \ref{fig:clumps}. The outer boundary of the clumps approximately 
corresponds to the 5\,$\sigma$ contour in Figure \ref{fig:CO} (upper panel). 

\subsection{Clump temperatures}

The unweighted mean value of temperature across all of the pixels in a clump in the W40 
complex is calculated and presented in Table \ref{tab:results2}. There are no temperature 
data for 21 clumps as they are not detected at 450\,$\micron$ above the 5\,$\sigma$ noise 
level (0.0035\,Jy per pixel). For these cases we assign a temperature of 15$\pm$2\,K, 
consistent with \cite{Rumble:2015vn}. Where temperature data only partially cover the 
850\,$\micron$ clump, we assume the vacant pixels have a temperature equal to the 
mean of the occupied pixels. 

Partial coverage tends to occur at the edges of clumps as a result of lower signal-to-noise at 450\,\micron\ flux, relative to 850\,\micron\ flux. Setting missing pixels to the clump average could introduce a temperature bias if clump edges are systematically warmer (or colder) than the clump centres. This was tested by replacing vacant pixels with the average of the top 20\% of pixels values in each clump, rather than the average of all the pixels, given the assumption that the edges of the clumps were warmer than their centres. We found the mean clump temperature increased by at most 2.2K (averaged over all clumps). From observation, only a few clumps have systematically warmer edges, whereas the majority of clumps have warm pixels randomly distributed within them. We therefore treat this value as an upper limit on any bias. 

Where a clump carries only a small number of temperature pixels, the recorded clump 
temperature is unlikely to be representative of the whole clump. We find that 20\,\% of clumps 
are missing more than 75\,\% of the total potential temperature pixels with the most prominent 
of this set being W40-SMM 35. The following discussion concerns only clumps with complete 
or partial temperature data.

Figure \ref{fig:hist} (lower left) shows the distribution of derived temperatures. The 
W40 complex has a mean clump temperature of 20$\pm$3\,K with a mean percentage 
error across all clumps of 16\% due to calibration uncertainty. The mean temperature of the 
peripheral clumps (i.e. those not attributed to W40-N, W40-S or the Dust Arc) is 15$\pm$2\,K, 
equal to that found in the Serpens MWC 297 region \citep{Rumble:2015vn} and the 
assumptions used by \cite{Johnstone:2000fk}, \cite{Kirk:2006vn} and \cite{Sadavoy:2010ve} 
for isolated clumps. These findings are consistent with those of \cite{Foster:2009ve} who 
found that isolated clumps in Perseus were systematically cooler than the those in clusters.  

We compare our clump temperatures to those calculated by \cite{Maury:2011ys} and \cite{Konyves:2015uq} using SED fitting between, 2\,$\micron$ and 1.2\,mm, for sources extracted using the \emph{getsources} algorithm. Our mean temperature of the 19 sources common to all three catalogues is 19.8$\pm$2.8\,K. This is comparable to the \cite{Maury:2011ys} value of 18.5$\pm$0.4\,K,  but higher than the \cite{Konyves:2015uq} value of 14.1$\pm$1.3\,K. Whilst all three methods calculate similar minimum source temperatures (10-12\,K), our method calculates the highest maximum source temperatures (33.6\,K, compared to 27.0\,K and 20.6\,K in \citealt{Maury:2011ys} and \citealt{Konyves:2015uq}, respectively). This is because the warmest dust lies in the low column density edges of filaments (see Figure 10) where it is likely to be omitted by the \emph{getsources} algorithm, which is optimised to find centrally condensed cores.

Figure \ref{fig:scatter1} shows clump temperature as a function of projected distance 
from OS1a. The clumps at distances greater than 1.2\,pc (marked) have 
near-constant temperatures (on average 16$\pm$3\,K), again consistent with those of 
isolated clumps (\citeauthor{Johnstone:2000fk} \citeyear{Johnstone:2000fk}, 
\citeauthor{Kirk:2006vn} \citeyear{Kirk:2006vn} and \citeauthor{Rumble:2015vn} 
\citeyear{Rumble:2015vn}). At distances of less than 1.2\,pc there is a strong negative 
correlation between temperature and projected distance to OS1a. The lower panel in 
Figure \ref{fig:scatter1} shows how clump temperature does not increase significantly 
given the presence of a protostar with all protostellar clumps having the same mean 
temperature (as a function of distance) as the starless clumps, within the calculated 
uncertainties. This suggests that internal heating from a protostar is not significant 
enough to raise the temperatures of clumps in the W40 complex. However, use of a 
constant $\beta$ = 1.8, consistent with the ISM, may mask heating in protostars where 
low values of $\beta$ have been observed (Chen et. al, submitted).

VLA 21\,cm continuum data trace free-free emission from the super-heated 
\HII\ region. Figures \ref{fig:21cm} and \ref{fig:freefree21} show the extent 
of the \HII\ region and where it coincides with several of the SCUBA-2 clumps 
in the Dust Arc and W40-N. The size of the \HII\ region corresponds to a 
0.17\,\hbox{ pc} radius, but Figure \ref{fig:scatter1} shows temperatures increasing 
inversely with radius from OS1a out to 1.2\,pc (8.25\arcmin). Figure \ref{fig:scatter2} 
shows none of the clumps within the \HII\ region has a temperature of less than 
21\,K (ignoring W40-SMM 19) and the mean clump temperature of 29\,K is 
almost twice the temperature of an isolated clump. Our conclusions support the 
\cite{Matzner:2002qf} model where radiative feedback from the OB association 
(including ionising and non-ionising photons) is the dominant external mechanism 
for heating clumps. 

\begin{table*}
\caption{The properties of a sample of submillimetre clumps in the W40 complex. The full table is available online.}
\begin{tabular}{@{}llllllllll}
Index	&	S$_{850}$$^{a}$	&	Mass$^{b}$	&	Temp$^{c}$.	&	Column density$^{d}$	&	YSO density$^{e}$	&	Protostars$^{e}$	&	M$_{\mathrm{J}}$$^{f}$	&	M/M$_{\mathrm{J}}$	&	Distance$^{g}$ \\
	&	(Jy/pixel)	&	(M$_{\odot}$)	&	(K)	&	(H$_{2}$ cm$^{-2}$)	&	(YSO$\hbox{ pc$^{-2}$}$)	&	(per clump)	&	(M$_{\odot}$)	&		&	(pc) \\
\hline
\hline
W40-SMM1	&	0.046	&	12.5$\pm$2.6	&	33.6$\pm$5.7	&	75$\pm$16 $\times$10$^{21}$	&	147	&	4	&	20.6$\pm$3.5	&	0.6$\pm$0.2	&	0.3\\
W40-SMM2	&	0.041	&	9.7$\pm$1.7	&	28.1$\pm$3.7	&	69$\pm$12 $\times$10$^{21}$	&	17	&	0	&	12.9$\pm$1.7	&	0.8$\pm$0.2	&	0.6\\
W40-SMM3	&	0.040	&	16.6$\pm$3.3	&	23.0$\pm$3.2	&	83$\pm$17 $\times$10$^{21}$	&	22	&	1	&	14.6$\pm$2.0	&	1.1$\pm$0.3	&	0.7\\
W40-SMM4	&	0.040	&	4.0$\pm$0.6	&	30.2$\pm$3.9	&	62$\pm$10 $\times$10$^{21}$	&	26	&	1	&	9.2$\pm$1.2	&	0.4$\pm$0.1	&	0.7\\
W40-SMM5	&	0.037	&	5.6$\pm$1.1	&	33.1$\pm$5.5	&	57$\pm$12 $\times$10$^{21}$	&	86	&	1	&	13.8$\pm$2.3	&	0.4$\pm$0.1	&	0.3\\
W40-SMM6	&	0.036	&	9.9$\pm$1.7	&	27.8$\pm$3.7	&	73$\pm$12 $\times$10$^{21}$	&	21	&	1	&	12.9$\pm$1.7	&	0.8$\pm$0.2	&	0.6\\
W40-SMM7	&	0.030	&	7.3$\pm$1.5	&	35.8$\pm$6.0	&	44$\pm$9 $\times$10$^{21}$	&	47	&	0	&	18.1$\pm$3.0	&	0.4$\pm$0.1	&	0.5\\
W40-SMM8	&	0.030	&	10.8$\pm$1.7	&	23.8$\pm$2.8	&	60$\pm$10 $\times$10$^{21}$	&	25	&	1	&	10.6$\pm$1.3	&	1.0$\pm$0.2	&	0.7\\
W40-SMM9	&	0.029	&	6.0$\pm$1.2	&	26.3$\pm$4.2	&	48$\pm$10 $\times$10$^{21}$	&	56	&	0	&	10.4$\pm$1.6	&	0.6$\pm$0.2	&	0.5\\
W40-SMM10	&	0.027	&	10.5$\pm$1.8	&	16.5$\pm$1.6	&	80$\pm$14 $\times$10$^{21}$	&	36	&	2	&	7.3$\pm$0.7	&	1.5$\pm$0.3	&	1.1\\
\hline
\end{tabular}\\
\raggedright
$^{a}$Peak SCUBA-2 850\,$\micron$ flux of each clump. The 850\,$\micron$ uncertainty is 0.0025\,Jy/pix. There is a systematic error in calibration of  3.4\,\%.\\
$^{b}$As calculated with Equation \ref{eqn:mass}. These results do not include the systematic error in distance (10\,\%) or opacity (up to a factor of two).\\
$^{c}$Mean temperature as calculated from the temperature maps. Uncertainty is the mean pixel standard deviation, as propagated through the method. Where no temperature data are available an arbitrary value of 15$\pm$2\,K is assigned that is consistent with previous authors (\citeauthor{Johnstone:2000fk} \citeyear{Johnstone:2000fk}, \citeauthor{Kirk:2006vn},\citeyear{Kirk:2006vn}, \citeauthor{Rumble:2015vn}\citeyear{Rumble:2015vn}).\\
$^{d}$Peak column density of the clump. These results do not include the systematic error in distance or opacity.\\
$^{e}$Calculated from the composite YSO catalogue outlined in Section 2.3.\\
$^{f}$As calculated with Equation \ref{eqn:sadavoy}. These results do not include the systematic error in distance.\\
$^{g}$Projected distance between clump and OS1a, the primary ionising star in the W40 complex OB association. \\
\label{tab:results2}
\end{table*}

\subsection{Clump column densities and masses}

\begin{figure*}
\begin{center}
\includegraphics[scale=0.9]{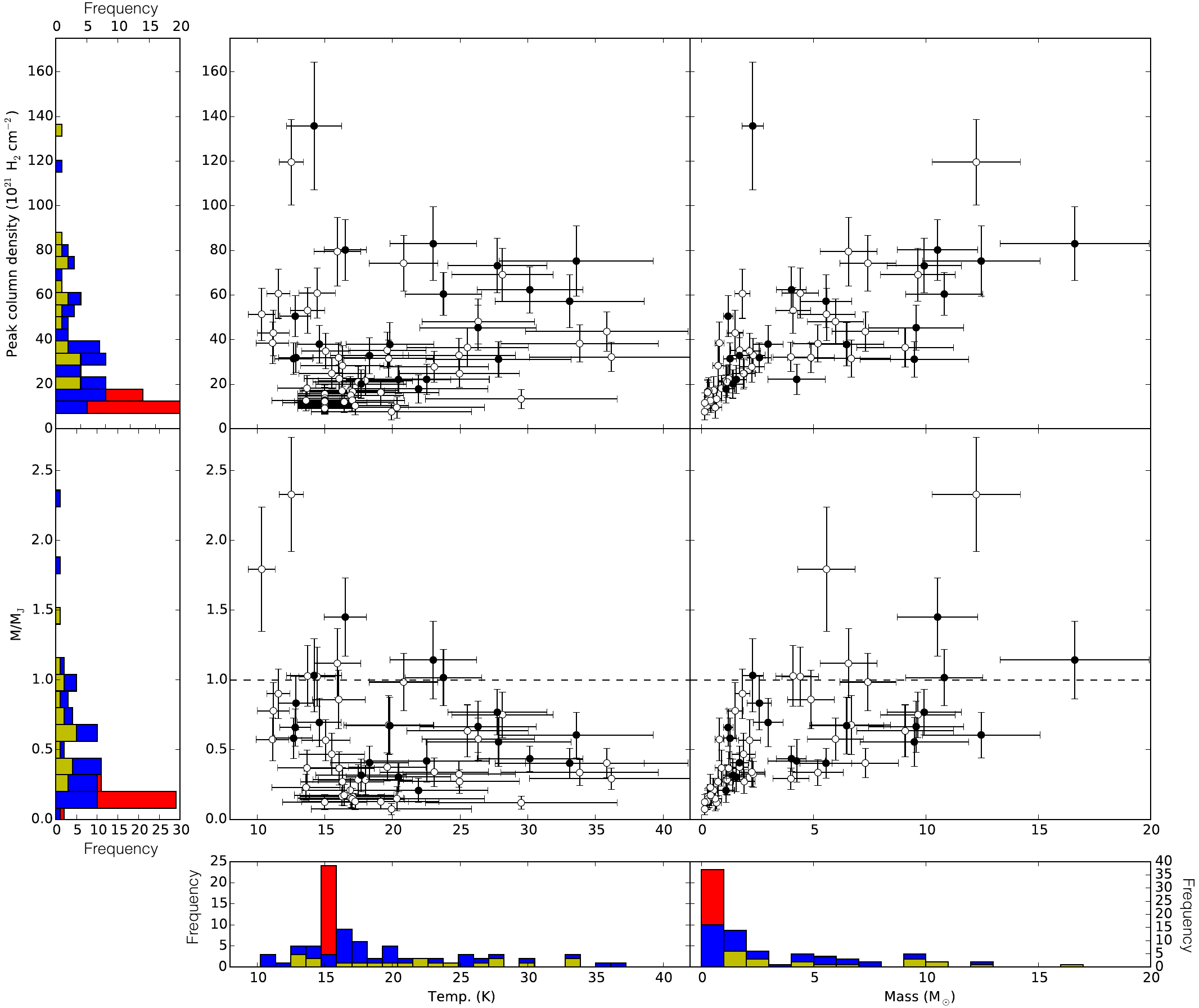}
\caption{Properties of clumps within the W40 complex. Histograms show the distribution of mass, temperature, $\mathrm{M/M_{J}}$ and peak column density. Total height of each bar represents the total number of clumps in each bin whereas each colour represents a specific subset: blue for clumps that have measured temperature, yellow for clumps with measured temperatures that contain at least one YSO, and red for clumps with no temperature data, for which the value 15$\pm$2\,K is assumed.
Scatter plots show any existing correlation between clump properties: top left shows column density as a function of temperature, top right shows column density as a function of mass, bottom left shows Jeans stability, M$_{850}$/M$_{\mathrm{J}}$, as a function of temperature and bottom right shows Jeans stability as a function of mass. Filled markers indicate protostellar clumps and hollow markers indicate starless clumps.}
\label{fig:hist}
\end{center}
\end{figure*}

Masses of the clumps in the W40 complex are calculated by assuming a single 
temperature grey body spectrum \citep{Hildebrand:1983fy}. We follow the standard 
method for calculating clump mass for a given distance, $d$, and dust opacity, 
$\kappa_{\mathrm{850}}$, (\citeauthor{Johnstone:2000fk} \citeyear{Johnstone:2000fk}; 
\citeauthor{Kirk:2006vn} \citeyear{Kirk:2006vn}; \citeauthor{Sadavoy:2010ve} 
\citeyear{Sadavoy:2010ve}; \citeauthor{Enoch:2011lh} \citeyear{Enoch:2011lh}). 
Clump masses are calculated by summing the SCUBA-2 850\,$\micron$ flux, per 
pixel $i$ (in Jy per pixel) using

\begin{eqnarray}
M & = 0.39 \sum_{i} S_{{\mathrm{850},i}}\left [\exp \left(\frac{17\,\mathrm{K}}{T_{\mathrm{d},i}} \right) - 1 \right ] & \nonumber \\
& \times \left(\frac{d}{250\,\mathrm{pc}} \right)^{2}\left(\frac{\kappa_{\mathrm{850}}}{0.012\,\mathrm{\hbox{cm}^{2} \hbox{g}^{-1}}} \right)^{-1}\,\mathrm{M_\odot}.
\label{eqn:mass}
\end{eqnarray}
The dust opacity, $\kappa_{\mathrm{850}}$, is given in Equation \ref{eqn:beta} and we assume a distance $d = 500\pm50\hbox{ pc}$ following \cite{Mallick:2013kx}, as outlined in Section 1.

We can also incorporate temperature measurements alongside the SCUBA-2 
850\,$\micron$ fluxes (Equation \ref{eqn:mass}) to calculate pixel column densities, 
N$_{i,H_{2}}$,  from pixel masses, M$_{i}$, using the pixel area, A$_{i}$, and the mean 
molecular mass per H$_{2}$, $\mu$=2.8 \citep{kauffmann08}, 

\begin{equation}
N_{i,H_{2}}=\frac{M_{i}}{\mu _{H_{2}}m_{p}A_{i}}.
\label{eqn:CD}
\end{equation}
Peak column densities per clump are also presented in Table \ref{tab:results2}.  Figure \ref{fig:CD} 
presents a map of column density. Figure \ref{fig:hist} (upper left) shows the distribution of peak 
column densities in the W40 complex as a function of temperature.



We find a minimum peak column density of 1.7$\times$10$^{22}$ H$_{\mathrm{2}}$ cm$^{-2}$ for clumps containing a protostar. Above this value there is no significant correlation between peak column density and temperature or protostellar occupancy. The upper right panel of Figure \ref{fig:hist} shows how peak column density is tightly correlated with mass in the clumps. Above 3\,M$_{\odot}$, the correlation is looser with several examples of clumps of similar column density having masses varying between 3 and 12\,M$_{\odot}$. 

The peak column density of \emph{Herschel} sources, detected by \cite{Konyves:2015uq}, are compared to 69 matching (within 15\arcsec, one JCMT beam width at 850\,$\micron$) SCUBA-2 clumps as this value is independent of clump size. Figure \ref{fig:CDcomp} shows that the two sets are loosely correlated. The mean peak column density of the SCUBA-2 clumps (3.2$\pm$0.7$\times$10$^{\mathrm{22}}$\,H$_{\mathrm{2}}$\,cm$^{-2}$) is comparable to that of the \emph{Herschel} sources (2.6$\times$10$^{\mathrm{22}}$\,H$_{\mathrm{2}}$\,cm$^{-2}$). It is notable that the majority of objects have a lower peak column density recorded by \emph{Herschel} than by SCUBA-2. This can be explained by the SED fitting method used by \cite{Konyves:2015uq} which can be biased towards higher temperature clouds, and has a lower resolution of 36.6\arcsec\ (consistent with the \emph{Herschel} 500\,$\micron$ beam size).

We calculate a lower limit on the average volume density along the line of sight for clumps from 
the ratio of peak column density and clump depth (assumed equal to the flux weighted clump 
diameter as calculated by \textsc{fellwalker}, which is equivalent to the clump FWHM, see 
Appendix E for more info). 



From our sample of clumps, Table \ref{tab:dense} lists 31 `dense cores' with a volume density greater 
than 10$^{5}$\,cm$^{-3}$, along with any protostars within these clump and their respective Jeans 
stabilities (see section 6.3). At densities greater than 10$^{4.5}$\,cm$^{-3}$ we can be confident 
the dust and gas temperatures are well coupled (\citealt{Goldreich:1974kx} and \citealt{Goldsmith:2001fk}). 
Dense cores account for approximately 63\% of the mass observed by SCUBA-2 at 850\,$\micron$. 
The Dust Arc has nine dense cores, W40-N has nine, W40-S has four, and there are 11 isolated dense 
cores. In total, 42\% of the dense cores contain at least one protostar, confirming that a significant 
proportion of clumps in the W40 complex is likely to be undergoing star-formation.

\subsection{Clump stability}

Jeans stability \citep{Jeans:1902dz} of the clumps is measured by a critical ratio, above 
which self-gravity will overwhelm thermal support in an idealised cloud of gas, causing 
it to collapse and begin star-formation.The condition for collapse is defined as when the 
mass of a clump, M$_{850}$, is greater than its Jeans mass, 

\begin{equation}
M_{\mathrm{J}} = 1.9\left(\frac{\bar{T}_{\mathrm{g}}}{10\,\mathrm{K}} \right)\left(\frac{R_{\mathrm{c}}}{0.07\,\mathrm{pc}} \right)\,\mathrm{M}_{{{\sun}}}, 
\label{eqn:sadavoy}
\end{equation}
where R$_{\mathrm{c}}$ is an effective radius produced by \textsc{fellwalker} from the clump area (in pixels) assuming spherical structure. This measure of R$_{\mathrm{c}}$ is typically twice the flux weighted clump radius and better represents the complete extent of the clump. The gas temperature $\bar{T_{g}}$ (assumed isothermal in the theory) is taken to equal the mean dust temperature of the clump. We note that the dust and gas may be poorly coupled for clumps not included in our dense cores list ($n > 10^5\hbox{ H}_2\hbox{ cm}^{-3}$; see Table \ref{tab:dense}); if gas temperatures drop below the dust temperature \citep{Tielens:1985uq}, their Jeans stability may be overestimated. Additional forces such as magnetism and turbulence could provide additional support against gravitational collapse (as explained in, \citealt{Sadavoy:2010ve} and \citealt{Mairs:2014zr}). 

Due to the high optical depth of our $^{12}$CO 3-2 line data, it is not possible to use these data to calculate the turbulent support.  \cite{Arzoumanian:2011kx} calculate a sonic scale of 0.05--0.15\,pc below which the sound speed is comparable to the velocity dispersion and turbulent pressure dominates over thermal pressure. This scale range includes 83\% of all the clumps in the W40 complex so our working assumption (in the absence of direct measurements) is that the majority of the W40 complex is subsonic or transonic. This is supported by observations of starless cores in Ophiuchus, which is similarly located on the edge of an OB association; these cores are transonic or mildly supersonic \citep{Pattle:2015ys}. Providing the W40 complex is similar, then clumps could be stable against gravitational collapse to a few times the Jeans mass. Magnetic support is poorly characterised and also left out of our stability analysis.



Jeans stability (M$_{850}$/M$_{\mathrm{J}}$) values for each clump are presented in Table \ref{tab:results2}, Table \ref{tab:dense} and Figure \ref{fig:hist} (lower left). Out of 82 clumps in the W40 complex, we find that 10 are unstable with M$_{850}$/M$_{\mathrm{J}} \geq$ 1. Given the likely variety in clump morphologies, only those with M$_{850}$/M$_{\mathrm{J}} \geq$ 2 can be considered truly unstable \citep{Bertoldi:1992rt}; the stability threshold is raised further if turbulent support is significant.

As with column density, we find M$_{850}$/M$_{\mathrm{J}}$ is tightly correlated with mass below 3\,M$_{\odot}$ and more loosely correlated above 3\,M$_{\odot}$. We can also determine that M$_{850}$/M$_{\mathrm{J}}$ is loosely negatively correlated with temperature; see bottom left panel of Figure \ref{fig:hist}. Below 24\,K there is a mixture of stable and unstable clumps but above this temperature all clumps are stable.

W40-SMM 14 is an example of a clump with high M$_{850}$/M$_{\mathrm{J}}$ (0.3) and high 
temperature (36\,K). Measuring the clump radius as 0.05\,pc and flux as 3.96\,Jy, we can estimate 
using Equations \ref{eqn:mass} and \ref{eqn:sadavoy} that if this clump had a typical temperature 
of 15\,K, it would have a mean M$_{850}$/M$_{\mathrm{J}}$ of 6.5. Whilst this number is only an 
estimate, it is over 20 times the measured value and therefore we are confident that the raised 
temperature of this clump is reducing M$_{850}$/M$_{\mathrm{J}}$ and potentially suppressing 
collapse. The two most unstable clumps are W40-SMM 16 (2.3$\pm$0.4) and 35 (1.8$\pm$0.4) 
which are cold, isolated clumps on the periphery of the W40 complex. W40-SMM 16 contains a 
protostar whereas 35 is currently starless.

The eastern Dust Arc is positioned on the edge of the \HII\ region. Raised temperatures 
mean that clumps here have a mean Jeans mass of 17\,M$_{\odot}$ for clumps in the 
eastern Dust Arc, compared to 12\,M$_{\odot}$ in the western Dust Arc, and 5\,M$_{\odot}$ 
for the average clump in the W40 complex. Likewise the median M$_{850}$/M$_{\mathrm{J}}$ 
is 0.4 compared to 0.8 in the western Dust Arc which is considered outside of the \HII\ region 
and consequently has lower temperatures. Note that both filaments have similar mean clump 
masses of these regions (7 and 8\,M$\odot$, respectively). Given its common CO gas velocity 
(Figure \ref{fig:CO} upper) the Dust Arc, as a whole, is likely a continuous filament, and therefore 
we might expect its clumps to evolve at a similar rate due to similar initial conditions along the 
length of the filament. Significant differences in stabilities along the length of the filament as a 
result of heating by the OB association, however, hint that star-formation may take place there 
at different rates. 

We further examine the impact of radiative heating by the OB association on the global 
sample of clumps in the W40 complex by comparing the stabilities of the population inside 
the nebulosity to those on the outside. The limit of the nebulosity is defined as where the 
mean 70\,$\micron$ flux from \emph{Herschel} is less than 1000\,MJy/Sr. The M$_{850}$/M$_{\mathrm{J}}$ of 
interior and exterior populations, as a function of column density, is plotted in Figure 
\ref{fig:jeans}. A degree of correlation is expected as both M$_{850}$/M$_{\mathrm{J}}$ and column 
density are derived from our SCUBA-2 data. Two correlations are observed, with a clear 
divergence between the two clump populations. A clump found within the nebulosity is 
more likely to be stable than one with the same peak column density on the outside.

Figure \ref{fig:jeans} provides direct evidence that radiative heating from the OB association 
is directly influencing the Jeans stability of clumps and star-formation within Sh2-64 (the red 
population). We note that whilst this divergence is prominent amongst clumps with high 
column densities, the two populations have similar distributions below 55$\times$10$^{21}$ 
H$_{2}$ cm$^{-2}$ (within the uncertainties). 

\begin{figure}
\begin{center}
\includegraphics[scale=0.4]{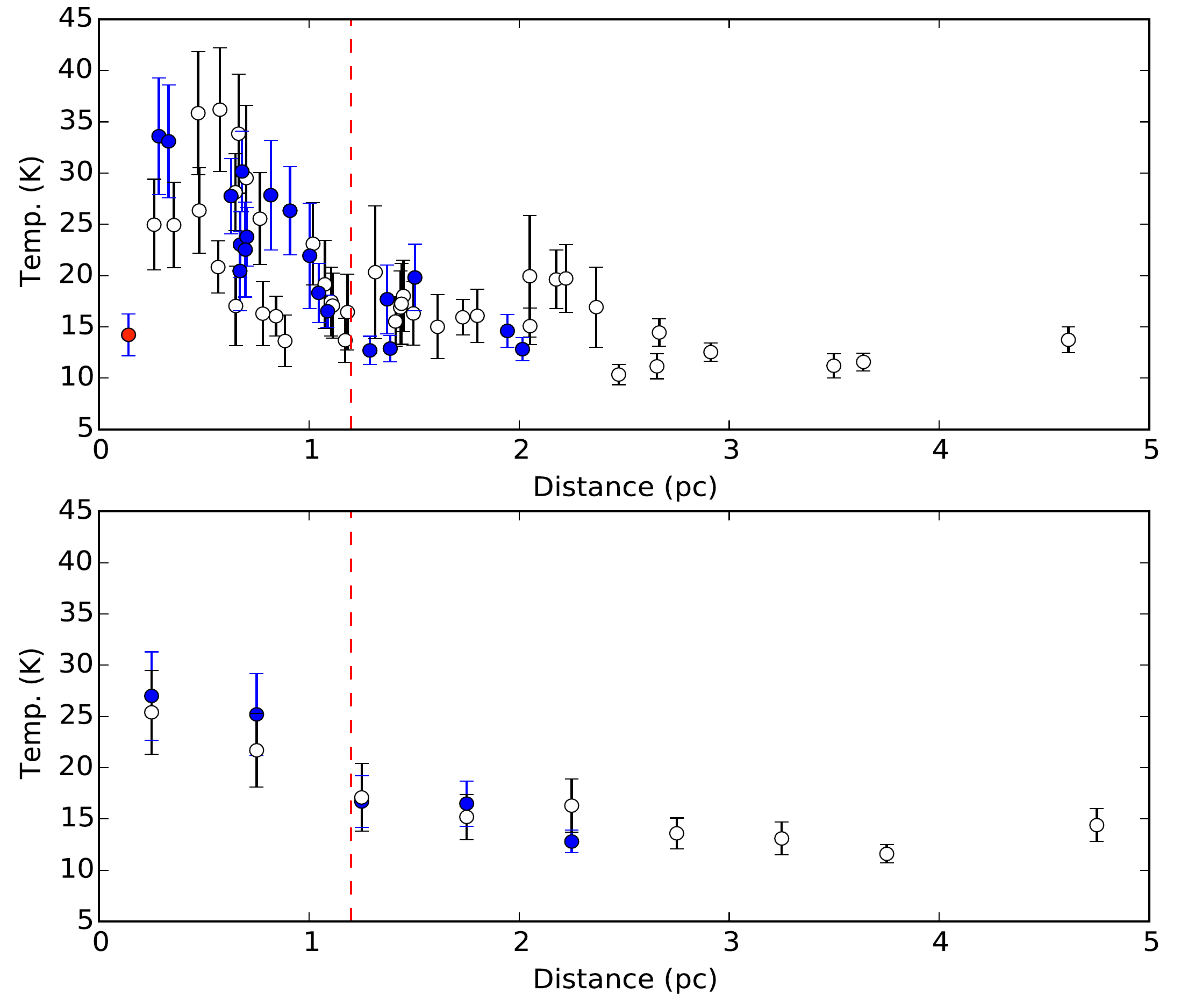}
\caption{Clump temperature as a function of projected distance, in pc, to OS1a, the most luminous star in the W40 complex. The upper plot shows all clump individually plotted. The lower plot shows the mean temperature (and uncertainty) of clumps in 0.5\,pc bins. In both plots filled markers represent protostellar clumps whereas hollow markers represent starless clumps.  W40-SMM 19 is considered anomalous and its data are flagged as a red point. Red lines mark the partition between clump temperature regimes.}
\label{fig:scatter1}
\end{center}
\end{figure}

\begin{figure}
\begin{center}
\includegraphics[scale=0.4]{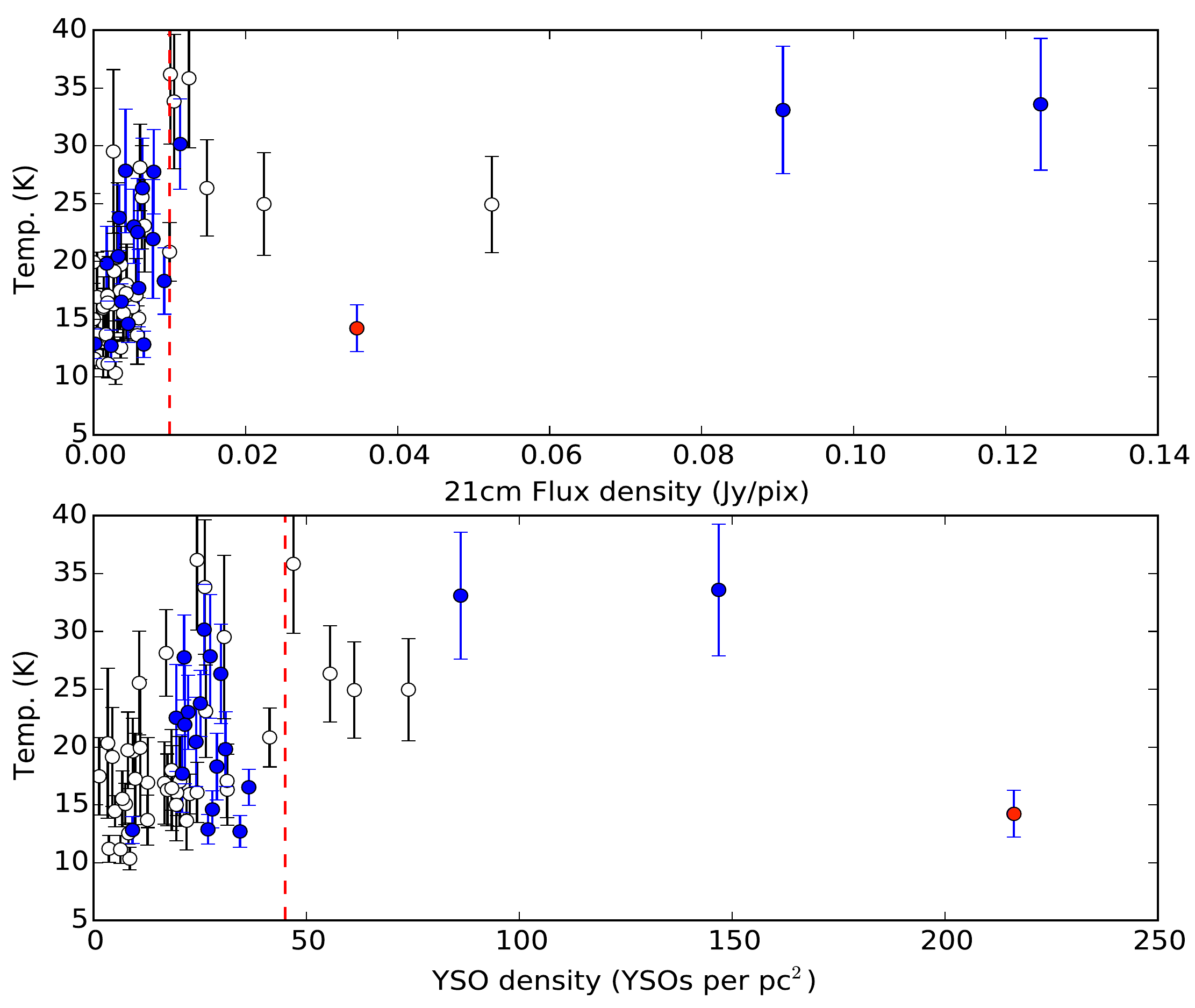}
\caption{Clump temperature as a function of: (upper) mean VLA 21\,cm flux detected in the area of each clump; and (lower) YSO surface density. In both plots, filled markers represent protostellar clumps whereas hollow markers represent starless clumps.  W40-SMM 19 is considered anomalous and its data are flagged as a red point. Red lines mark the partition between clump temperature regimes.}
\label{fig:scatter2}
\end{center}
\end{figure}

\begin{table}
\caption{Dense cores in the W40 complex.}
\begin{tabular}{|llllcl}
Clump ID & Region$^{a}$	&Diameter$^{b}$ & Density$^{c}$                      & Proto- & M/M$_{\mathrm{J}}$ \\
(W40-SMM)&	&  (pc)   & (cm$^{-3}$) & stars   &                    \\
\hline
\hline
1        &	E-DA	& 0.20 & 1.2$\times$10$^{5}$                          & 4          & 0.6$\pm$0.2        \\
2        & 	W-DA	& 0.14 &  1.5$\times$10$^{5}$                          & -          & 0.8$\pm$0.2        \\
3        &	W-DA	& 0.21 & 1.2$\times$10$^{5}$                          & 1          & 1.1$\pm$0.3        \\
4        & 	W-DA	& 0.10 & 2.1$\times$10$^{5}$                          & 1          & 0.4$\pm$0.1        \\
5        & 	E-DA	& 0.16 & 1.2$\times$10$^{5}$                          & 1          & 0.4$\pm$0.1        \\
6        & 	W-DA	& 0.17 &  1.4$\times$10$^{5}$                          & 1          & 0.8$\pm$0.2        \\
8        & 	W40-N	& 0.14 &  1.4$\times$10$^{5}$                          & 1          & 1.0$\pm$0.2        \\
9        & 	W40-N	& 0.13 & 1.2$\times$10$^{5}$                          & -          & 0.6$\pm$0.2        \\
10       & 	ISO		& 0.16 &  1.7$\times$10$^{5}$                          & 2          & 1.5$\pm$0.3        \\
12       & 	W40-N	&  0.13 & 1.8$\times$10$^{5}$                          & -          & 1.0$\pm$0.2        \\
15       & 	W40-S	& 0.13 & 2.0$\times$10$^{5}$                          & -          & 1.1$\pm$0.2        \\
16       & 	ISO		& 0.15 &  2.6$\times$10$^{5}$                          & -          & 2.3$\pm$0.4        \\
18       & 	W40-S	&  0.09 & 2.1$\times$10$^{5}$                          & -          & 1.0$\pm$0.2        \\
20       & 	W40-N	& 0.10 & 1.1$\times$10$^{5}$                          & -          & 0.3$\pm$0.1        \\
22       &	ISO		& 0.1 & 1.7$\times$10$^{5}$                            & n/a$^{d}$          & 1.0$\pm$0.2        \\
26       & 	W40-S	& 0.08 & 1.5$\times$10$^{5}$                          & -          & 0.4$\pm$0.1        \\
28	  &	ISO 		& 0.05 & 3.9$\times$10$^{5}$			     & n/a$^{d}$  	& 0.9$\pm$0.2 \\
29	  & 	ISO		& 0.04 & 3.3$\times$10$^{5}$			     & 1	&	0.7$\pm$0.1	\\
30       &	ISO		& 0.08 & 1.3$\times$10$^{5}$                          & -          & 0.6$\pm$0.1        \\
33       &	ISO		& 0.10 & 1.2$\times$10$^{5}$                          & 2          & 0.7$\pm$0.2        \\
34       & 	W40-N	& 0.09 & 1.2$\times$10$^{5}$                          & 1          & 0.4$\pm$0.1        \\
35       &	ISO		& 0.11 & 1.5$\times$10$^{5}$                          & -          & 1.8$\pm$0.4        \\
37       &	ISO		& 0.09 & 1.2$\times$10$^{5}$                          & 1          & 0.8$\pm$0.2        \\
38	  &	ISO		& 0.06 & 2.4$\times$10$^{5}$			     & -	& 0.8$\pm$0.2 		\\
43	  & 	ISO		& 0.04 & 2.5$\times$10$^{5}$			     & -	&0.6$\pm$0.2 		\\
45	  &	ISO		& 0.07 & 1.5$\times$10$^{5}$			     & 1	&0.6$\pm$0.1 		\\
47	  &	W40-S	& 0.06 & 1.5$\times$10$^{5}$			     & - 	&0.3$\pm$0.1 		\\
52	  &	ISO		& 0.04 & 1.1$\times$10$^{5}$			     & -	&0.2$\pm$0.1 		\\
53	  &  	W40-N	& 0.03 & 1.1$\times$10$^{5}$			     &-		&0.2$\pm$0.1 		\\
62	  &	W40-N	& 0.04 & 1.1$\times$10$^{5}$			     &-		&0.1$\pm$0.1 		\\
\hline
\end{tabular}\\
\raggedright
$^{a}$ Region key: eastern Dust Arc (E-DA), western Dust Arc (W-DA), W40-N, W40-S and isolated clumps (ISO).\\
$^{b}$ Flux weighted effective diameter as calculated by the clump-finding algorithm \textsc{fellwalker} (values were not deconvolved with respect to the JCMT beam).
\\
$^{c}$ The average volume density of a dense core along the line of sight. Each observation is a lower limit as effective size of the cloud is typically larger than a core. A dense core is defined where the density limit is greater than 10$^{5}$\,cm$^{-3}$, a value five times greater than the typical density of a star forming filament (2$\times$10$^{4}$\,cm$^{-3}$, \citeauthor{Andre:2014kx} \citeyear{Andre:2014kx}) and where the gas and dust temperatures are well coupled \citep{Goldsmith:2001fk}.\\       
$^{d}$ Clumps beyond the coverage of our composite YSO catalogue.\\
\label{tab:dense}
\end{table}

\subsection{YSO distribution}

\begin{figure*}
\begin{center}
\includegraphics[scale=1]{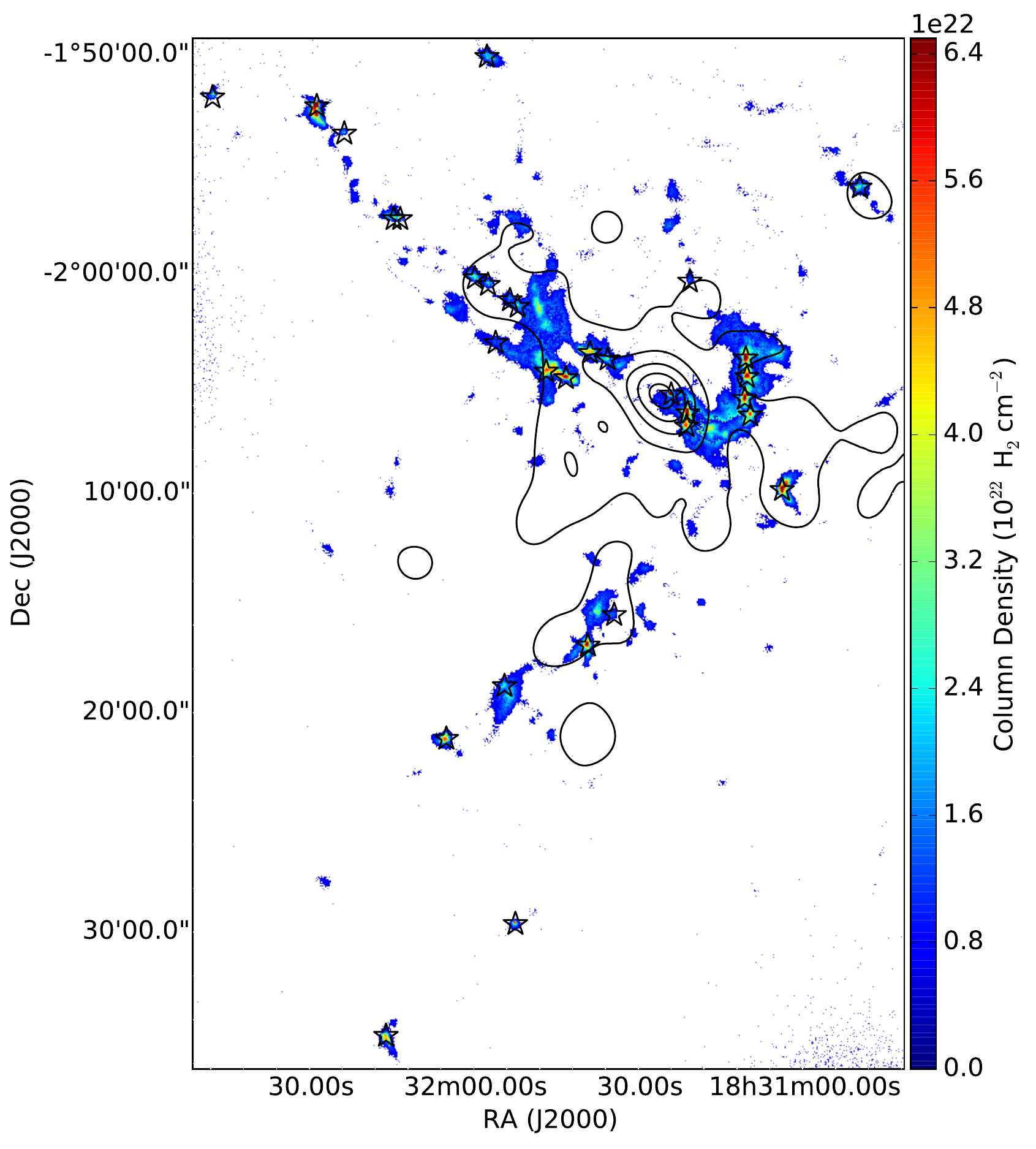}
\caption{SCUBA-2 column density map of the W40 complex. Column density is shown for areas where emission is detected at or above S/N = 3 at 850\,$\micron$. Column density is calculated using temperatures as presented in Figure \ref{fig:temp} where available and a constant value of 15\,K where a derived temperature is not. Contours describe YSO surface density with levels at 20, 60, 110, 160 and 210 YSO$\hbox{ pc$^{2}$}$. Stars mark the location of each dense core identified in Table \ref{tab:dense}.}
\label{fig:CD}
\end{center}
\end{figure*}

In this section we consider the YSO distribution based on the composite YSO catalogue 
produced from the SGBS list merged with the catalogues published by \cite{Kuhn:2010kl}, 
\cite{Rodriguez:2010bs}, \cite{Maury:2011ys} and \cite{Mallick:2013kx} (see Section 2.3 
for full details). The locations of YSOs in our composite catalogue are plotted 
in Figure \ref{fig:H70} with Class 0/I protostars and Class II/III PMS-stars denoted separately. 

The YSO distribution was mapped by convolving the YSO positions with a 2\arcmin\ FWHM 
Gaussian to produce a surface density map with units of YSOs pc$^{-2}$ as shown in 
Figure \ref{fig:CD}. The stellar cluster is visible in Figure \ref{fig:CD} and has a FWHM size 
of approximately 3.5\arcmin $\times$ 2.5\arcmin. The Dust Arc has its eastern end located 
towards the centre of the star cluster where the density peaks at 232 YSOs$\hbox{ pc$^{-2}$}$. 
However, this value quickly drops off to 20 YSOs$\hbox{ pc$^{-2}$}$ at its western edge 
near W40-SMM 31. 

An increase in clump temperatures is also observed when the YSO surface density 
is greater than 45 YSO$\hbox{ pc$^{-2}$}$ (Figure \ref{fig:scatter2} lower, marked).
Our YSO surface density map does not distinguish between embedded protostars 
and free-floating PMS-stars and therefore the YSO densities will be over-estimates 
of the densities of objects embedded within clumps. Given these uncertainties, we 
conclude that the radiative feedback from OS1a is dominating over any potential 
heating by the embedded YSO within this region.

The absolute number of protostars located within each clump was recorded. 
A total of 21/82 clumps have at least one Class 0/I protostar. Figure \ref{fig:hist} shows 
how the distribution of clumps with protostars, compared to those without, is shifted to 
greater values in mass (5.5$\pm$1.3 from 2.0$\pm$0.6\,M$_{\odot}$), column 
density (49$\pm$11 from 26$\pm$7$\times$10$^{21}$\,H$_{2}$\,cm$^{-2}$), temperature 
(21$\pm$3 from 18$\pm$3\,K) and M$_{850}$/M$_{\mathrm{J}}$ (0.6$\pm$0.2 from 0.4$\pm$0.1). 
These results are consistent with those of \cite{Foster:2009ve} who observed that 
protostellar clumps appear warmer, more massive and more dense than starless 
clumps in Perseus. 

For all cases in Figure \ref{fig:hist}, a threshold is observed, below which 
protostars are not found in clumps. These results argue that more massive, dense 
clumps are more likely to be unstable and contain a Class 0/I object. They also suggest 
that these clumps may be warmer, though the significant overlap in the temperature 
range renders this result inconclusive. 



\begin{figure}
\begin{center}
\includegraphics[scale=0.35]{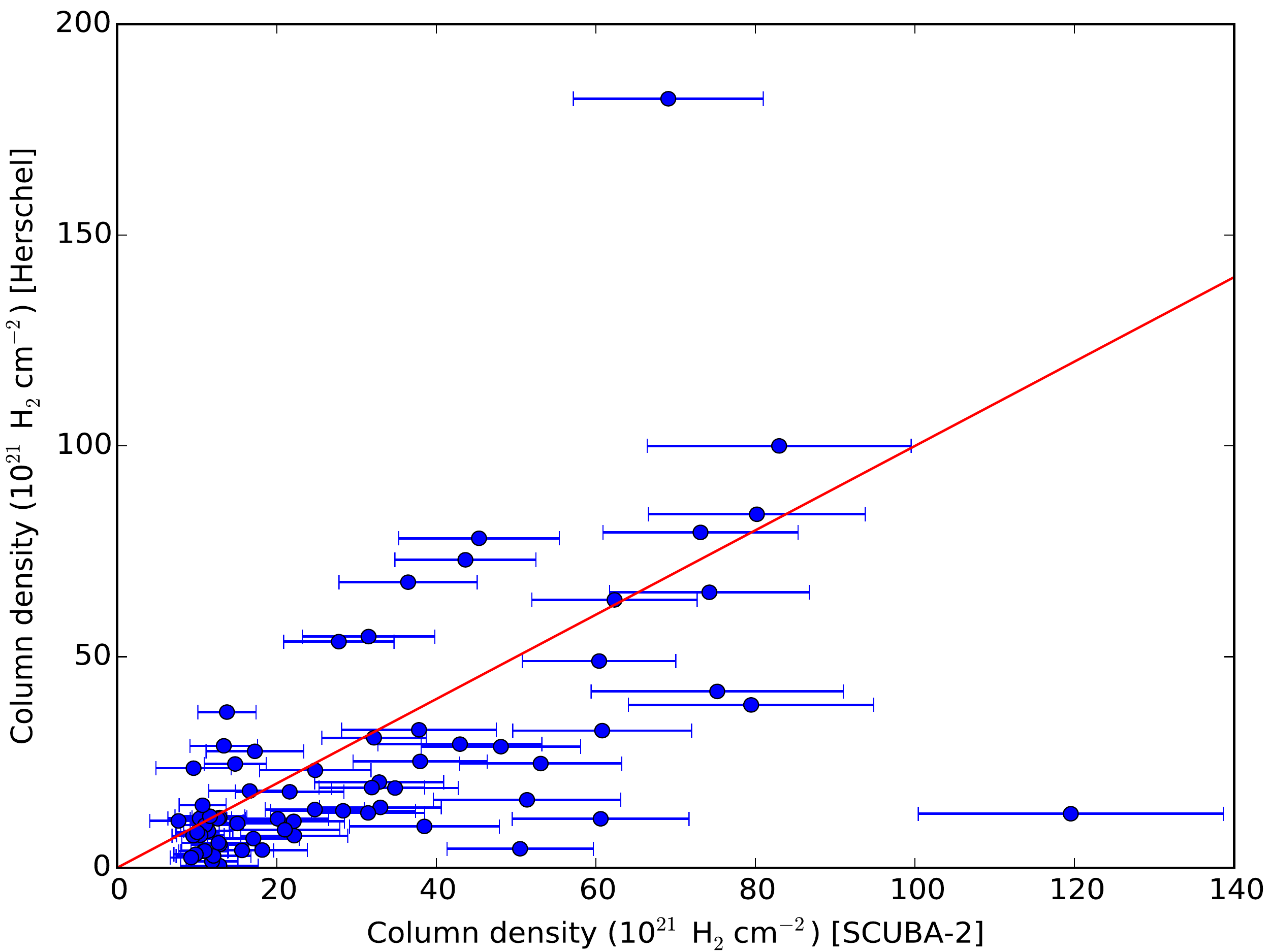}
\caption{The peak column density of SCUBA-2 clumps compared to \emph{Herschel} sources, as published by K$\mathrm{\ddot{o}}$nyves et al. (2015). A parity line is marked in red. Clump pairs are matched within a separation of 15\arcsec. }
\label{fig:CDcomp}
\end{center}
\end{figure}

\begin{figure}
\begin{center}
\includegraphics[scale=0.8]{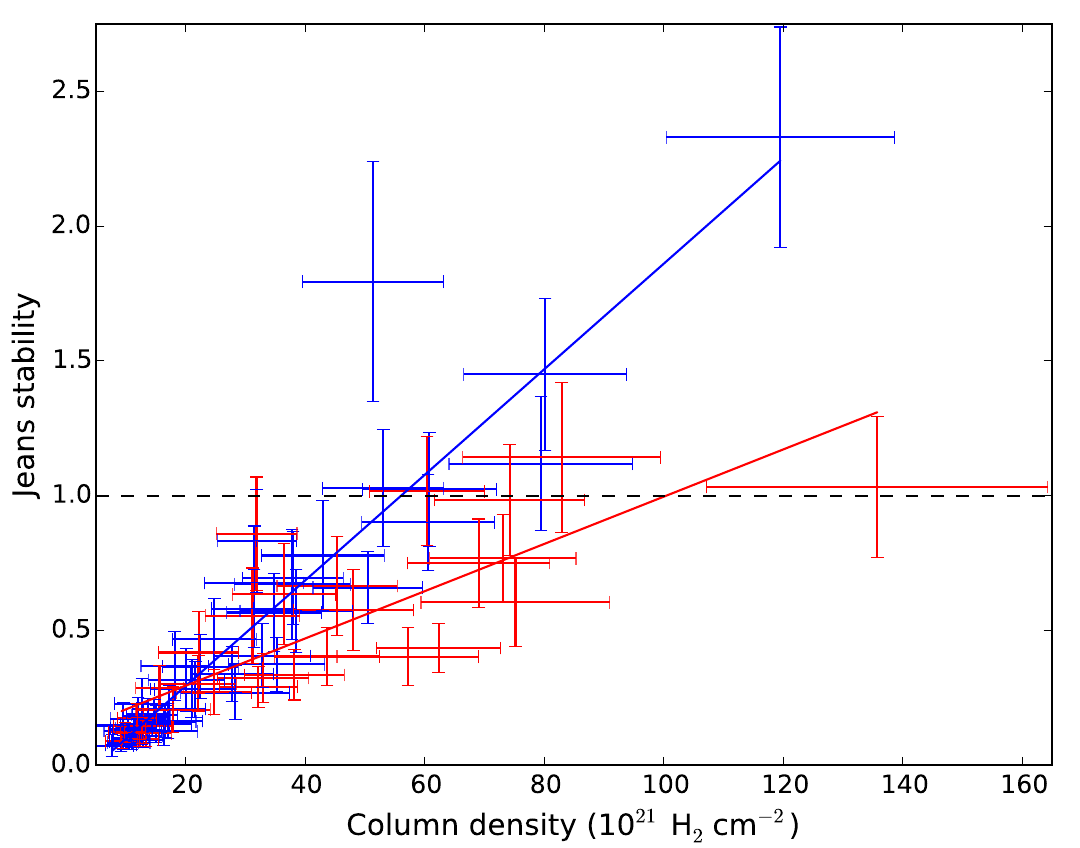}
\caption{Jeans stability, M$_{850}$/M$_{\mathrm{J}}$, as a function of column density. Exterior peripheral clumps (defined as having a mean \emph{Herschel} 70\,$\micron$ flux of less than 1000\,MJy/Sr) are marked in blue, interior clumps within the nebulosity Sh2-64 are marked in red. The unweighted linear regression fit to each population is marked as a line of the same colour.}
\label{fig:jeans}
\end{center}
\end{figure}

\section{Evidence for radiative heating}

\begin{figure*}
\begin{center}
\includegraphics[scale=0.75]{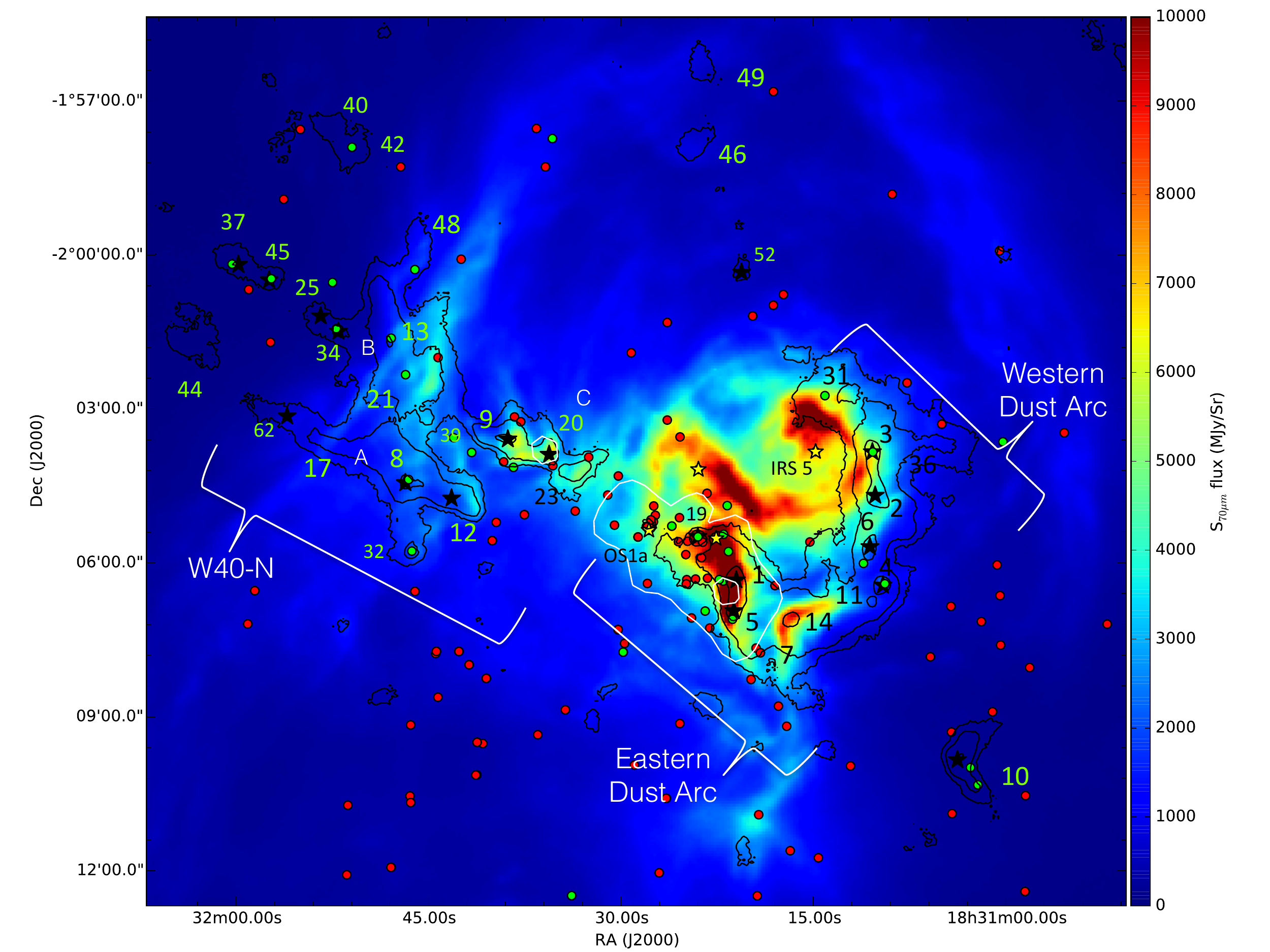}
\caption{Archival \emph{Herschel} 70\,$\micron$ flux density map of the W40 complex (colour scale). Morphological features of W40-N and the Dust Arc are labelled alongside major clumps detected in SCUBA-2 850\,$\micron$ emission (see Figure \ref{fig:clumps} for accurate clump positions). Black contours show SCUBA-2 850\,$\micron$ at the 5\,$\sigma$, 15\,$\sigma$ and 50\,$\sigma$ level.  White contours show archival VLA 21\,cm emission at 5\,$\sigma$ and 25\,$\sigma$ \citep{Condon:1998kx}. The four OB stars OS1a, 2a, 3a and IRS 5 are marked as yellow stars. YSOs from our composite catalogue are displayed. Protostars (Class 0/I) are marked as green and PMS-stars (Class II/III) are marked as red. Dense cores identified in Table \ref{tab:dense} are marked as black stars in their associated clump.}
\label{fig:H70}
\end{center}
\end{figure*}

The W40 complex contains OB stars that radiate photons with sufficient energy to heat and ionise a portion of the ISM.  Direct evidence of this heating is observed in the dust temperature maps presented and discussed in Sections 5 and 6. 

\begin{figure}
\begin{center}
\includegraphics[scale=1.15]{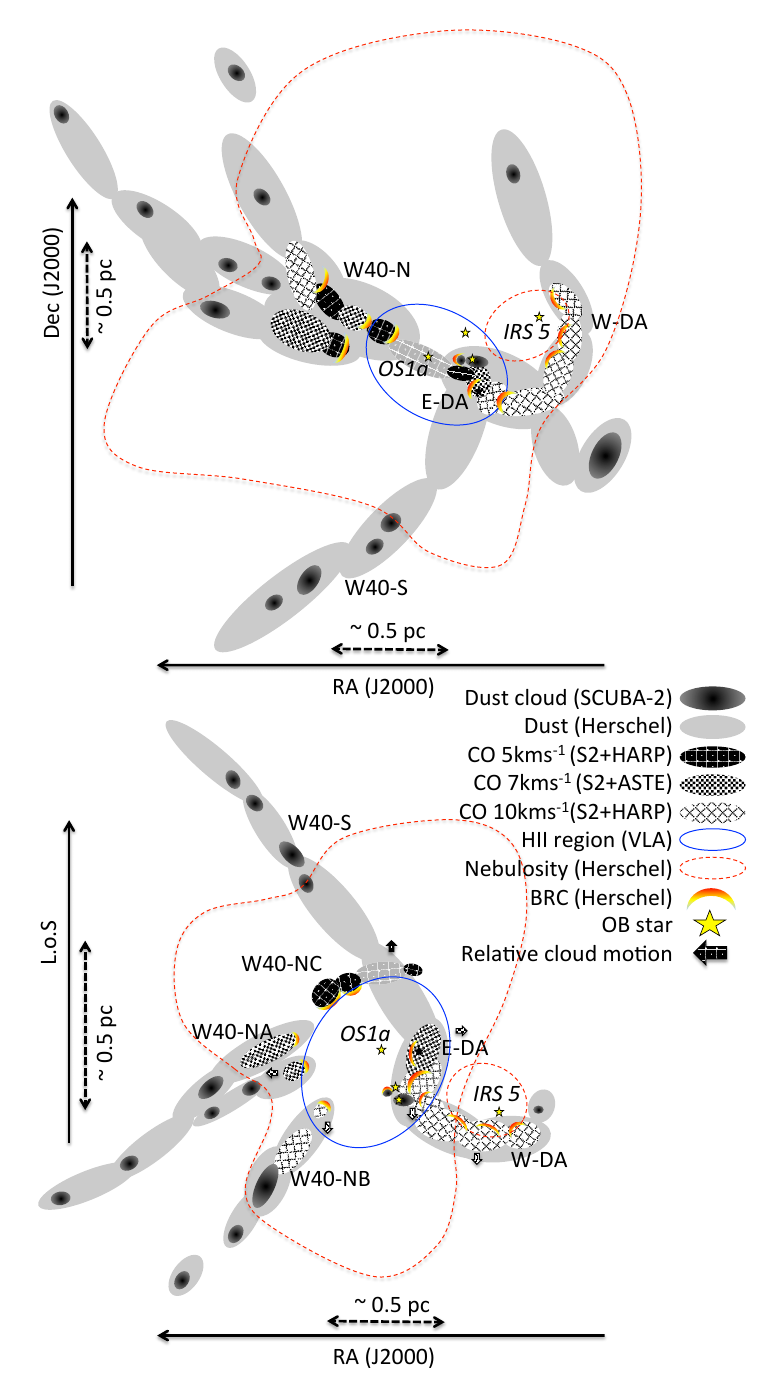}
\caption{A schematic diagram of the location of clouds and features within the W40 complex. 
The \textbf{upper} figure represents the RA/Dec view of the \emph{Herschel}/SCUBA-2 data presented in Figures \ref{fig:findingchart} and \ref{fig:maps} with cloud velocities observed by HARP and ASTE (Shimoikura et. al. 2015). Other features are explained in the key. The \textbf{lower} sketch shows the same region, this time viewed in line-of-sight vs RA space. 
The 5\,km\,s$^{-1}$ blue-shifted clouds are considered to be on the far side of the expanding \HII\ bubble and at a further distance. Likewise the 10\,km\,s$^{-1}$ red-shifted clouds are considered to be on the near side and at a closer distance. The 7\,km\,s$^{-1}$ clouds observed by ASTE are taken to be moving perpendicular and at an intermediate distance. 
Where clouds have no velocity information, distance is inferred through attachment to clouds with a known distance.
The location of the massive stars is given by Shuping et. al. (2012).}
\label{fig:schematic}
\end{center}
\end{figure}

The most prominent heating in the W40 complex occurs along the eastern Dust Arc, a very complex region of star-formation running from W40-SMM 19 to 14, and is the direct result of external heating from the nearby OS1a. The O9.5v star is primarily responsible for a mean clump temperature of 35$\pm$6\,K in this filament. By comparison, a number of isolated clumps, with or without protostars, found well outside of the nebulosity have a mean temperature of 15\,K, consistent with those derived for cores in Perseus using Bonnor-Ebert models (\citealt{Johnstone:2000fk}, \citealt{Kirk:2006vn}). Bright free-free emission observed in the eastern Dust Arc, as shown in Figures \ref{fig:21cm} and \ref{fig:SMM1}, is considered evidence of an interaction between the eastern Dust Arc and the \HII\ region. Figure \ref{fig:schematic} shows a possible configuration for this interaction. 





The western Dust Arc leads from W40-SMM 31 southeast towards W40-SMM 11, and includes the B1 star IRS 5 which appears to be producing a secondary nebulosity visible in \emph{Herschel} 70\,$\micron$ data (Figure \ref{fig:H70}) that is consistent with H$\alpha$ emission (\citeauthor{Mallick:2013kx} \citeyear{Mallick:2013kx}). A population of Class 0/I protostars is observed in the western Dust Arc by \cite{Maury:2011ys}, some of which coincide with dense cores W40-SMM 2, 3, 4, and 6. This filament lies well outside of the main stellar cluster associated with OS1a and has a YSO density of 22 YSO$\hbox{ pc$^{-2}$}$ which is comparable to W40-N. We observe a mean clump temperature of 26$\pm$3\,K for the western Dust Arc. Though this is warmer than the average clump in the W40 complex, it is notably cooler that the eastern Dust Arc (35$\pm$6\,K).

\cite{Shimoikura:2015kx} argues that the western Dust Arc is a shell of material forming around the \HII\ region of OS1a. However, our temperature maps lead us to believe that the Dust Arc is located significantly outside of the \HII\ region as we do not observe heating and free-free emission along its length to the extent of that observed in the eastern Dust Arc. Figure \ref{fig:schematic} presents a schematic layout of the W40 complex in RA/Dec/Line-of-sight space (by assimilating 3D information from the CO maps presented by us and \citeauthor{Shimoikura:2015kx} \citeyear{Shimoikura:2015kx}, and the distance measurements of \citealt{Shuping:2012ly}). 

Evidence from the Serpens MWC 297 region suggests that radiative heating from a primary generation of high-mass stars can raise clump temperature, and potentially suppress any subsequent star-formation in the neighbouring clumps \citep{Rumble:2015vn}. In Section 6.3 we have shown how heating from OS1a in the eastern Dust Arc is making these clumps, in particular W40-SMM 14, more stable to gravitational collapse (due to the increase of thermal support) than those in the western Dust Arc. By increasing the Jeans mass, the heating has the potential to skew the initial mass function to larger masses. However, as none of the clumps in the eastern Dust Arc have sufficient mass to exceed their enlarged Jeans mass, fragmentation under gravitational collapse is less likely to occur and the star formation rate may be suppressed. Given the continued radiative feedback from OS1a/expansion of the \HII\ region, it seems unlikely that clumps in the eastern Dust Arc will cool sufficiently to allow self-gravity to overwhelm thermal support and fragmentation to occur. We therefore conclude that it is likely that the eastern Dust Arc is less active in star-formation than the western Dust Arc.


In addition to the OB stars, there is an association of low-mass PMS-stars, observed by \cite{Kuhn:2010kl}, that also produces photons that may further externally heat the ISM. However, it is not possible to draw conclusions about the general significance of this mechanism given the dominance of the OB star heating in this region.  



Addressing internal sources of radiative feedback, and their influence on the ISM, requires an assessment of embedded star formation occurring within the clumps. 
The majority of Class 0/I YSOs, presented in Figure \ref{fig:H70}, are associated with SCUBA-2 dust emission. Those associated with a local peak are Class 0/I protostars and those aside from a local peak are either very low-mass protostars or potentially mis-identified (due to IR contamination from the OB association or additional dust along the line-of-sight) PMS-stars. Those protostars found outside of SCUBA-2 850\,$\micron$ 5$\sigma$ emission level are considered to be misidentified: false detections or edge-on  disks \citep{Heiderman:2015qf}. 
Considering clumps containing protostars (21$\pm$3\,K) compared to those without protostars (18$\pm$3\,K), we find there is no significant difference between the mean temperatures of the two populations (within the uncertainties). Figure \ref{fig:scatter1} shows how this trend is independent of distance from OS1a, though we note that some of the temperatures calculated for distances less than 1.2\,pc are likely influenced by OS1a.  

We consider the specific case of OS2b. The B4 star appears embedded in the tip of W40-SMM1 (see Figures \ref{fig:freefree3_6} and \ref{fig:H70}), where a peak in SCUBA-2 emission at 450 and 850\,$\micron$ is detected, suggesting that we are observing a Class 0/I YSO. The temperature at the position of OS2b is 21$\pm$2\,K (insert Figure \ref{fig:temp}). This value is comparable to the mean temperature of the dense cores in the Dust Arc (21\,K, W40-SMM1, 2, 3, 4, 5, 6). No significant variation in temperature is noted amongst this sample irregardless of whether or not they contain a protostar. These findings suggest there is no evidence that embedded stars, up to B4 in spectral type, significantly heat their immediate clump environment (given the resolution of JCMT and a constant $\beta$). This finding supports the conclusions of \cite{Foster:2009ve} that mid-B or later type stars have a relatively weak impact on their environment.

\section{Summary and conclusions}

We observed the W40 complex as part of the James Clerk Maxwell Telescope (JCMT) Gould Belt 
Survey (GBS) of nearby star-forming regions with SCUBA-2 at 450 and 850\,$\micron$. The 
$^{12}$CO 3\hbox{--}2 line at 345.796\,GHz was observed separately using HARP. The HARP 
data were used to subtract CO contributions to the SCUBA-2 850\,$\micron$ map. In addition, archival 
radio data from \cite{Condon:1998kx} and \cite{Rodriguez:2010bs} were examined to assess the 
large- and small-scale free-free flux contributions to both SCUBA-2 bands from the high-mass stars 
in the W40 complex OB association.

We produced maps of dust temperature and column density and estimated the Jeans stability, M$_{850}$/M$_{\mathrm{J}}$, of 
submillimetre clumps. In conjunction with a new composite YSO candidate catalogue, we probed 
whether dust heating is caused by internal or external sources and what implications this heating 
has for star-formation in the region. Throughout this paper we refer to the Dust Arc, W40-N, W40-S 
and isolated clumps as various morphological features of the W40 complex. A schematic layout of 
these features is presented in Figure \ref{fig:schematic}.

In addition to the catalogues presented here, all of the reduced datasets analysed in this paper are available at Doi xxxx. 


Our key results on the W40 complex are summarised as follows:

\begin{enumerate}
\item We find evidence for significant levels of $^{12}$CO 3\hbox{--}2 line emission in 
HARP data that contaminates the 850\,$\micron$ band by between 3 and 10\% of the flux 
seen in the majority of the filaments. In a minority of areas contamination reaches up to 20\%. 
Removing the $^{12}$CO 3\hbox{--}2 contamination significantly increases the 
calculated dust temperatures, beyond the calculated uncertainties. 

\item Free-free emission is observed on large and small scales. Large-scale emission from 
an existing \HII\ region (spectral index of $\alpha_{\mathrm{ff}}$ = -0.1) powered by the 
primary ionising star OS1a contributes 0.5\% of the peak flux at 450\,$\micron$ and 5\% at 850\,$\micron$. 
Small-scale emission from an \UCHII\ region around the Herbig star OS2a contributes 9 and 
12\% at 450 and 850\,$\micron$, respectively. Free-free emission for both large and small-scale 
sources was found to have a non-negligible, if limited, impact on dust temperature, often 
within the calculated uncertainties. 

\item 82 clumps were detected by \textsc{fellwalker} in the 850\,$\micron$ data and 21 of these 
have at least one protostar embedded within them. Clump temperatures range from 10 to 36\,K. 
The mean temperatures of clumps in the Dust Arc, W40-N and W40-S are 26$\pm$4, 21$\pm$4 
and 17$\pm$3\,K. The mean temperature of the isolated clumps is 15$\pm$2\,K. This result is 
consistent with temperatures observed in Serpens MWC 297 \citep{Rumble:2015vn} and other 
Gould Belt regions (\citeauthor{Sadavoy:2010ve} \citeyear{Sadavoy:2010ve} and Chen et al. 
submitted).

\item We find that clump temperature correlates with proximity to OS1a and the \HII\ region. 
We conclude that external radiative heating from the OB association is raising the temperature 
of the clumps. There is no evidence that embedded protostars are internally heating the 
filaments, though external influences may be masking such heating. As a result, the eastern Dust Arc 
has exceptionally high temperatures (mean 35$\pm$5\,K), Jeans masses (mean 17\,M$_{\odot}$), 
and Jeans stable clouds (mean M/M$_{\mathrm{J}}$ = 0.43). Partial radiative heating of the 
Dust Arc (internally or externally) has likely influenced the evolution of star-formation in the filament, 
favouring it in the cooler west, and potentially suppressing it in the warmer east. Globally, we find 
the population of clumps within the nebulosity Sh2-64 are more stable, as a function of peak 
column density, than that outside. 



\end{enumerate}

The W40 complex represents a high-mass star-forming region with a significant cluster of evolved 
PMS-stars and filaments forming new protostars from dense, starless cores. The region 
is complex and requires careful study to appreciate which radiative sources, from external 
and internal, are heating clumps of gas and dust. The region is dominated by an OB 
association that is powering an \HII\ region. In the near future we can expect this \HII\ region to 
expand and envelop many surrounding filaments. Within a few Myrs, we expect OS1a to go supernova. 
This event will likely have a cataclysmic impact on star-formation within the region. Any 
filament mass that has not been converted into stars, or eroded by the \HII\ region, may be 
destroyed at this point, bringing an end to star-formation in the W40 complex in its current form.  


\section{Acknowledgements}

The JCMT has historically been operated by the Joint Astronomy Centre on behalf of the Science and Technology Facilities Council of the United Kingdom, the National Research Council of Canada and the Netherlands Organisation for Scientific Research. Additional funds for the construction of SCUBA-2 were provided by the Canada Foundation for Innovation. The authors thank the JCMT staff for their support of the GBS team in data collection and reduction efforts. The program under which the SCUBA-2 data used in this paper were taken is MJLSG33. This work was supported by a STFC studentship (Rumble) and the Exeter STFC consolidated grant (Hatchell). The Starlink software \citep{Currie:2014ly} is supported by the East Asian Observatory. These data were reduced using a development version from December 2014 (version 516b455a). This research used the services of the Canadian Advanced Network for Astronomy Research (CANFAR) which in turn is supported by CANARIE, Compute Canada, University of Victoria, the National Research Council of Canada, and the Canadian Space Agency.  This research made use of APLpy, an open-source plotting package for Python hosted at http://aplpy.github.com, and Matplotlib, a 2D graphics package used for Python for application development, interactive scripting, and publication-quality image generation across user interfaces and operating systems. Herschel is an ESA space observatory with science instruments provided by European-led Principal Investigator consortia and with important participation from NASA. We would like to thank James Di Francesco for his internal review of this manuscript.


\bibliographystyle{mn2e}
\bibliography{bibliography}

\section{Affiliations}

$^{2}$Jeremiah Horrocks Institute, University of Central Lancashire, Preston, Lancashire, PR1 2HE, UK\\
$^{3}$NRC Herzberg Astronomy and Astrophysics, 5071 West Saanich Rd, Victoria, BC, V9E 2E7, Canada\\
$^{4}$Astrophysics Group, Cavendish Laboratory, J J Thomson Avenue, Cambridge, CB3 0HE\\
$^{5}$Kavli Institute for Cosmology, Institute of Astronomy, University of Cambridge, Madingley Road, Cambridge, CB3 0HA, UK\\
$^{6}$Joint Astronomy Centre, 660 N. A`oh\={o}k\={u} Place, University Park, Hilo, Hawaii 96720, USA\\
$^{7}$Department of Physics and Astronomy, University of Victoria, Victoria, BC, V8P 1A1, Canada\\
$^{8}$Department of Physics and Astronomy, University of Waterloo, Waterloo, Ontario, N2L 3G1, Canada  \\
$^{9}$LSST Project Office, 933 N. Cherry Ave, Tucson, AZ 85719, USA\\
$^{10}$Leiden Observatory, Leiden University, PO Box 9513, 2300 RA Leiden, The Netherlands\\
$^{11}$Max-Planck Institute for Astronomy, K{\"o}nigstuhl 17, 69117 Heidelberg, Germany\\
$^{12}$School of Physics and Astronomy, Cardiff University, The Parade, Cardiff, CF24 3AA, UK\\
$^{13}$European Southern Observatory (ESO), Garching, Germany\\
$^{14}$Jodrell Bank Centre for Astrophysics, Alan Turing Building, School of Physics and Astronomy, University of Manchester, Oxford Road, Manchester, M13 9PL, UK\\
$^{15}$Current address: Institute for Astronomy, ETH Zurich, Wolfgang-Pauli-Strasse 27, CH-8093 Zurich, Switzerland\\
$^{16}$Universit\'e de Montr\'eal, Centre de Recherche en Astrophysique du Qu\'ebec et d\'epartement de physique, C.P. 6128, succ. centre-ville, Montr\'eal, QC, H3C 3J7, Canada\\
$^{17}$James Madison University, Harrisonburg, Virginia 22807, USA\\
$^{18}$School of Physics, Astronomy \& Mathematics, University of Hertfordshire, College Lane, Hatfield, HERTS AL10 9AB, UK\\
$^{19}$Astrophysics Research Institute, Liverpool John Moores University, Egerton Warf, Birkenhead, CH41 1LD, UK\\
$^{20}$Imperial College London, Blackett Laboratory, Prince Consort Rd, London SW7 2BB, UK\\
$^{21}$Dept of Physics \& Astronomy, University of Manitoba, Winnipeg, Manitoba, R3T 2N2 Canada\\
$^{22}$Dunlap Institute for Astronomy \& Astrophysics, University of Toronto, 50 St. George St., Toronto ON M5S 3H4 Canada\\
$^{23}$Dept. of Physical Sciences, The Open University, Milton Keynes MK7 6AA, UK\\
$^{24}$The Rutherford Appleton Laboratory, Chilton, Didcot, OX11 0NL, UK.\\
$^{25}$UK Astronomy Technology Centre, Royal Observatory, Blackford Hill, Edinburgh EH9 3HJ, UK\\
$^{26}$Institute for Astronomy, Royal Observatory, University of Edinburgh, Blackford Hill, Edinburgh EH9 3HJ, UK\\
$^{27}$Centre de recherche en astrophysique du Qu\'ebec et D\'epartement de physique, de g\'enie physique et d'optique, Universit\'e Laval, 1045 avenue de la m\'edecine, Qu\'ebec, G1V 0A6, Canada\\
$^{28}$Department of Physics and Astronomy, UCL, Gower St, London, WC1E 6BT, UK\\
$^{29}$Department of Physics and Astronomy, McMaster University, Hamilton, ON, L8S 4M1, Canada\\
$^{30}$Department of Physics, University of Alberta, Edmonton, AB T6G 2E1, Canada\\
$^{31}$Max Planck Institute for Astronomy, K\"{o}nigstuhl 17, D-69117 Heidelberg, Germany\\
$^{32}$University of Western Sydney, Locked Bag 1797, Penrith NSW 2751, Australia\\
$^{33}$National Astronomical Observatory of China, 20A Datun Road, Chaoyang District, Beijing 100012, China

%

\newpage

\appendix

\section{SCUBA-2 data}

\begin{figure*}
\begin{center}
\includegraphics[scale=1.5]{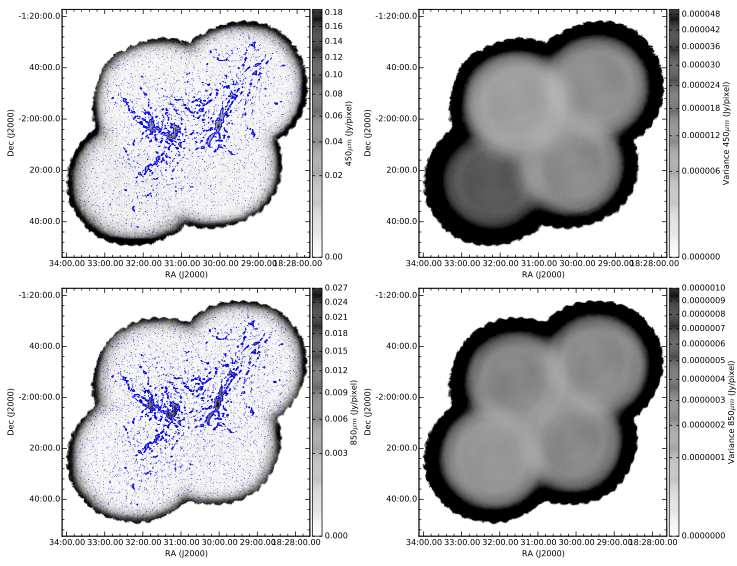}
\caption{SCUBA-2 IR2 reductions of Aquila region (\textbf{east} the W40 complex, \textbf{west} Serpens South) featuring data and variance maps at 450 and 850\,$\micron$. Contours show the outline of the SCUBA-2 external mask used in the reduction process.}
\label{fig:SCUBA-2}
\end{center}
\end{figure*}
Figure \ref{fig:SCUBA-2} shows complete SCUBA-2 450 and 850\,$\micron$ mosaics that include the W40 complex and the neighbouring Serpens South region (not considered to be physically connected to W40 complex). In addition we also present variance maps for the observations where the four PONG of varying noise levels have been mosaiced together. Plotted on the data maps are the SCUBA-2 external mask that are outlined in Section 2.

\section{Method for temperature maps}



A convolution kernel, $K\left ( A\Rightarrow B \right )$, is constructed from an analytical beam map or published 
beam model such that a point spread function (PSF$_A$) is mapped onto a PSF$_B$. Full details of the application of the kernel are given in \cite{Aniano:2011fk} and \cite{Pattle:2015ys}.

The post-CO and free-free subtracted SCUBA-2 450\,$\micron$ map is convolved to the resolution of the 850\,$\micron$ map using the SCUBA-2 convolution kernel using analytical beam map PSFs \citep{Dempsey:2013uq}. Pre- and post-convolution kernel maps are prepared to ensure a common pixel grid and clipping at a level of 5\,$\sigma$ at 450\,$\micron$, the details of which are covered in more detail by \cite{Aniano:2011fk}. Ratio maps are then prepared by dividing 450\,$\micron$ flux by 850\,$\micron$ flux as in the form of Equation \ref{eqn:temp}. Further details of these steps are discussed in \cite{Rumble:2015vn}. 

Noise on the input maps is propagated through the flux ratio algorithm to calculate the error per pixel 
on the temperature. \cite{Rumble:2015vn} calculate this solution analytically for their dual-beam cross-convolution 
method. Here, we calculate uncertainty through a Monte-Carlo process. Namely, random numbers are generated from a Gaussian distribution of width $\sigma_{\mathrm{input}}$, equivalent to the noise level of the map. Each number is added to a 
unique pixel on an otherwise flat 450\,$\micron$ map. The input maps are then convolved through the kernel algorithm and the spread of the distribution of pixels across the output map, $\sigma_{\mathrm{output}}$, is measured to recover the propagated uncertainty. 

A cut on temperature is made based on the propagated uncertainties. Pixels where the fractional error is 
greater than one standard deviation of a normal distribution were considered bad. This criterion had the effect at removing uncertain edge pixels (a known problem with this method; see \citeauthor{Rumble:2015vn} \citeyear{Rumble:2015vn}).

\begin{figure}
\begin{center}
\includegraphics[scale=0.35]{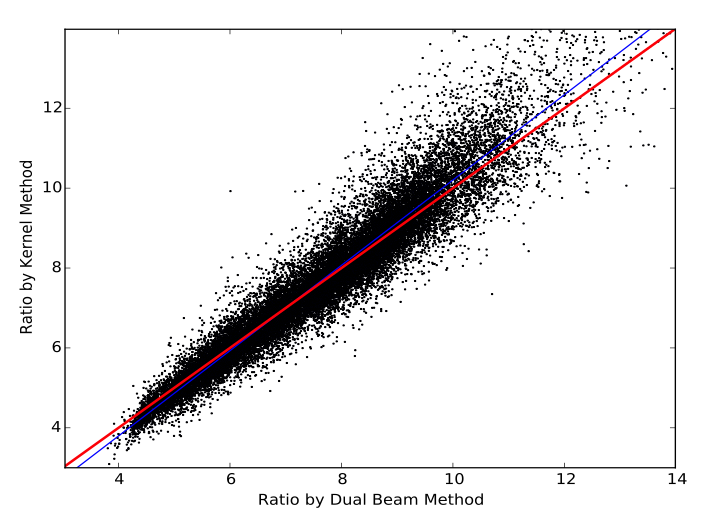}
\caption{Comparison of the pixel flux ratio values from the dual beam and kernel convolution methods. The blue line shows a 
linear regression fit to the distribution with a Pearson-correlation coefficient of 96\% to parity (red line). }
\label{fig:ratiocomp}
\end{center}
\end{figure}

We find that the ratio uncertainties produced by the kernel method of \cite{Aniano:2011fk} and \cite{Pattle:2015ys} 
are systematically higher than those produced by the dual-beam method in \cite{Rumble:2015vn} by a factor of 
approximately two. Increasing resolution inevitably increases uncertainty as the beam area samples fewer 
data points. The ratio of beam areas between the kernel method (14.8\arcsec) and the dual-beam method 
(19.9\arcsec) is 1.9, consistent with the observations. Appendix B (Figure \ref{fig:ratiocomp}) shows that, despite the increase 
in uncertainty, the absolute values of the dust flux ratios calculated by the two methods are found to be comparable 
with a Pearson-correlation co-efficent of 96\%.

We therefore conclude that the two methods produce consistent results for ratio and temperature, with the kernel 
method improving the resolution at the expense of the error in the results. 

\section{Spatial filtering}

Spatial filters necessary to reduce excess large-scale structure in the SCUBA-2 reductions 
were applied to the data using the {\sc cupid} tool \textsc{findback} (\citeauthor{Berry:2007vn} 
\citeyear{Berry:2007vn}, \citeauthor{Berry:2013uq} \citeyear{Berry:2013uq}). This tool works 
by twice filtering pixels (firstly to a minimum and secondly a maximum value) with respect to 
those pixels within a pre-defined spatial scale or `box' parameter which we refer to as the 
filter size. Using this method a lower envelope of the data is estimated. 

\begin{figure}
\begin{centering}
\includegraphics[scale=0.35]{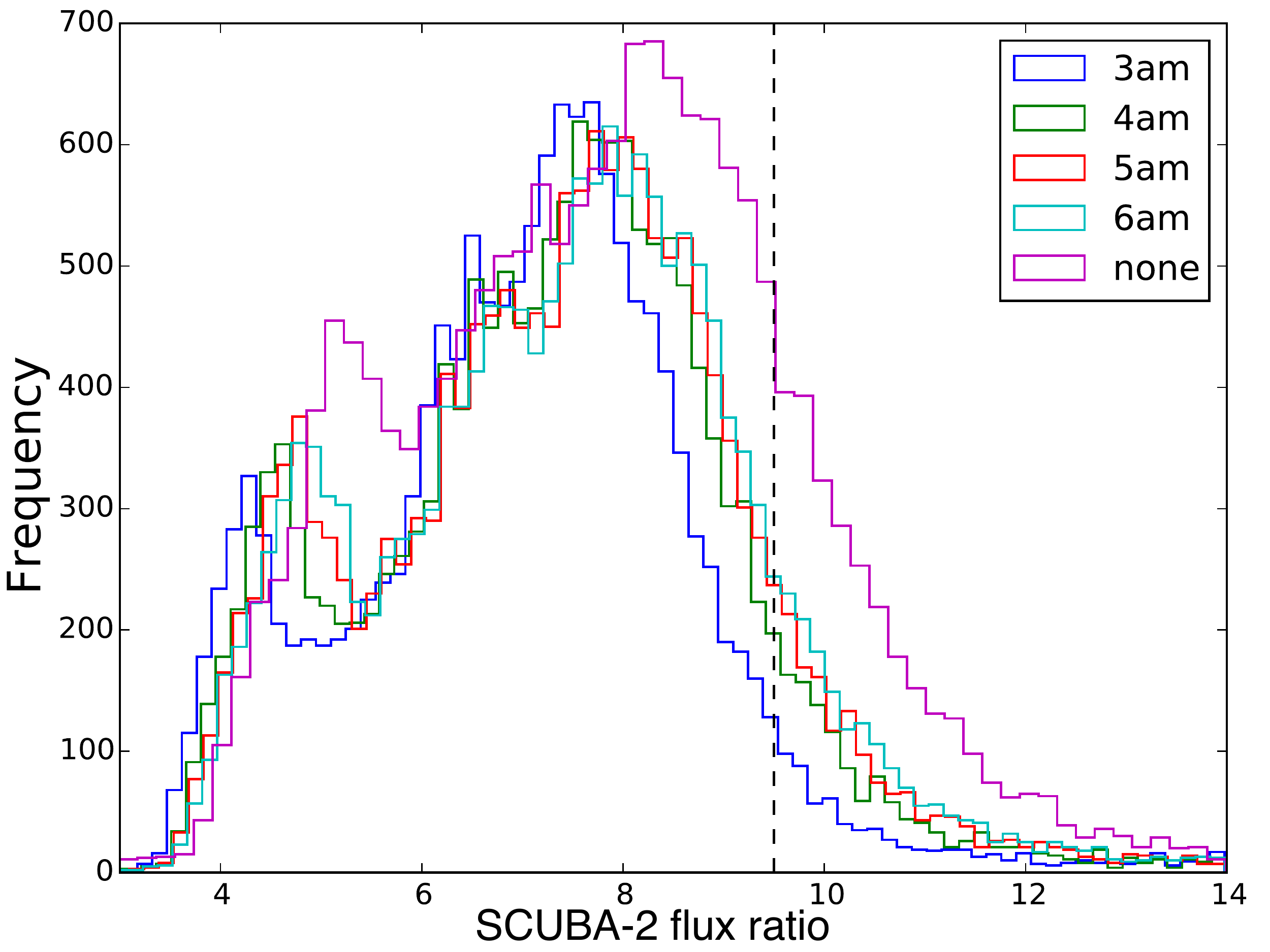}
\caption{Histograms of flux ratio for the Aquila CO subtracted SCUBA-2 reductions with varying spatial filters applied. Filter sizes, in arcminutes, are labelled on the plot. The flux ratios calculated with no filter are also included for comparison. Histograms are examined for total volume below 9.5, an arbitrary value above which ratios are considered unphysical.} 
\label{fig:filters}
\end{centering}
\end{figure}

The results of tests of various filter sizes are presented in Figure \ref{fig:filters}, alongside 
the unfiltered data, as the distribution of flux ratios for a control area of the W40 complex 
(W40-N). Larger filters were found to maximise the number of pixels in the data at the expense 
of excessive flux ratios. We set a criterion for the ideal filter that was a compromise 
between the highest absolute number of pixels retained (post filtering and 5\,$\sigma$ cut), 
in a sample area of the map and the lowest percentage of `bad' pixels (ratios higher than 
9.5). A scale size of 4\arcmin\ is the optimum solution with 5.7\% of pixels returning unphysical 
ratios (down from 28.8\% in the unfiltered data). See Section 4 for details of how flux ratio is 
calculated. The filtered maps of the W40 complex are shown in Figure \ref{fig:maps}.

\section{Temperature method comparison}

As outlined in Section 5, the degeneracy between temperature and $\beta$ can be removed with the inclusion of additional submillimeter wavelengths and through the fitting of a SED model to the data to independently determine these free parameters. 
\emph{Herschel} SPIRE and PACS bands have been used by \cite{Bontemps:2010fk}, \cite{Konyves:2010oq} and \cite{Konyves:2015uq} to calculate temperatures in the Aquila region using this method. 

SCUBA-2 bands have been incorporated alongside \emph{Herschel} bands to improve completeness at longer wavelengths by \cite{Sadavoy:2013qf} and Chen et al. (2016). The SCUBA-2 data reduction process explicitly filters large-scale structure that may trace lower density, warmer material so that dense cores and filaments can be targeted. This large-scale structure is retained in \emph{Herschel} observations. \cite{Sadavoy:2013qf} outlines additional spatial filtering and corrections to be made to \emph{Herschel} images so that they can be combined with SCUBA-2 images.

Incorporating additional wavelengths requires all the data to take the lowest common resolution. Using all available \emph{Herschel} bands requires a common resolution of 36.4\arcsec\ associated with the 500\,$\micron$ band. At this resolution, analysis is limited to a general overview of temperatures in the region. The 14.6\arcsec\ resolution achieved using the SCUBA-2 450/850\,$\micron$ flux ratio method (as outlined in Appendix B) allows temperatures to be studied down to the length scale of a protostellar core (0.05\,pc, \citealt{Rygl:2013ve}) under the assumption of constant $\beta$.

\begin{figure}
\begin{centering}
\includegraphics[scale=0.35]{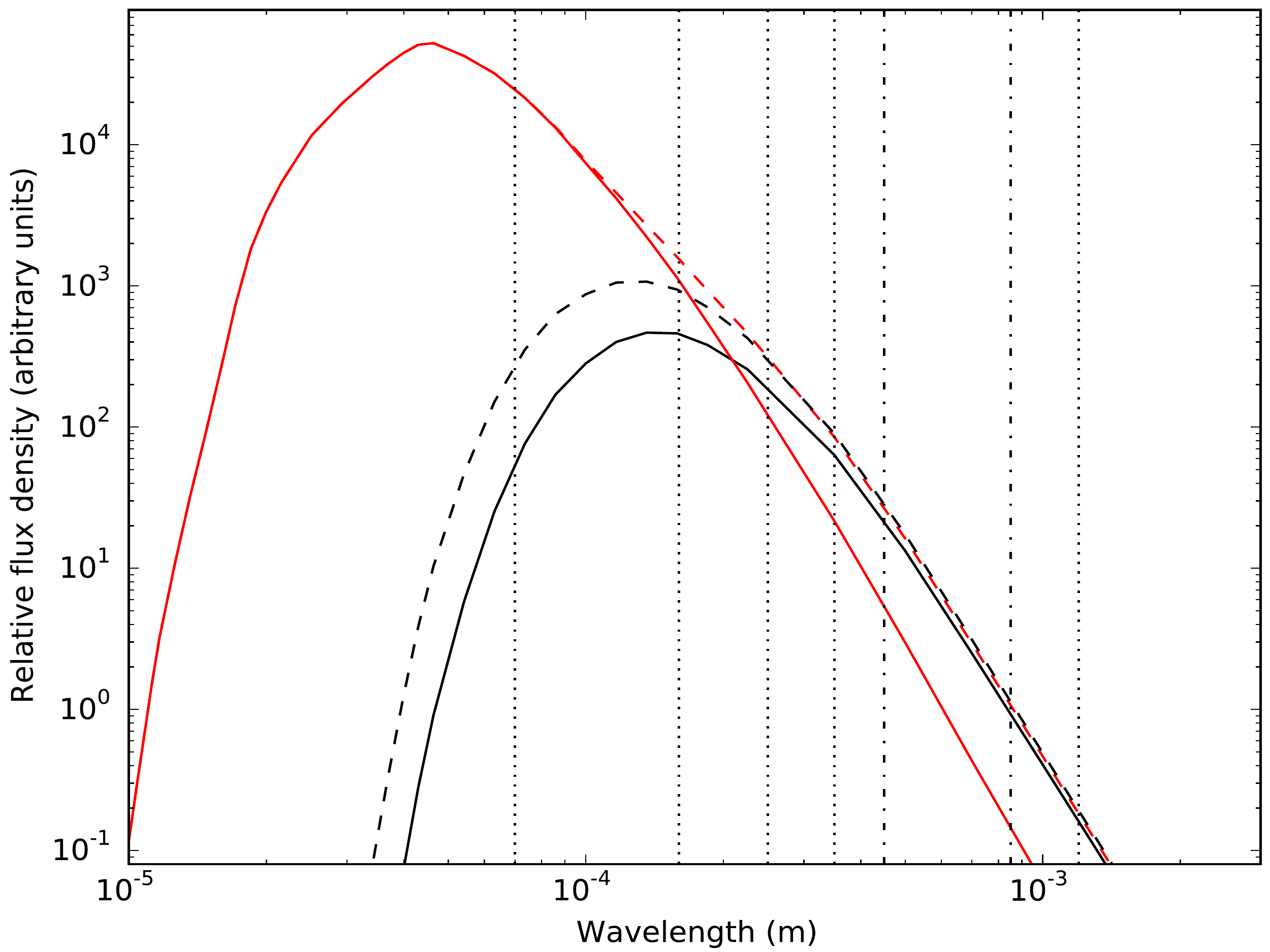}
\caption{Model opacity-modified blackbody SEDs for a cold (black, 15\,K) cloud and a hot (grey, red online, 50\,K) cloud with 3\% of the mass of the cold cloud. The red dashed SED represents the two cloud components combined and the black dashed SED represents how the combined SED resembles a single 17\,K cloud at SCUBA-2 450 and 850\,$\micron$ bands (dot-dashed vertical lines). The \emph{Herschel} 70, 160, 250 and 350\,$\micron$ bands, and the MAMBO 1.2\,mm band are also marked (dotted vertical lines). Note how the 50\,K component dominates the fluxes at 70 and 160 microns.} 
\label{fig:SED}
\end{centering}
\end{figure}

By using a more complete spectral range, SED fitting can discern multiple temperature cloud components in a way that the ratio method cannot. Should a target contain both hot ($\sim$50\,K or greater) and cold ($\sim$15\,K) cloud components, emission from the hot dust can dominate emission at shorter wavelengths (70 and 160\,$\micron$) at even a small fraction of the mass of the cold component, as illustrated in Figure \ref{fig:SED}. Fitting only the wavelengths longwards of the SED peak will preferentially sample the colder cloud components that dominate in mass \citep{Shetty:2009kx}. The SCUBA-2 450/850\,$\micron$ flux ratio is well positioned on the SED to measure the temperature of these colder components for the purposes of calculating their Jeans masses.

\section{Clumpfinding analysis}

We use the clump-finding algorithm \textsc{fellwalker} \citep{Berry:2015uq} to extract a catalogue 
of irregular clumps from the SCUBA-2 850\,$\micron$ data (Figure \ref{fig:maps}). Each clump then forms the 
basis of a mask for the temperature map (Figure \ref{fig:temp}) so that a single clump temperature can be 
found and used to calculate various additional properties. 

Details of how we apply \textsc{fellwalker}  to SCUBA-2 data are given in \cite{Rumble:2015vn}. Briefly, by 
setting the parameter MinDip = 3$\sigma$, \textsc{fellwalker} is tuned to break up large-scale continuous 
clouds with multiple bright cores into discrete clumps. Also, the Noise and MinHeight parameters are set to 5\,$\sigma$ 
and MaxJump is set to one pixel ensuring that all extracted clumps are significant detections but allowing 
for fragmentation peaks. Beam deconvolution of clump sizes was set to false. However, the flux weighted beam 
sizes were calculated using a FWHMbeam parameter of 1 pixel, or 3\arcsec. This was found to be equivalent to 
the FWHM given high SNR. At low SNR (less than 10) \textsc{fellwalker} will underestimate the size of the clump. 
A reduced beam size was used to ensure that point-like sources, such as isolated protostellar envelopes or discs, 
were not omitted by the algorithm. Finally, by setting MinPix to four pixels, a large number of single-pixel objects 
that were likely noise artefacts were removed from the catalogue.   

The original observations also include objects that are part of Serpens South which is located near to the W40 
complex on the sky. There is no physically defined point in the SCUBA-2 data that describes where the W40 complex ends 
and Serpens South begins and so we define an arbitrary cut off along the meridian RA(J2000) = 18:30:40 with all objects 
eastward belonging to the W40 complex and westward to Serpens South. Whilst this approach may risk associating some clumps with the wrong region, we estimate this will affect less than 5\% of members overall. 


\end{document}